\documentclass[11pt]{article}
\usepackage[dvipdfmx]{graphicx}   
\usepackage{amssymb}
\usepackage{array} 
\usepackage[authoryear]{natbib}
\usepackage{bibentry}
\usepackage{amsmath}
\usepackage{verbatim}
\usepackage{color}
\usepackage{makeidx}
\usepackage{bm}
\usepackage{appendix}
\usepackage{lscape}
\usepackage{floatrow}
\usepackage{subcaption}
\usepackage{rotating} 
\usepackage{tabularx}
\usepackage{booktabs}

\usepackage{titling}
\begin{document}
\title{Dynamic Factor Stochastic Volatility-in-Mean VAR for Large Macroeconomic Panels}
\author{\textsc{Daichi Hiraki} \\
%EndAName
\textit{Graduate School of Economics, University of Tokyo, Tokyo 113-0033, Japan} \\
\texttt{hdaichi397@gmail.com}
\and \textsc{Siddhartha Chib} \\
%EndAName
\textit{Olin Business School, Washington University, St Louis, USA} \\
\texttt{chib@wustl.edu} 
\and \textsc{Yasuhiro Omori} \\
%EndAName
\textit{Faculty of Economics, University of Tokyo, Tokyo 113-0033, Japan}\\
\texttt{omori@e.u-tokyo.ac.jp} 
}
\date{\today}
%\date{October 2024}
% Revised ??? 2024
\maketitle

\begin{abstract}
We develop a dynamic factor stochastic volatility-in-mean (SVM) specification for vector autoregressions (VARs) that embeds an SVM component within a dynamic factor stochastic volatility structure. A small number of latent volatility factors capture common movements in conditional variances, while volatility enters the conditional mean of the VAR. This specification allows time-varying uncertainty to influence macroeconomic dynamics through both second moments and expected outcomes while preserving tractability in large panels. We construct an efficient Markov chain Monte Carlo algorithm for estimation in this high-dimensional, non-Gaussian setting. Using quarterly data on twenty variables from the FRED-QD database, we compare predictive performance with the benchmark stochastic volatility VAR model. The dynamic factor SVM specification delivers superior forecasts for more variables during major macroeconomic disruptions such as the 2008 global financial crisis. The results indicate that allowing volatility to enter the mean captures an important transmission channel in macroeconomic dynamics.
\end{abstract}


\newpage
\section{Introduction}
\label{sec:introduction}

Time-varying volatility and volatility clustering are central features of macroeconomic and financial data. These features have led to a class of time series models in which conditional variances evolve stochastically. The stochastic volatility (SV) formulation of \cite{Taylor(08)}, which models log-volatility as a latent autoregressive process, underlies much of this work. Its practical relevance was firmly established by \cite{KimShephardChib(98)}, who developed a tractable estimation method and demonstrated that SV models provide a flexible alternative to deterministic specifications such as GARCH.

Over the past two decades, the availability of large macroeconomic panel datasets has motivated multivariate SV models. To address dimensionality and identification challenges, these models typically impose a dynamic factor structure in which a small number of latent volatility processes capture comovement across many variables. Factor SV models, beginning with \cite{AguilarWest(00)} and developed further by \cite{ChibNadrdariShephard(06)}, provide a tractable framework for large systems. Subsequent contributions incorporate leverage effects and high-dimensional estimation strategies (\cite{IshiharaOmori(17)}, \cite{YamauchiOmori(23)}).

In parallel, another strand of research allows volatility to enter the conditional mean equation. These stochastic volatility-in-mean (SVM) models capture risk--return trade-offs (\cite{Black(76)}) and have been used to study excess returns and volatility feedback effects (\cite{KoopmanHol(02)}, \cite{HirakiChibOmori(25)}). Extensions to multivariate vector autoregressions (SVMVAR) permit volatility to influence macroeconomic dynamics directly (\cite{MumtazZanetti(13)}, \cite{CarrieroClarkMarcellino(18)}, 
\cite{AriasRamirezShin(23)},
\cite{CrossHouKoopPoon(23)}, \cite{DavidsonHouKoop(25)}). Empirical evidence in these studies shows that allowing volatility to enter the mean improves forecasting performance during periods of elevated uncertainty.

These two developments, dynamic factor stochastic volatility models and volatility-in-mean models, have progressed largely independently. Factor SV models scale to large panels but restrict volatility to affect only second moments. SVM and SVMVAR models allow volatility to affect conditional means but do not provide a scalable structure for high-dimensional systems. A unified framework that combines these features for large macroeconomic panels has not yet been developed. The goal of this paper is to provide such a framework. We develop a dynamic factor stochastic volatility-in-mean (DFSVM) model for VAR models that embeds an SVM component within a dynamic factor SV structure. A small number of latent volatility factors captures common movements in conditional variances, while volatility enters the conditional mean of the VAR. This specification allows time-varying uncertainty to influence macroeconomic dynamics through both second moments and expected outcomes while preserving tractability in large panels.

For generality, we further incorporate leverage by allowing correlation between innovations in the conditional mean and innovations in volatility. The resulting specification, denoted DFSVML, captures the asymmetric responses frequently observed in macro-financial data. The joint presence of dynamic factors, volatility-in-mean effects, and leverage yields a high-dimensional non-Gaussian likelihood. We develop an efficient Markov chain Monte Carlo (MCMC) algorithm that extends mixture-based samplers for stochastic volatility models to this setting.

We evaluate the proposed framework using quarterly data on twenty variables from the FRED-QD database and compare its predictive performance to benchmark stochastic volatility VAR models. The results show that the DFSVM model adapts more effectively to sharp monetary policy shifts and major macroeconomic shocks during the 2008 global financial crisis (GFC). 
%and the post-COVID-19 period. 
In addition, the static FSVM specification outperforms benchmark models during the COVID-19 pandemic period. These findings indicate that restricting volatility to affect only conditional variances omits an important transmission channel in macroeconomic dynamics.

The remainder of the paper proceeds as follows. Section~\ref{sec:DFSVML model} presents the model and estimation strategy. Section~\ref{sec:Data} describes the data and reports in-sample results. Section~\ref{sec:out-of-sample result} evaluates predictive performance across economic regimes and across models. Section~\ref{sec:conclusion} concludes. Supplementary material provides computational details.

\section{Dynamic Factor SV–In–Mean VAR Model}
\label{sec:DFSVML model}
\subsection{DFSVM and DFSVML Models}
We first propose a dynamic factor stochastic volatility–in–mean VAR (DFSVM) model. The DFSVM model relates the $p$–dimensional observed vector $\bm{y}_t = (y_{1t},\dots,y_{pt})'$ to a $q$–dimensional latent factor $\bm{f}_t = (f_{1t},\dots,f_{qt})'$ and a $(p+q)$–dimensional log–volatility state $\bm{h}_t = (h_{1t},\dots,h_{p+q,t})'$, with $q<p$.  Let us partition $\bm{h}_t = (\bm{h}_{1t}',\bm{h}_{2t}')'$ where the $p\times 1$–vector $\bm{h}_{1t}$ for $\bm{y}_t$ and the $q\times 1$–vector $\bm{h}_{2t}$ for $\bm{f}_t$. 
Further, we let $N_{m}(\bm{\mu},\,\mathbf{\Sigma})$ denote $m$–dimensional normal distribution with mean $\bm{\mu}$ and covariance matrix $\mathbf{\Sigma}$.
The observation and factor equations are given jointly by
\begin{align}
\bm{y}_{t}
  &= \sum_{\ell=1}^{L}\mathbf{B}_{\ell}\,\bm{y}_{t-\ell}
    + \mathbf{B}\bm{f}_{t}
    + \mathbf{\Lambda}_{t}\,\bm{\beta}
    + \mathbf{V}_{1t}^{1/2}\bm{\epsilon}_{1t}, 
    &\bm{\epsilon}_{1t} &\sim  \mbox{\it i.i.d.} \ N_{p}(\bm{0},\,\mathbf{I}_{p}), \notag\\
\bm{f}_{t}
  &= \bm{\gamma} +  \mathbf{\Psi}\,(\bm{f}_{t-1} - \bm{\gamma})
    + \mathbf{V}_{2t}^{1/2}\,\bm{\epsilon}_{2t},
    &\bm{\epsilon}_{2t} &\sim  \mbox{\it i.i.d.} \ N_{q}(\bm{0},\,\mathbf{I}_{q}).
\label{eq:dfs_observation_factor}
\end{align}
where the abbreviation $i.i.d.$ stands for independent and identically distributed, and $\mathbf{I}_{p}$ denotes a $p$-dimensional identity matrix. For simplicity, we assume $\bm{f}_0 \equiv \bm{\gamma} \in \mathbb{R}^q$. The $\mathbf{B}_{\ell}$ is a $p\times p$ autoregressive coefficient matrix at lag~$\ell$, and $\mathbf{B}$ is the full $p\times q$ factor loading matrix, which is identified up to column sign and permutation (see \cite{ChanEisenstatYu(22)}).  
The diagonal matrix $\mathbf{\Psi} = \mathrm{diag}(\psi_{1},\dots,\psi_{q})$ with $|\psi_{j}|<1$ imposes AR(1) persistence on each factor $f_{jt}$.  The stochastic volatility–in–mean effect enters through 
\begin{equation}
\mathbf{\Lambda}_{t}
  = \mathrm{diag}\bigl(\exp(h_{1t}/2),\dots,\exp(h_{pt}/2)\bigr),
\label{eq:dfs_lambda}
\end{equation}
so that the $i$-th element of $\bm{\beta}\in\mathbb{R}^{p}$ scales the impact of its corresponding inherent log–volatility on the dependent variable. The in-mean effect from one volatility to other means will be captured through the stochastic volatility of the factor $\bm{f}_t$. 
The time–varying covariance matrices are 
\begin{align}
\mathbf{V}_{1t} &= \mathrm{diag}\bigl(\exp(h_{1t}),\dots,\exp(h_{pt})\bigr), 
&\mathbf{V}_{2t} &= \mathrm{diag}\bigl(\exp(h_{p+1,t}),\dots,\exp(h_{p+q,t})\bigr),
\label{eq:dfs_V}
\end{align}
and log–volatilities are assumed to follow an AR(1) process:
\begin{align}
\bm{h}_{t+1}
  &= \bm{\mu}
    + \mathbf{\Phi}\,(\bm{h}_{t}-\bm{\mu})
    + \bm{\eta}_{t},
    &\bm{\eta}_{t} &\sim  \mbox{\it i.i.d.} \ N_{p+q}(\bm{0},\,\mathbf{\Sigma}).
\label{eq:dfs_volatility}
\end{align}
In Equation \eqref{eq:dfs_volatility}, $\bm{\mu}=(\mu_{1},\dots,\mu_{p+q})'$ is the unconditional log-volatility vector --- here we set $\mu_{p+1}=\dots=\mu_{p+q}=0$ for identification --- $\mathbf{\Phi}=\mathrm{diag}(\phi_{1},\dots,\phi_{p+q})$ with $|\phi_{i}|<1$ governing the persistence, and $\mathbf{\Sigma}=\mathrm{diag}(\sigma_{1}^{2},\dots,\sigma_{p+q}^{2})$ is the covariance matrix of $\bm{\eta}_{t}$. Initial states follow the stationary distribution, $h_{i1}\sim N\bigl(\mu_{i},\sigma_{i}^{2}/(1-\phi_{i}^{2})\bigr)$ for $i=1,\ldots,p+q$.
Thus, DFSVM model is defined by Equations (\ref{eq:dfs_observation_factor})--(\ref{eq:dfs_volatility}).

%\noindent
Secondly, we extend the above DFSVM model to incorporate the leverage effect, which we call DFSVML model. Each pair $(\epsilon_{i t},\eta_{i t})$ for $i=1,\dots,p+q$ is assumed to follow a bivariate normal distribution with a correlation parameter $\rho_{i}\in(-1,1)$ in the DFSVML model:
\begin{align}
\begin{pmatrix}\epsilon_{i t}\\\eta_{i t}\end{pmatrix}
  &\sim \mbox{\it i.i.d.} \ N_{2}  \left(\bm{0},\,
    \begin{pmatrix}
      1 & \rho_{i}\,\sigma_{i}\\
      \rho_{i}\,\sigma_{i} & \sigma_{i}^{2}
    \end{pmatrix}\right).
\label{eq:dfs_leverage}
\end{align}

\noindent
{\it Remark}. There are several alternative specifications for $\mathbf{\Lambda}_{t}$. We use the standard deviation $\exp(h_{it}/2)$ in the mean equation, rather than the variance $\exp(h_{it})$, for $\mathbf{\Lambda}_{t}$, to match the units of the outcome variable as considered in \cite{HirakiChibOmori(25)}. We could also consider the stochastic volatility-in-mean in the factor equation, but it is found to be unsupported in terms of forecasting performance in our empirical studies in Section \ref{sec:out-of-sample result}.
\subsection{Prior Distributions}
Let $\bm{\rho} = (\rho_1, \ldots, \rho_{p+q})'$, $\bm{f} = \{\bm{f}_t\}_{t=1,\ldots,n}$, $\bm{h} = \{\bm{h}_t\}_{t=1,\ldots,n}$, $\bm{y} = \{\bm{y}_t\}_{t=1,\ldots,n}$, and let $\mathbf{B}_{i\cdot}$ denote the $i$-th row of $\mathbf{B}$. Define $\mathbf{\bar{B}}$ to be the $p \times (pL)$ coefficients matrix, $\mathbf{\bar{B}} = (\mathbf{B}_1,...,\mathbf{B}_L)$, and $\mathbf{\bar{B}}_{i\cdot}$ to be the $i$-th row of $\mathbf{\bar{B}}$. The prior distributions for model parameters are assumed as follows.
\begin{eqnarray*}
    &&
    \mathbf{B}_{i\cdot} \sim N_q(\bm{0}, \mathbf{I}_q), 
    \quad \mathbf{\bar{B}}_{i\cdot} \sim N_{pL}(\bm{m}_{\mathbf{\bar{B}}_{i\cdot}}, \mathbf{S}_{\mathbf{\bar{B}}_{i\cdot}}), 
\quad i=1, \ldots, p, \\
    &&\bm{\gamma} \sim N_q(\bm{m}_{\bm{\gamma}}, \mathbf{S}_{\bm{\gamma}}), \quad \frac{\psi_j+1}{2} \sim Beta(a_{\psi}, b_{\psi}), \quad j=1, \ldots, q, \quad \bm{\beta} \sim N_p(\bm{m}_{\bm{\beta}}, \mathbf{S}_{\bm{\beta}}), \\
    &&\mu_k \sim N(m_\mu, v_\mu^2), \quad \sigma_k^2 \sim IG(n_{\sigma^2}/2, d_{\sigma^2}/2), \\
    &&\frac{\phi_k+1}{2} \sim Beta(a_\phi, b_\phi),\quad \rho_k \sim U(-1,1), \quad k=1, \ldots , p+q.
\end{eqnarray*}
where $Beta(\cdot,\cdot)$, $IG(\cdot,\cdot)$, and $U(\cdot,\cdot)$ denote beta, inverse gamma, and uniform distributions.
The prior distributions for $\psi_j$ and $\phi_k$ ensure the stationarity of the latent factor and volatility processes, respectively.

To identify $\mathbf{B}$, we may need further constraints, such as sign restrictions. However, our focus here is on forecasting, not on estimating the impulse response function. Although it is theoretically true that we face the label switching problem, this rarely occurs in practice, and we do not impose any constraints for $\mathbf{B}$ (see Supplementary Material \ref{sec:posterior_summary} for checking the possible label switching problem in our empirical studies). 

For the VAR coefficients $\mathbf{\bar{B}} = (\mathbf{B}_1, \dots, \mathbf{B}_L)$, we adopt the Minnesota prior as described in \cite{BanburaGiannoneReichlin(10)}, \cite{CarrieroClarkMarcellino(16)}, \cite{CarrieroClarkMarcellino(19)}, \cite{CrossHouPoon(20)}, \cite{CrossHouKoopPoon(23)}, %\cite{ChanKoopYu(24)}, \cite{ChanYuZhang(25)} 
and so on. Specifically, for each $\ell = 1, \dots, L$ and each $i,j = 1,\dots,p$, the $(i,j)$ element of $\mathbf{B}_\ell$ is given the prior:
\begin{align*}
    \text{Var} ( (\mathbf{B}_\ell)_{i,j} ) \sim 
    \begin{cases}
        \displaystyle \frac{\pi_1}{\ell^2} & \text{if } i = j, \\
        \displaystyle \frac{\pi_1 \pi_2 s_i^2}{\ell^2 s_j^2} & \text{if } i \neq j,
     \end{cases}
\end{align*}
where $s_i^2$ is the residual variance of the $i$-th series estimated from an AR($L$) model. This specification encourages the shrinkage of higher-order lags and cross-variable effects, reflecting prior beliefs in sparsity and parsimony in large VAR systems. The shrinkage parameters $\pi_1$ (overall) and $\pi_2$ (cross-variable) are treated as hyperparameters with uniform priors. The superiority of this Minnesota prior formulation over alternative shrinkage priors, such as the Dirichlet–Laplace or Horseshoe priors, has been demonstrated in forecasting exercises by \citet{CrossHouPoon(20)}.

\subsection{Posterior Distribution and MCMC algorithm}
Let $\bm{\theta} = (\mathbf{\bar{B}}, \mathbf{B}, \bm{\gamma}, \bm{\psi}, \bm{\beta}, \bm{\alpha})$ and $\bm{\alpha} = (\bm{\mu}, \bm{\phi}, \bm{\sigma^2}, \bm{\rho}$), with corresponding prior probability density functions $\pi(\bm{\theta})$ and $\pi(\bm{\alpha})$.
Noting that
\begin{equation*}
    \begin{pmatrix}
        \bm{\epsilon}_{1t} \\
        \bm{\epsilon}_{2t} \\
        \bm{\eta}_{t} \\
    \end{pmatrix}
 %   \overset{i.i.d.}{\sim} N_{2(p+q)}(\bm{0}, \mathbf{\Sigma}^*), \quad 
   \sim \ \mbox{\it i.i.d.} \ N_{2(p+q)}(\bm{0}, \mathbf{\Sigma}^*), \quad 
    \mathbf{\Sigma}^* = 
    \begin{pmatrix}
        \mathbf{I}_{p+q} & \mathbf{\Sigma}_{\epsilon \eta} \\
        \mathbf{\Sigma}_{\epsilon \eta} & \mathbf{\Sigma}
    \end{pmatrix},
    \quad t = 1,\ldots,n.
\end{equation*}
where $\mathbf{\Sigma}_{\epsilon \eta} = \text{diag}(\rho_1 \sigma_1, \ldots, \rho_{p+q} \sigma_{p+q})$,
the joint posterior probability density conditioned on $\bm{y} = (\bm{y_1}, \ldots, \bm{y}_n)$ is given by
\begin{eqnarray*}
\pi(\bm{\theta}, \bm{\alpha}, \bm{f}, \bm{h}|\bm{y}) 
    &\propto &|\mathbf{\Sigma}|^{-\frac{n-1}{2}} \exp \left[ -\frac{1}{2} \sum_{t=1}^n \Big\{ \sum_{j=1}^{p+q} h_{jt} + \bm{\nu}_t' (\mathbf{P}_t \mathbf{\Sigma}^* \mathbf{P}_t')^{-1} \bm{\nu}_t \Big\} \right] \\
    & &\times \prod_{j=1}^{p+q} \frac{\sqrt{1-\phi_j^2}}{\sigma_j} \exp \left\{ -\frac{(1-\phi_j^2)(h_{j1}-\mu_j)^2}{2 \sigma_j^2} \right\} 
    \times \pi(\bm{\theta}) \pi(\bm{\alpha}),
\end{eqnarray*}
where
\begin{align*}
    &\bm{\nu}_t =
    \begin{pmatrix}
        \tilde{\bm{y}}_t \\
        \tilde{\bm{f}}_t \\
        \bm{h}_{t+1} - \bm{\gamma} - \mathbf{\Phi} (\bm{h}_t - \bm{\gamma})
    \end{pmatrix},
    \quad t = 1, \ldots,  n-1, \quad \bm{\nu}_n = 
    \begin{pmatrix}
        \tilde{\bm{y}}_n \\
        \tilde{\bm{f}}_n
    \end{pmatrix}, \\
    &\mathbf{P}_t =
    \begin{pmatrix}
        \mathbf{V}_{1t}^{1/2} & \mathbf{O} & \mathbf{O} \\
        \mathbf{O} & \mathbf{V}_{2t}^{1/2} & \mathbf{O} \\
        \mathbf{O} & \mathbf{O} & \mathbf{I}_{p+q}
    \end{pmatrix},
    \quad t = 1, \ldots,  n-1, \quad \mathbf{P}_n =
    \begin{pmatrix}
        \mathbf{V}_{1t}^{1/2} & \mathbf{O} & \mathbf{O} \\
        \mathbf{O} & \mathbf{V}_{2t}^{1/2} & \mathbf{O}
    \end{pmatrix},
    \\
    &\tilde{\bm{y}}_t = \bm{y_t} - \sum_{\ell=1}^{L}\mathbf{B}_{\ell}\,\bm{y}_{t-\ell} - \mathbf{B}\bm{f}_{t} - \mathbf{\Lambda}_{t}\,\bm{\beta}, \quad t=1,\ldots,n, \\
    & \tilde{\bm{f}}_t = \bm{f}_t - \bm{\gamma} - \mathbf{\Psi} (\bm{f}_{t-1} - \bm{\gamma}), \quad t = 2,\ldots,n,
    \quad \tilde{\bm{f}}_1 = \bm{f}_1 - \bm{\gamma}.
\end{align*}

We implement the MCMC algorithm to estimate the posterior distributions of the parameters and latent variables in the following seven blocks. 
Let $\bm{\theta}_{\backslash \bm{\beta}}$, for example, denote $\bm{\theta}$ excluding $\bm{\beta}$.
\begin{enumerate}
    \item Generate $\bm{\beta} \sim \pi(\bm{\beta} | \bm{\theta}_{\backslash \bm{\beta}}, 
    %= (\mathbf{\bar{B}}, \mathbf{B}, \bm{\gamma}, \bm{\psi}, \bm{\alpha}),
    \bm{f}, \bm{h}, \bm{y})$.
    
    \item Generate $(\bm{h}, \bm{\alpha}) \sim \pi(\bm{h}, \bm{\alpha}|\bm{\theta}_{\backslash \bm{\alpha}}, 
    \bm{f}, \bm{y})$.

    \item Generate $(\mathbf{B}, \mathbf{\bar{B}}) \sim \pi(\mathbf{B}, \mathbf{\bar{B}}| \bm{\theta}_{\backslash (\mathbf{B}, \mathbf{\bar{B}})},
    \bm{f}, \bm{h}, \bm{y})$.

    \item Generate $\bm{\psi} \sim \pi(\bm{\psi}| \bm{\theta}_{\backslash \bm{\psi}}, 
    \bm{f}, \bm{h}, \bm{y})$.

    \item Generate $\bm{\gamma} \sim \pi(\bm{\gamma} | \bm{\theta}_{\backslash \bm{\gamma}},
    \bm{f}, \bm{h}, \bm{y})$.

    \item Generate $\bm{f} \sim \pi(\bm{f}| \bm{\theta}, \bm{h}, \bm{y})$.

    \item Go to Step 2.
\end{enumerate}
Details are described in Supplementary Material \ref{sec:MCMC algorithm}. 

\section{Application to U.S. Macroeconomic data}
\label{sec:Data}
\subsection{Data}

\begin{table}[H]
%    \footnotesize
    \scriptsize
    \centering
    \begin{tabularx}{\textwidth}{rl X l}
        \toprule
        \textbf{\#} & 
        \textbf{FRED-ID} & \textbf{Series Name} & \textbf{Transformation} \\
        \midrule
        1 & GDPC1 & Real Gross Domestic Product & $\Delta \log$ \\
        2 & PCECTPI & Personal Consumption Expenditures: Chain-type Price Index & $\Delta \log$ \\
        3 & FEDFUNDS & Effective Federal Funds Rate & No transformation \\
        4 & PCECC96 & Real Personal Consumption Expenditures & $\Delta \log$ \\
        5 & CMRMTSPLx & Real Manufacturing and Trade Industries Sales & $\Delta \log$ \\
        6 & INDPRO & Industrial Production Index & $\Delta \log$ \\
        7 & CUMFNS & Capacity Utilization: Manufacturing & No transformation \\
        8 & UNRATE & Civilian Unemployment Rate & No transformation \\
        9 & PAYEMS & All Employees: Total Nonfarm & $\Delta \log$ \\
        10 & CES0600000007 & Average Weekly Hours of Production and Nonsupervisory Employees: Goods-Producing & $\log$ \\
        11 & CES0600000008 & Average Hourly Earnings of Production and Nonsupervisory Employees: Goods-Producing & $\Delta \log$ \\
        12 & WPSFD49207 & Producer Price Index by Commodity for Final Demand: Finished Goods & $\Delta \log$ \\
        13 & PPIACO & Producer Price Index for All Commodities & $\Delta \log$ \\
        14 & AMDMNOx & Real Manufacturers’ New Orders: Durable Goods & $\Delta \log$ \\
        15 & HOUST & Housing Starts: Total: New Privately Owned Housing Units Started & $\log$ \\
        16 & S\&P 500 & S\&P’s Common Stock Price Index: Composite & $\Delta \log$ \\
        17 & EXUSUKx & U.S./U.K. Foreign Exchange Rate & $\Delta \log$ \\
        18 & TB3SMFFM & 3-Month Treasury Constant Maturity Minus Federal Funds Rate & No transformation \\
        19 & T5YFFM & 5-Year Treasury Constant Maturity Minus Federal Funds Rate & No transformation \\
        20 & AAAFFM & Moody’s Seasoned Aaa Corporate Bond Minus Federal Funds Rate & No transformation \\
        \bottomrule
    \end{tabularx}
    \caption{Description of the variables. The number, the FRED mnemonic, the series names and the transformation applied to each series. The difference in logarithms was multiplied by 100 when transformation is ``$\Delta \log$".}
    \label{tab:FRED-QD}
\end{table}
This section details the dataset used for the empirical application. 
We use a balanced panel of 20 quarterly U.S. macroeconomic and financial time series, sourced from the Federal Reserve Economic Data (FRED-QD) \footnote{Available from the website of the Federal Reserve Bank of St. Louis.} (see e.g. \cite{MccrackenNg(16)}). The dataset includes a wide range of macroeconomic indicators, including real activity, the labor market, prices, and financial conditions. The sample spans from 1960Q1 to 2024Q3, which is widely used in large-scale Bayesian vector autoregressions with stochastic volatility (e.g. \citet{CarrieroClarkMarcellino(19)}, \citet{CrossHouKoopPoon(23)}).
%, and \citet{ChanKoopYu(24)}).

Table \ref{tab:FRED-QD} lists all 20 variables, their FRED-ID mnemonics, and the transformations applied to ensure stationarity, as discussed in the previous literature for the dynamic factor models. The dataset includes key indicators for real economic activity (e.g. \#1 GDPC1, \#6 INDPRO), price levels (\#2 PCECTPI, \#12 WPSFD49207, \#13 PPIACO), and monetary policy variables such as \#3 FEDFUNDS and various interest rate spreads (\#18,\#19,\#20). Most real activity and price series are transformed using the first difference of logarithms (multiplied by 100), while interest rates and utilization rates are generally left in levels.

\subsection{Hyperparameters for Prior Distributions}
We assume the following prior distributions for $(\bm{\gamma}, \mathbf{B}_{i\cdot},\psi_j, \bm{\beta}, \mu_k,\phi_k,\sigma_k^2, \rho_k)$:
\begin{eqnarray*}
    &&\bm{\gamma} \sim N_p(\bm{0}, 10^2\mathbf{I}_p), \quad \mathbf{B}_{i\cdot} \sim N_q(\bm{0}, \mathbf{I}_q), \quad i=1, \ldots, p, \\
    &&\psi_j\sim U(-1,1), \quad j=1, \ldots, q, \quad \bm{\beta} \sim N_p(\bm{0}, 0.5^2\mathbf{I}_{p}), \\
    &&\mu_k \sim N(0, 10^2), \quad \sigma_k^2 \sim IG(0.005, 0.005),\\
    &&  \frac{\phi_k+1}{2} \sim Beta(20,1.5), \quad \rho_k \sim U(-1,1), \quad k = 1, \ldots, p+q.
\end{eqnarray*}
The prior distribution for $\phi_k$ reflects past empirical studies of stochastic volatility models, where the prior mean is 0.86 and the prior standard deviation is 0.11.
The inverse-gamma prior on $\sigma_k^2$ is chosen to be weakly informative but proper, while the uniform prior on $\rho_k$ allows for full flexibility in the leverage effect.
The number of factors is set to $q=3$ since our preliminary factor analysis shows a cumulative contribution rate exceeding 70\%, which is consistent with the findings of \cite{CrossHouKoopPoon(23)}. That is, this dataset can be characterized by three primary components: real economic activity, price indices, and monetary policy variables.
 We set the number of VAR lags to $L = 4$, a conventional choice for quarterly time series data, which is used for the Minnesota prior of $\bar{\mathbf{B}}$.

\subsection{Estimation Results}
\label{sec:in-sample result}

This section reports the posterior estimation results of the proposed DFSVML model using the full dataset (1960Q1--2024Q3).
All time series are standardized prior to estimation. This preprocessing improves the numerical stability of the MCMC algorithm and ensures that the relative scale of the series does not influence parameter inference. 
The posterior inference is conducted via the MCMC algorithm using a generalized mixture sampler for the SV-in-mean components (\cite{KimShephardChib(98)}; \cite{HirakiChibOmori(25)}) and simulation smoothing introduced by \cite{DeShephard(95)} and \cite{DurbinKoopman(02)} for the latent states (see Supplementary Material \ref{sec:MCMC algorithm} for more details). We sampled 170,000 draws after a 30,000 burn-in period and kept every 5th sample. The algorithm shows good mixing, as reflected in the reported inefficiency factors (IF)\footnote{IF
is calculated by $1 + 2\sum_{s=1}^{\infty}\rho_s$, where $\rho_s$ is the sample autocorrelation at lag $s$. This is interpreted as the ratio of the numerical variance of the posterior mean from the chain to the variance of the posterior mean from hypothetical uncorrelated draws. They are overall small, as expected, which means that the MCMC sampling is close to the uncorrelated sampling.
} in subsequent tables. 
\subsubsection{VAR coefficients matrix}
\begin{figure}[H]
    \centering
    \includegraphics[width=1.0\linewidth]{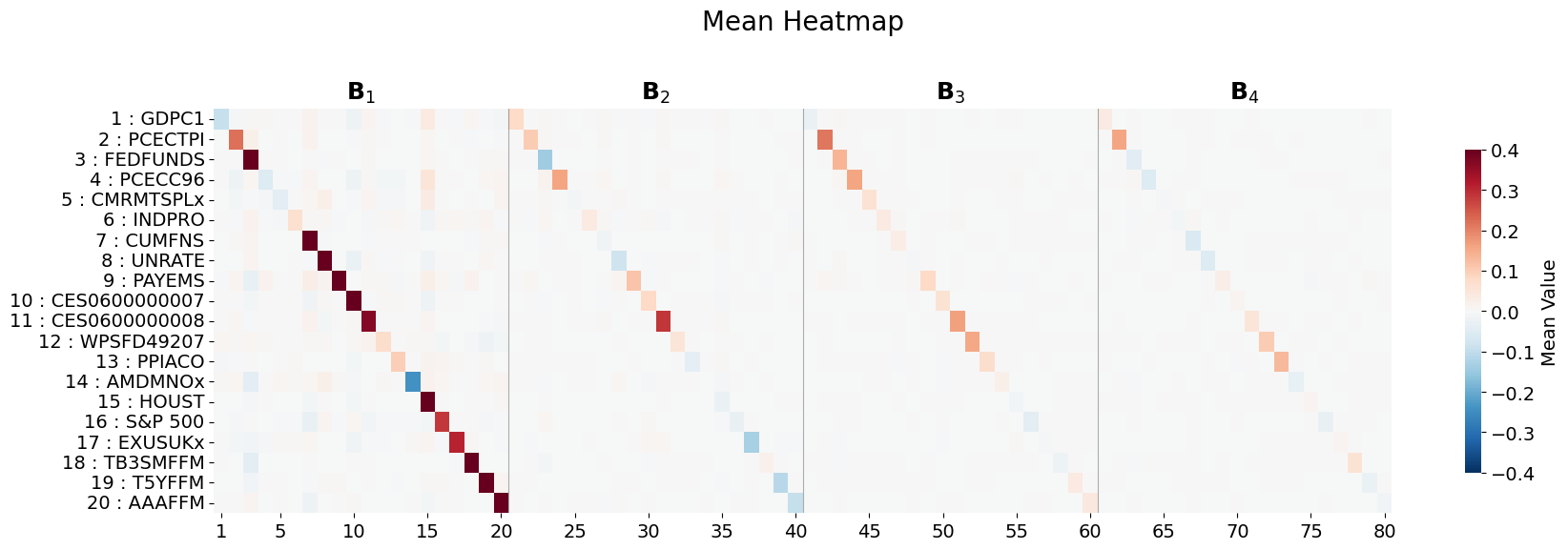}
    \caption{A heatmap of the posterior mean of $20 \times 80$ matrix $(\mathbf{B}_{1}, \ldots, \mathbf{B}_4)$. }
    Left: $\mathbf{B}_{1}$. Center left:$\mathbf{B}_{2}$. Center right:$\mathbf{B}_{3}$. Right:$\mathbf{B}_{4}$.
    \label{fig:placeholder}
\end{figure}

Figure \ref{fig:placeholder} shows a heatmap of the posterior mean of the $20 \times 80$ matrix $(\mathbf{B}_{1}, \ldots, \mathbf{B}_4)$. Detail estimation results are omitted due to space limitations. We found that the dark red and dark blue colors are concentrated on the diagonal elements of the matrix for $\mathbf{B}_i$ ($i=1,2,3,4$), while the light colors are scattered across the non-diagonal elements. This suggests that the individual dependent variable is mostly explained by its own lagged values, especially by the first lagged dependent variable.  

\subsubsection{Stochastic Volatility}
The persistence parameters ($\phi_i$) are generally high across all variables, with most posterior means ranging from 0.682 to 0.980 for twenty individual variables, and from 0.962 to 0.991 for three factors,  $i=21, 22, 23$ (see Table \ref{tab:result_phi_rho} of Supplementary Material \ref{sec:estimation_results} in detail). 
This confirms the widely documented high persistence of macroeconomic and financial volatility. The estimates for the leverage parameter ($\rho_i$) show more heterogeneity. The strong negative leverage  is found for \#6 INDPRO ($Pr(\rho_{6}<0|\bm{y}) = 0.995$), and the first factor ($Pr(\rho_{21}<0|\bm{y}) = 0.997$) as well as \#16 S\&P 500 ($Pr(\rho_{16}<0|\bm{y}) = 0.999$) and \#18 TB3SMFFM ($Pr(\rho_{18}<0|\bm{y}) = 0.951$), which is consistent with financial market behavior. For most other variables, we found no strong idiosyncratic leverage effect in the sense that 95\% credible intervals include zeros.

The estimates $\mu_i$ indicate the baseline level of idiosyncratic volatility for each series, and the highest level is found for \#16 S\&P 500 index returns. The estimates $\sigma_i$, which capture the magnitude of volatility shocks, and the largest magnitude is found for the total nonfarm payrolls,  \#9 PAYEMS (see Table \ref{tab:result_mu_sigma} of Supplementary Material \ref{sec:estimation_results} in detail).
\begin{figure}[H]
    \centering    \includegraphics[width=0.95\linewidth]{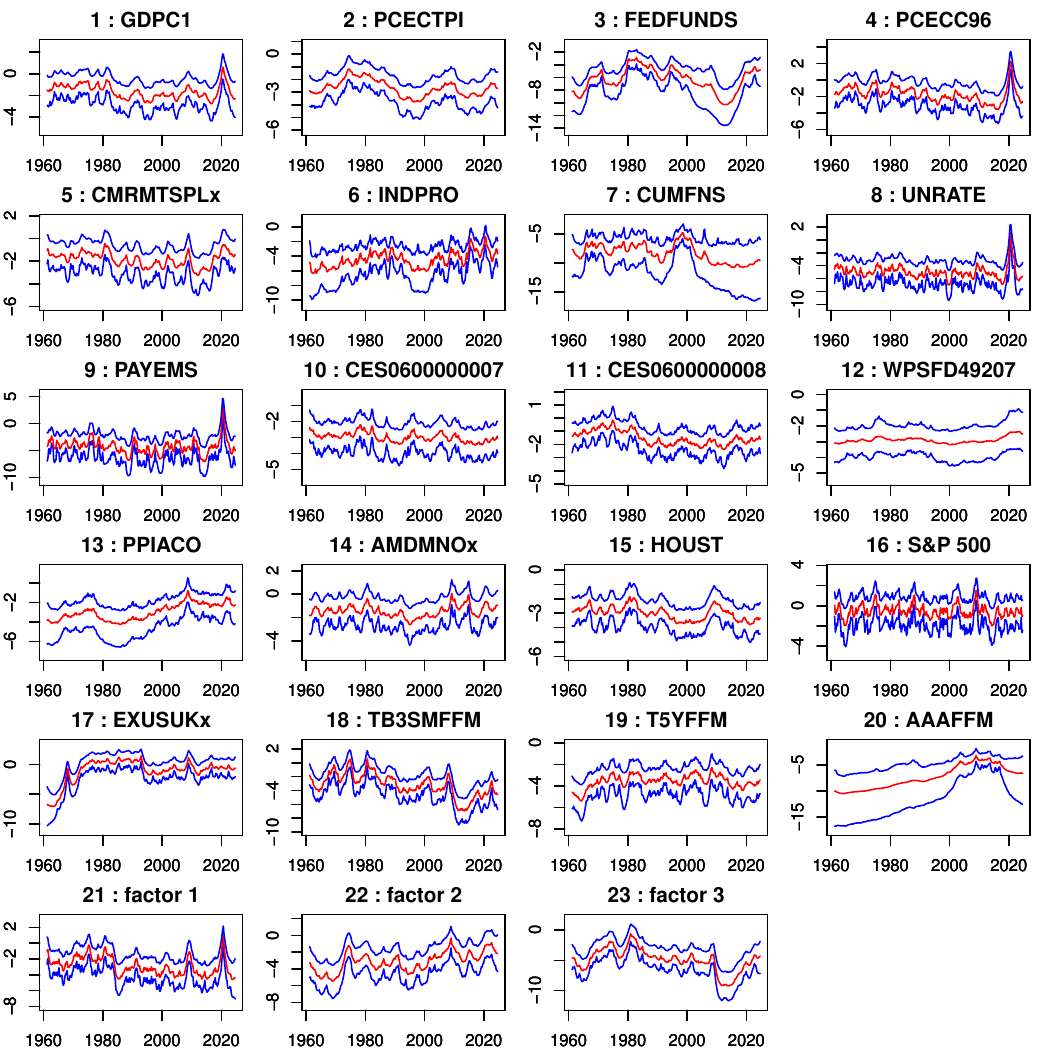}
    \caption{Time series plots of estimated log-volatilities. Posterior mean (red) of log volatility $h_t$ with 95\% credible interval (blue).}
    \label{fig:result_h}
\end{figure}
Figure \ref{fig:result_h} plots the posterior means and 95\% credible intervals for the idiosyncratic log-volatilities ($h_{it}$) for all twenty variables and the three factors. The plots visually confirm the high persistence indicated by the estimates $\phi_i$. They also capture key historical episodes of high uncertainty, such as the COVID-19 pandemic period (2020-2021), the GFC (around 2008) and the Volcker Shock (early 1980s). 
\subsubsection{Factor and Loading Matrix}
\begin{figure}[H]
    \centering
    \includegraphics[width=0.5\linewidth]{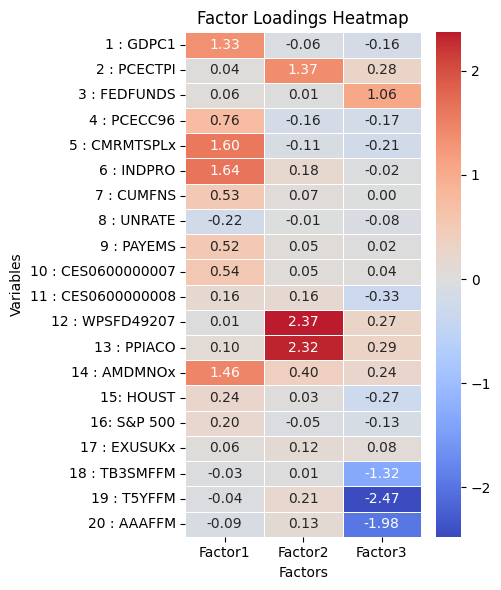}
    \caption{Estimated posterior mean of $\mathbf{B}$.}
    \label{fig:heatmap}
\end{figure}

Figure \ref{fig:heatmap} shows the posterior means of the factor loading matrix $\mathbf{B}$ with the heatmap (see Table \ref{tab:result_B} of Supplmentary Material \ref{sec:estimation_results} for more details). The first factor, exhibiting consistently high positive loadings across key real economic indicators such as \#1, \#5, \#6 and \#14. This factor clearly represents real economic activity and growth and comprehensively captures the overall level of economic output, consumption, production, and employment, thus reflecting the cyclical fluctuations of the macroeconomy.

The second factor is characterized by exceptionally high positive loadings on major price indices, including \#2, \#12, \#13. 
This strong association with inflation-related variables leads to its interpretation as the price levels and the inflation factor, signifying its role in capturing inflationary pressures and the general movement of prices within the economy.

Finally, the third factor, primarily influenced by a positive loading on \#3, and by strong negative loadings on various interest rate spreads (\#18, \#19, \#20). It is interpreted as a financial conditions factor and effectively captures the stance of monetary policy, the dynamics of the yield curve, and credit spreads. The factor reflects how changes in policy rates impact the term structure of interest rates and market liquidity.

\begin{figure}[H]
    \centering
    \includegraphics[width=0.6\linewidth]{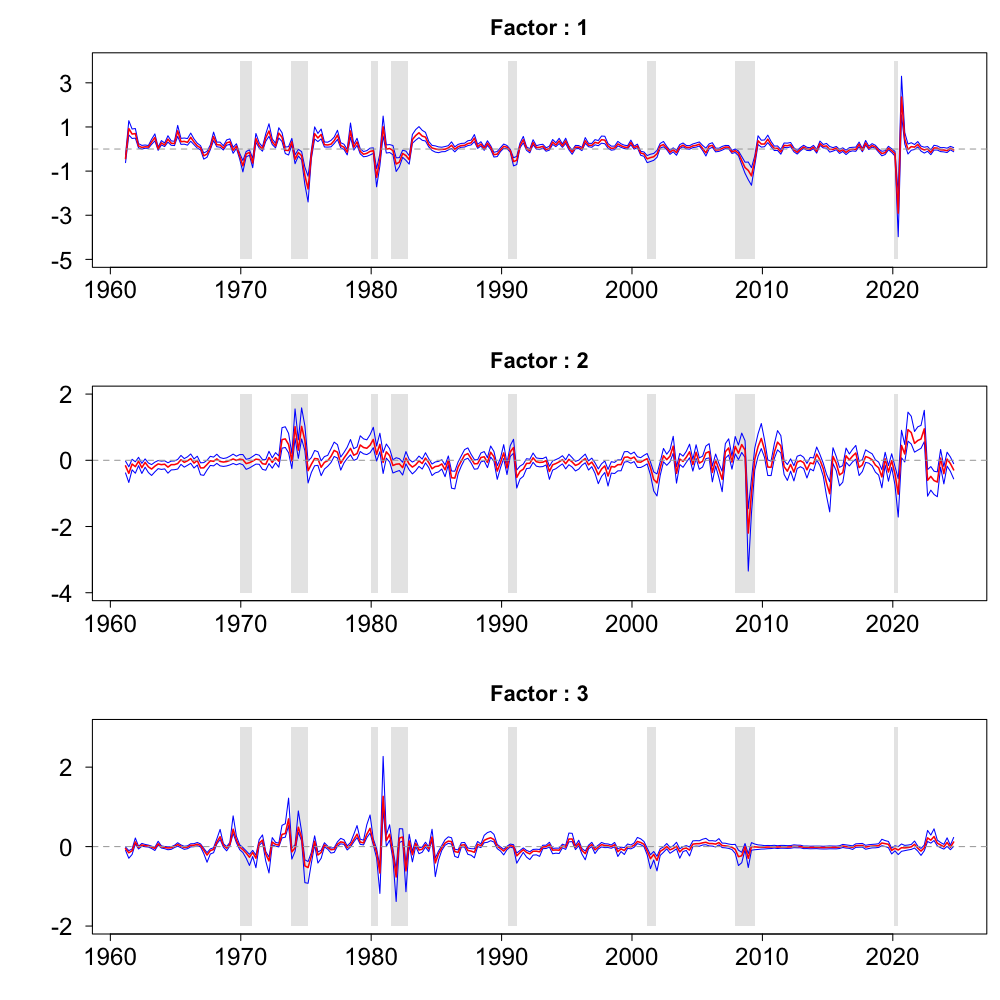}
    \caption{Posterior mean (red) with 95\% credible interval (blue) of dynamic factor $\bm{f}_t$. Shaded areas indicate NBER-defined recession periods.}
    \label{fig:result_f}
\end{figure}

Figure \ref{fig:result_f} presents the time series plots of three factors. For the first factor (which represents real economic activity), it exhibits continuous and significant volatility, with sharp downturns consistently reflecting periods of economic recession. Particularly notable are the drops corresponding to the 1970s energy crisis, the early 1980s Volcker Shock, and the 2008 global financial crisis. A unique exception is the 2020 COVID-19 pandemic period; following the unprecedented contraction in early 2020, the first factor shows an extraordinary spike around 2020Q3, reflecting the exceptionally high growth rates during the rapid economic reopening. The impact of the GFC is also distinctly evident in the second factor (which corresponds to the inflation), reflecting the heightened deflationary pressure experienced during that financial turmoil. Similar to the first factor, the second factor exhibits a large estimated value in the post-pandemic phase, capturing the volatile recovery process. For the third factor (which is interpreted as the financial conditions), the pronounced fluctuations observed from the late 1970s to the early 1980s are clearly attributable to the Volcker Shock, indicating the severe monetary policy tightening during that period.

\subsubsection{Stochastic Volatility in Mean Effect}
\label{sec:sv-in-mean}
\begin{table}[H]
    \scriptsize
    \centering
    \begin{tabular}{rlrcrr}
        \hline
        \# & Variable & Mean & 95\% interval & IF & Pr(+) \\
        \hline
        1 & GDPC1 & -0.188 & (-0.354, -0.025) & 59 & \textbf{0.011} \\
        2 & PCECTPI & 0.120 & (-0.255,  0.544) & 40 & 0.717 \\
        3 & FEDFUNDS & -0.005 & (-0.282,  0.322) & 43 & 0.465 \\
        4 & PCECC96 & -0.148 & (-0.301,  0.002) & 30 & \textbf{0.027} \\
        5 & CMRMTSPLx & -0.225 & (-0.416, -0.045) & 83 & \textbf{0.006} \\
        6 & INDPRO & -0.645 & (-1.184, -0.245) & 167 & \textbf{0.000} \\
        7 & CUMFNS & -1.368 & (-2.126, -0.672) & 122 & \textbf{0.000} \\
        8 & UNRATE & 0.089 & (-0.068,  0.250) & 44 & 0.861 \\
        9 & PAYEMS & -0.186 & (-0.385, -0.001) & 77 & \textbf{0.025} \\ 
        10 & CES0600000007 & -0.085 & (-0.235,  0.060) & 43 & 0.131 \\
        11 & CES0600000008 & -0.019 & (-0.153,  0.116) & 6 & 0.388 \\
        12 & WPSFD49207 & 0.278 & (-0.454,  1.105) & 46 & 0.758 \\
        13 & PPIACO & 0.234 & (-0.447,  0.942) & 40 & 0.752 \\
        14 & AMDMNOx & -0.155 & (-0.328,  0.010) & 57 & \textbf{0.033} \\ 
        15 & HOUST & -0.038 & (-0.176,  0.099) & 15 & 0.295 \\
        16 & S\&P 500 & 0.012 & (-0.124,  0.149) & 2 & 0.568 \\
        17 & EXUSUKx & 0.058 & (-0.081,  0.198) & 2 & 0.795 \\
        18 & TB3SMFFM & 0.030 & (-0.140,  0.197) & 12 & 0.643 \\
        19 & T5YFFM & -0.020 & (-0.284,  0.209) & 42 & 0.458 \\
        20 & AAAFFM & -0.036 & (-0.637,  0.528) & 52 & 0.458 \\
        \hline
    \end{tabular}
    \caption{Estimated $\bm{\beta}$. The last column shows the posterior positive probability, $Pr(\beta_i>0|\bm{y})$. Bold figures indicate the strong negative effect (Pr(+) $<$ 0.05).}
    \label{tab:result_beta}
    \normalsize
\end{table}
Finally, the results in Table \ref{tab:result_beta} provide evidence for the in-mean effect, $\bm{\beta}$. Specifically, strong negative effects are observed for the real economic activity and growth variables, where the posterior probability of negative $\beta_i$ is greater than 95\% for $i=1, 4, 5, 6, 7, 9, 14$. The negative sign suggests that the risk premium may tend to decrease during periods of economic instability.
\subsubsection{Alternative Specification for Stochastic Volatility  in Mean}
\label{sec:sv-in-mean_alt}
Finally, we also consider an alternative model specification where the in-mean component enters the factor equation rather than the observation equation, as in 
\begin{equation}
\label{alternative_specification}
    \bm{f}_{t} = \bm{\gamma} + \mathbf{\Psi}(\bm{f}_{t-1} - \bm{\gamma}) + \mathbf{V}_{2t}^{1/2}\bm{\beta} + \mathbf{V}_{2t}^{1/2}\bm{\epsilon}_{2t}.
\end{equation}
This specification imposes a common, factor-driven risk premium on all variables. 
\begin{figure}[H]
    \centering
    \includegraphics[width=0.5\linewidth]{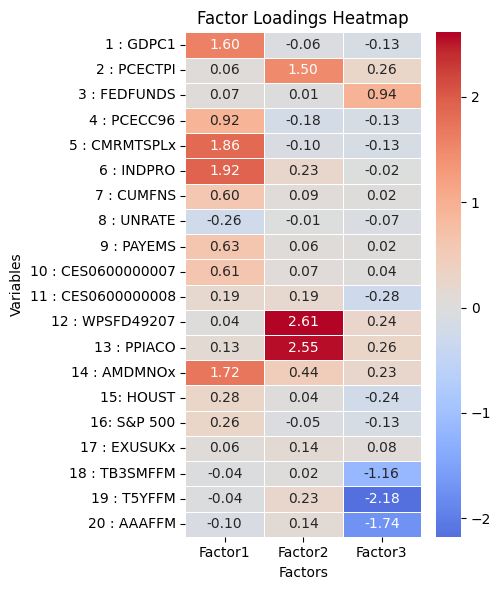}
    \caption{Estimated posterior mean of $\mathbf{B}$.}
    \label{fig:heatmap_alt}
\end{figure}
\noindent
The heatmap of the estimated factor loading matrix is shown in Figure \ref{fig:heatmap_alt}. It indicates that the three latent factors are interpreted similarly to the results in Figure \ref{fig:heatmap} for our main model. Specifically, the first factor represents real economic activity and growth, the second captures price levels and inflation, and the third reflects financial conditions.
\begin{table}[H]
%    \footnotesize
    \scriptsize
%    \small
    \centering
    \begin{tabular}{ccccc}
        \hline
        Factor & Mean & 95\% interval & IF  & Pr(+)  \\
        \hline
        1 & -0.099 & (-0.432,  0.221) & 12 & 0.282 \\
        2 & 0.241 & (-0.062,  0.567) & 12 & 0.939 \\
        3 & 0.168 & (-0.077,  0.410) & 73 & 0.914 \\
        \hline
    \end{tabular}
    \caption{Estimation result of $\beta_i$ for alternative SV in mean specification.}
    \label{tab:result_beta_alt}
    \normalsize
\end{table}
\noindent
Table \ref{tab:result_beta_alt} shows the posterior estimates for the SV in-mean coefficients in this alternative specification. Although we found high posterior probabilities, $Pr(\beta_2>0|\bm{y})=0.939$ and $Pr(\beta_3>0|\bm{y})=0.914$, the 95\% credible intervals still include zero. Taking into account that we have much higher posterior probabilities (that parameters are negative) in Table \ref{tab:result_beta} for our proposed models, this indicates relatively less evidence for this alternative structure. 
\section{Comparison of Predictive Performances}
\label{sec:out-of-sample result}
\subsection{Models}
This section evaluates the predictive performance of the proposed and benchmark models. Although the models are estimated using standardized data, as described in the previous section, all forecasting results are transformed back to their original units to ensure an intuitive interpretation of the forecast errors and likelihoods. 
The benchmark model is the Bayesian vector autoregressions with stochastic volatility (SVVAR) model, as utilized in \cite{CrossHouKoopPoon(23)}. Since we consider a model with twenty variables ($p=20$), it is referred to as a Large-SVVAR (LSVVAR) model, which is specified as:
\begin{align*}
    \mathbf{B}_0 \bm{y}_t &= \bm{b} + \sum_{\ell=1}^L\mathbf{B}_\ell \bm{y}_{t-\ell} + \bm{\epsilon}_{t}^y, \quad \bm{\epsilon}_{t}^y \sim N(\bm{0}_p, \mathbf{\Sigma}_t), \quad
    \mathbf{\Sigma}_t = \operatorname{diag}(\exp(h_{1t}), \ldots,  \exp(h_{pt})), \\
    \bm{h}_{t+1} &= \bm{\mu} + \mathbf{\Phi} (\bm{h}_{t} - \bm{\mu}) + \bm{\epsilon}_t^h, \quad \bm{\epsilon}_{t}^h \sim N(\bm{0}_p, \mathbf{\Sigma}_h),
\end{align*}
where $\mathbf{B}_0$ is a $p \times p$ matrix with ones on its diagonal, $\bm{b}$ is a $p \times 1$ vector of intercepts, and $\mathbf{B}_\ell$ for $\ell = 1,...,L$ is a $p \times p$ matrix of VAR coefficients. We set the number of lags to $L=4$ as in our proposed models. In the state equation for the log-volatility, $\mathbf{\Phi}$ is a $p \times p$ coefficient matrix, and $\mathbf{\Sigma}_h$ is a time-invariant $p \times p$ error covariance matrix. Following the common practice in macroeconometric literature (\cite{CrossHouKoopPoon(23)}), we assume the latent volatility processes are independent (i.e. $\mathbf{\Phi} = \operatorname{diag}(\phi_1,...,\phi_p)$ and $\mathbf{\Sigma}_h = \operatorname{diag}(\sigma_1^2,...,\sigma_p^2)$) and do not include a leverage effect. 
\begin{table}[H]
    \scriptsize
    \centering
    \begin{tabular}{lccc}
    \hline
    Model & Factor Dynamics ($\mathbf{\Psi}$) & In-Mean ($\bm{\beta}$) & Leverage ($\rho$) \\
    \hline
    DFSV & \checkmark &  &  \\
    DFSVL & \checkmark &  & \checkmark \\
    DFSVM & \checkmark & \checkmark &  \\
    DFSVML & \checkmark & \checkmark & \checkmark \\
    \hline
    FSV &  &  &  \\
    FSVL &  &  & \checkmark \\
    FSVM &  & \checkmark &  \\
    FSVML &  & \checkmark & \checkmark \\
    \hline
    \end{tabular}
    \caption{Model specifications. A \checkmark indicates the inclusion of the corresponding component. Models prefixed with `D' have a dynamic factor structure for the levels, while `F' indicates a standard factor model without dynamics in the observation equation.}
    \label{tab:models}
\end{table}
Table \ref{tab:models} lists all the model variations we compare in our out-of-sample forecasting exercise. The leverage effect $\rho$ is included only for the factor equations: the idiosyncratic leverage parameters $\rho_i$ for $i = 1,\ldots,20$ are set to $0$ in all cases for simplicity, taking into account that most of the 95\% credible intervals include zeros except for some $\rho_{i}$'s ($i=6,16,18$, negative effect). The exercise focuses on eight financial and macroeconomic indicators: \#2 PCECTPI,  \#3 FEDFUNDS (which is widely recognized as a representative financial variable in the existing literature, e.g., \cite{CarrieroClarkMarcellino(16)}, \cite{CrossHouKoopPoon(23)}, %\cite{ChanDoucetLeonRodney(25)},
\cite{DavidsonHouKoop(25)}), \#11 CES0600000008, and \#12 WPSFD49207,  \#13  PPIACO which are primarily associated with the second factor. Additionally, we evaluate \#18 TB3SMFFM, \#19 T5YFFM and \#20 AAAFFM. These spreads are primarily captured by the third latent factor and, as shown in our results, exhibit notable gains in predictive accuracy over the benchmark LSVVAR specification.
\subsection{Expanding Window Forecast Performance}
We employ an expanding window forecasting scheme, with an initial estimation period of 1960Q1--1999Q4. Using the parameters estimated from this sample, we produce forecasts up to 4 steps ahead. Subsequently, we expand the estimation window by one quarter, re-estimate the models, and generate a new set of forecasts. This process is repeated until the end of the evaluation sample (see Supplementary Material 
\ref{sec:prediction} for the prediction procedure in detail). We assess both point and density forecasts using the cumulative squared forecast error (CSFE) and the cumulative log predictive likelihood (CLPL), respectively. To focus our discussion on the immediate predictive ability and adaptivity of the models, the following analysis primarily presents the results for the 1-step-ahead horizon (see Supplementary Material \ref{sec:details-predictive-performance} for the 4-step-ahead horizon). The results are shown in 
%The results are presented in Figures \ref{fig:result_csfe_pre}--\ref{fig:result_csfe_post} (CSFE) and Figures \ref{fig:result_clpl_pre}--\ref{fig:result_clpl_post} (CLPL). 
Figures \ref{fig:result_csfe_gfc_1step}--\ref{fig:result_clpl_covid19_1step} for \#2 PCECPI, \#3 FEDFUNDS, \#11 CES0600000008, \#12 WPSFD49207, \#13 PPIACO, \#18 TB3SMFFM, \#19 T5YFFM and \#20 AAAFFM (for other varibales, see Figures \ref{fig:result_csfe_1step}--\ref{fig:result_clpl_4step} in Supplementary Material \ref{sec:details-predictive-performance}). These figures plot the cumulative metrics for each model relative to the LSVVAR benchmark, where the zero-line represents the benchmark's performance. A value below zero for CSFE or above zero for CLPL indicates that the model outperforms the LSVVAR.
\subsubsection{During Global Financial Crisis Period}
\begin{figure}[H]
    \centering
    \includegraphics[width=0.8\linewidth]{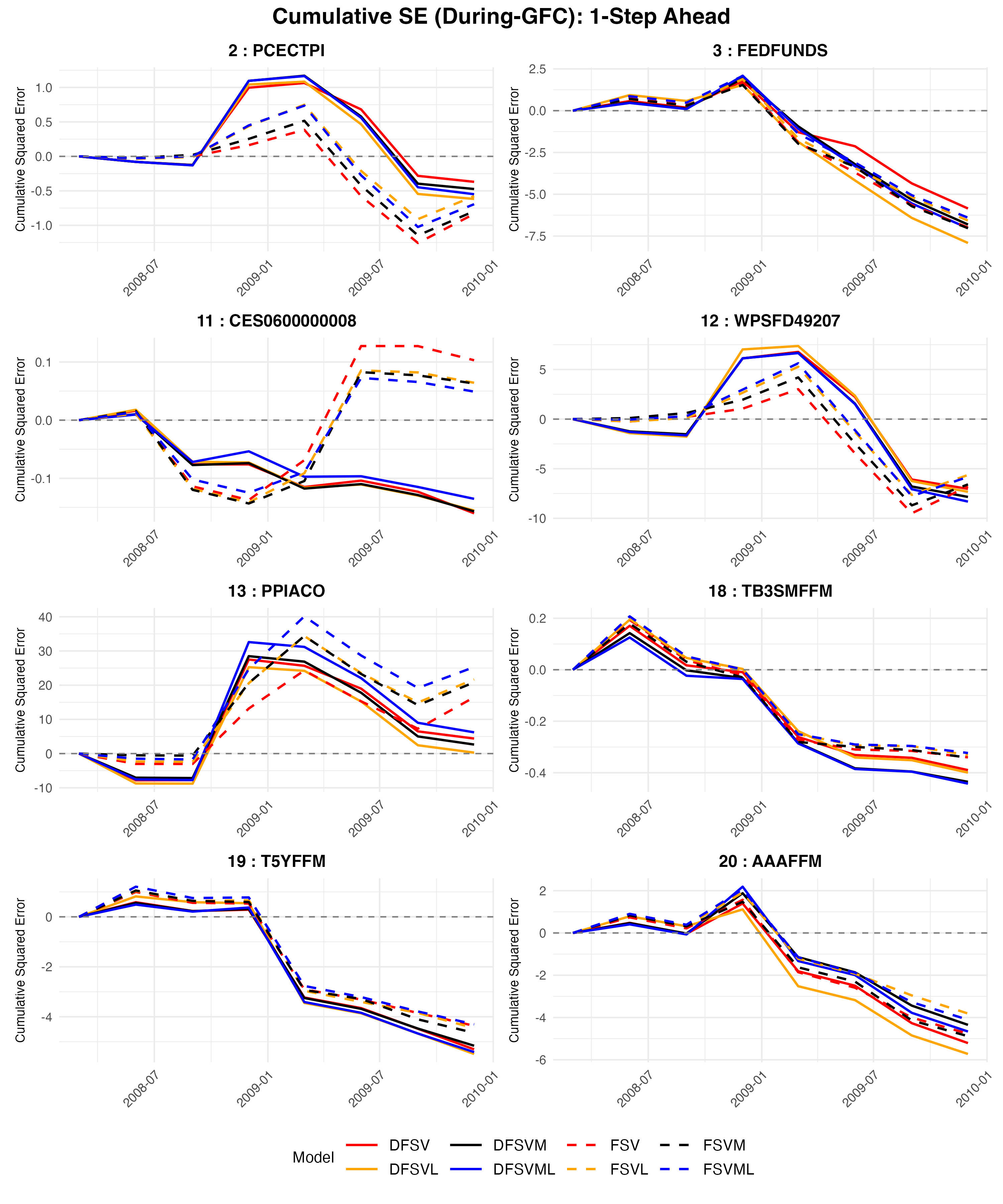}
    \caption{Cumulative squared forecast error (CSFE) relative to the LSVVAR benchmark (zero-line). 2008Q1-2009Q4.}
    \label{fig:result_csfe_gfc_1step}
\end{figure}

\begin{figure}[H]
    \centering
    \includegraphics[width=0.8\linewidth]{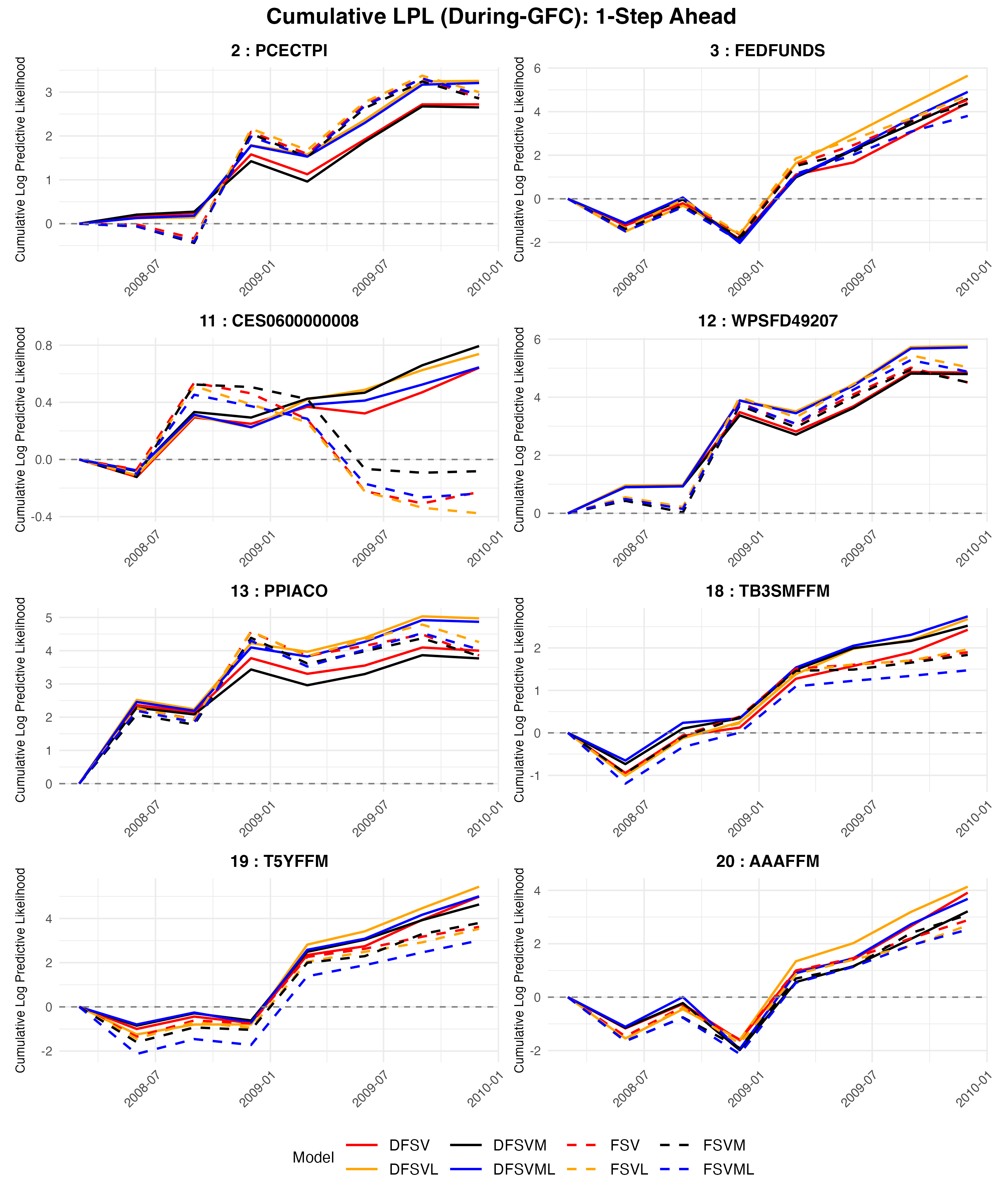}
    \caption{Cumulative log predictive likelihood (LPL) relative to the LSVVAR benchmark (zero-line). 2008Q1-2009Q4.}
    \label{fig:result_clpl_gfc_1step}
\end{figure}

Figures \ref{fig:result_csfe_gfc_1step} and \ref{fig:result_clpl_gfc_1step} show the results for the GFC period (2008Q1--2009Q4). For 1-step ahead forecast, following the onset of the crisis in 2008Q1, most factor model specifications initially exhibit larger CSFEs and smaller CLPLs compared to the LSVVAR benchmark. However, for the majority of focal variables, their predictive performance recovers rapidly and
\noindent
surpasses the benchmark within a few quarters. 
This recovery is particularly prominent in variables associated with the third factor, including interest rates and spreads (\#3, \#18, \#19, and \#20), as well as price-related variables linked to the second factor (\#2, \#12, and \#13). For these groups, the factor structure effectively captures the systemic financial distress and the subsequent monetary policy response, regaining accuracy more quickly
\noindent
than the LSVVAR model, which remains affected by persistent disruptions. In contrast, the wage-related index (\#11) exhibits a distinct pattern. While the dynamic factor specifications (DFSVM and DFSVML) follow a recovery path similar to other variables, the static factor models show a U-shaped deterioration in performance around the end of the GFC period, with their CSFEs increasing again relative to the benchmark toward 2010Q1. This suggests that for this labor market variable, modeling the dynamic persistence of common volatility factors is crucial for maintaining the predictive stability.
Regarding the choice between dynamic and static factor structures for the remaining variables, the distinction is less definitive during this crisis. For the price-related group (\#2, \#12, and \#13), both specifications eventually yield broadly comparable predictive performance, particularly in terms of CLPL. Conversely, for the interest rate spreads \#18 and \#19, dynamic factor models demonstrate a slight edge over their static counterparts. Consistent with the observations for the wage index (\#11), these findings suggest that the dynamic factor structure provides a more robust framework for capturing the persistence of uncertainty shocks and their transmission into the macroeconomy.
The predictive performance on the 4-step horizon presents a more varied picture among focal variables (see Supplementary Material \ref{sec:details-predictive-performance} for details). Regarding point forecasts, as measured by the CSFEs, the static factor specifications are generally supported for the price-related variables (specifically \#2, \#12 and \#13) and the variables related to the third factor (\#3,\#18,\#19,\#20) during the GFC period. 

In contrast, dynamic factor models continue to be favored for the wage index (\#11). For density forecasts evaluated by the CLPLs, however, there appears to be little difference between the dynamic and static specifications at this longer horizon. These results suggest that while explicit factor dynamics remains beneficial for specific indicators like wages, their impact on overall density forecast accuracy tends to be less pronounced as the prediction horizon increases.

\subsubsection{During COVID-19 Pandemic Period}
\begin{figure}[H]
    \centering
    \includegraphics[width=0.8\linewidth]{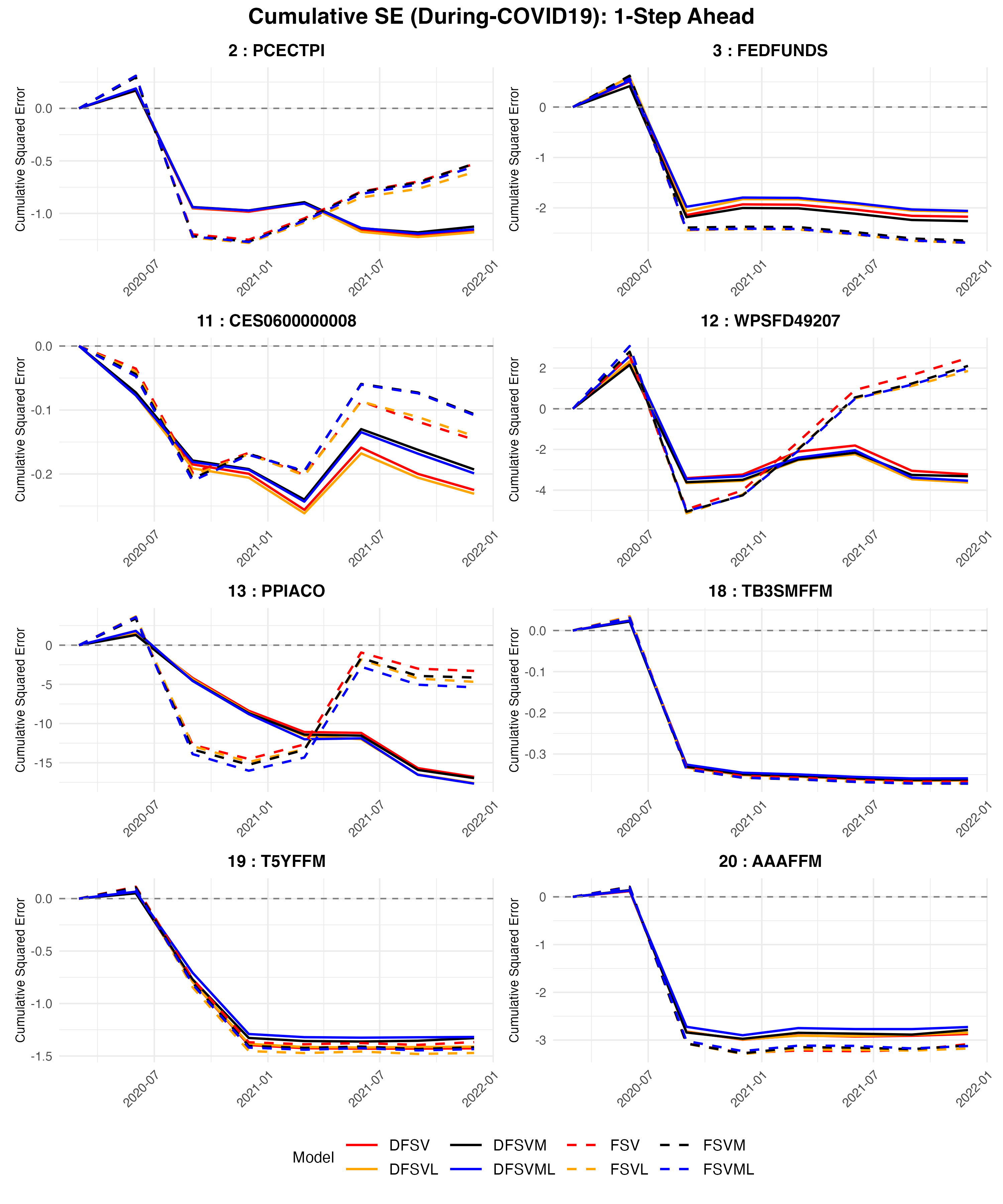}
    \caption{Cumulative squared forecast error (CSFE) relative to the LSVVAR benchmark (zero-line). 2020Q1-2021Q4.}
    \label{fig:result_csfe_covid19_1step}
\end{figure}

\begin{figure}[H]
    \centering
    \includegraphics[width=0.8\linewidth]{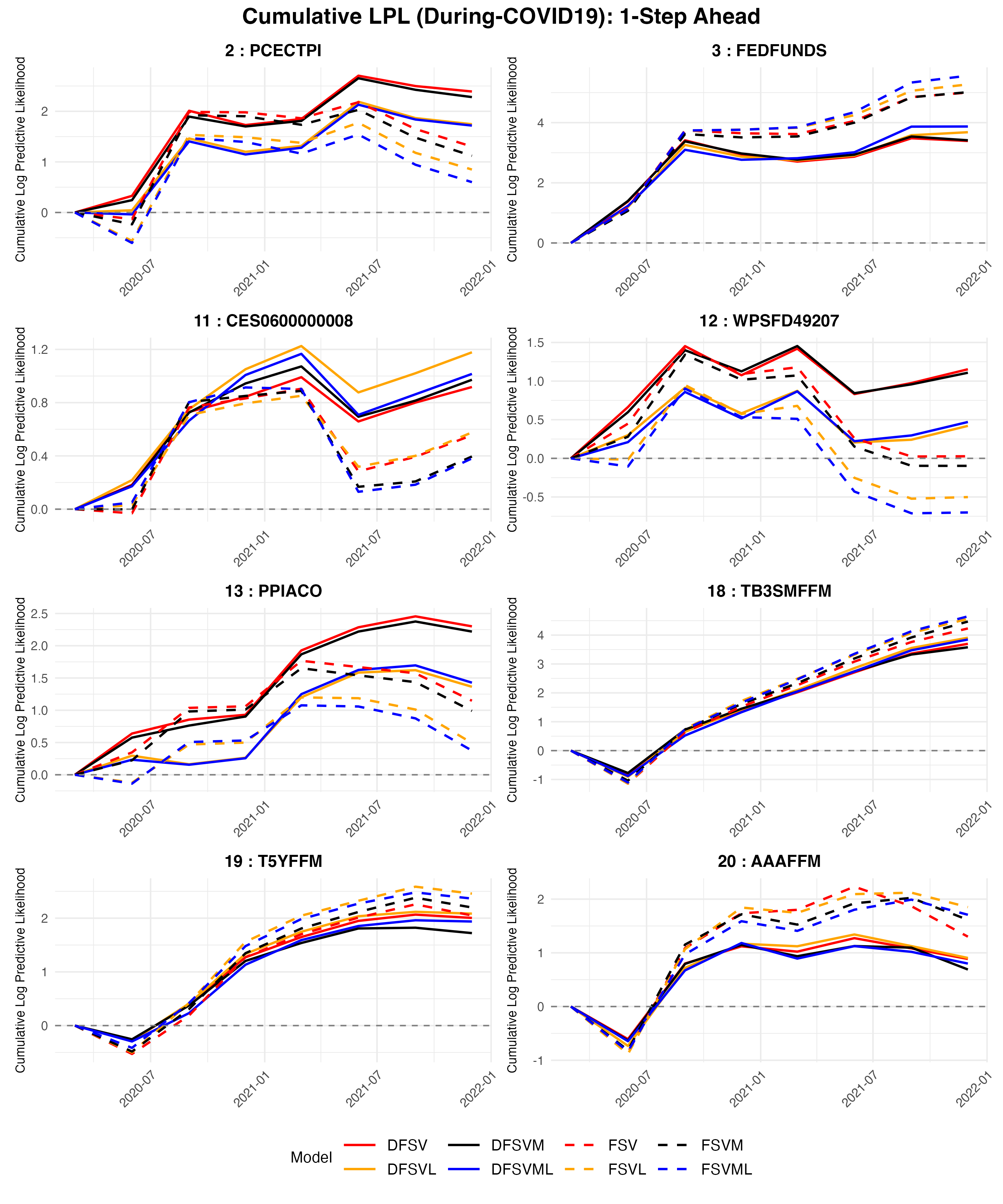}
    \caption{Cumulative log predictive likelihood (LPL) relative to the LSVVAR benchmark (zero-line). 2020Q1-2021Q4.}
    \label{fig:result_clpl_covid19_1step}
\end{figure}

For 1-step ahead forecast, a notable shift occurs during the initial pandemic period (2020Q1--2021Q4), as depicted in Figures \ref{fig:result_csfe_covid19_1step} and \ref{fig:result_clpl_covid19_1step}. Similar to the patterns observed during the GFC
\noindent
period, most variables, particularly those associated with the third factor (\#3, \#18, \#19, and \#20), exhibit an initial slight increase in CSFEs followed by a rapid and significant decrease. Throughout this period, all factor model specifications generally maintain superior predictive accuracy over the LSVVAR benchmark.

Regarding the model specifications for the price-related variables (\#2, \#12, and \#13), 
\noindent
static factor models initially provide superior forecasts compared to their dynamic counterparts during this highly volatile phase. Although the improvement is moderate, their consistent performance reinforces the advantage of adaptivity during crises, suggesting that the unprecedented structural breaks favored models capable of more flexibly adjusting to a rapidly changing environment. However, as the period progresses into 2021, the dynamic factor specifications (DFSVM and DFSVML) begin to outperform the static ones, yielding lower CSFEs and higher CLPLs. This reversal indicates that while the initial pandemic shock was idiosyncratic, the subsequent inflationary pressure retained a degree of persistence that is better captured by explicit factor dynamics.

The 4-step ahead forecasts during this period exhibit distinct characteristics compared to the 1-step horizon. As detailed in Supplementary Material \ref{sec:details-predictive-performance}, the point forecast performance for the third factor group (\#3, \#18, \#19, and \#20) and the price index (\#2) follows a similar trajectory, although the primary shifts in CSFEs are lagged by three quarters relative to the 1-step results. In contrast, for the wage index (\#11), the improvement offered by dynamic factor models is less pronounced at this longer horizon, with static specifications providing more accurate point forecasts. Furthermore, for variables \#12 and \#13, the proposed factor models consistently underperform the LSVVAR benchmark in terms of CSFE at the 4-step horizon.

Regarding density forecasts, the relative advantages remain more stable across horizons. For \#2, \#12, and \#13, dynamic factor models consistently yield higher CLPLs regardless of the forecast horizon. Conversely, for the third factor group (\#3, \#18, \#19, and \#20), static factor specifications tend to be superior. For the wage index (\#11), however, the comparison between dynamic factor and static factor structures remains ambiguous for density forecasting at the 4-step horizon.
\\

\noindent
{\it Remark}. To provide a rigorous statistical comparison, we also conduct the Diebold-Mariano test and the Model Confidence Set (MCS) procedure. For the sake of brevity, the comprehensive results of these tests are provided in Supplementary Material \ref{sec:appendix_prediction_tests}.

\subsubsection{Effect of Stochastic Volatility in Mean}
A key finding relates to the volatility-in-mean component ($\bm{\beta}$). While the posterior estimates in Section \ref{sec:sv-in-mean} indicated strong evidence of an in-sample effect, its out-of-sample benefit appears state-dependent, yet crucial. In the pre-pandemic period, encompassing the GFC, the gains from including the in-mean term were modest, although DFSVM often outperformed DFSV with respect to CSFE. 

However, its importance becomes more evident in turbulent regimes. During the COVID-19 pandemic period, the FSVML model was the top performer regarding CLPL for \#3 and \#18, suggesting that the risk-return trade-off captured by $\bm{\beta}$ was important during periods of extreme uncertainty. Furthermore, as detailed in the comprehensive results in Supplementary Material \ref{sec:details-predictive-performance}, this strength continues into the post-pandemic recovery phase, where the DFSVML model re-emerges as a best-in-class performer for short-term forecasts of these financial indicators.
\subsubsection{Effect of Stochastic Volatility in Mean in Alternative Specification}
As discussed in Section \ref{sec:sv-in-mean_alt}, we could 
consider an alternative model specification where the in-mean component enters the factor equation rather than the observation equation. The predictive performances under the alternative specification are shown in Figures \ref{fig:result_csfe_gfc_y_vs_f_1step} through \ref{fig:result_clpl_covid19_y_vs_f_4step} of Supplementary Material \ref{sec:sv-in-mean-comparison}.
In out-of-sample forecasting exercises, this alternative model exhibited dynamics similar to our main specification but yielded inferior results especially during the GFC period with respect to the CSFE for the 4-step ahead forecast. This finding provides empirical support for our chosen model structure. It suggests that the in-mean effect is not a homogeneous, factor-level phenomenon, but that it is instead better modeled as an idiosyncratic effect specific to each variable. This aligns with the economic intuition that different macroeconomic variables may respond heterogeneously to aggregate uncertainty.
\bigskip

\subsubsection{Comparison of Normal, Global Financial Crisis, and COVID-19 Pandemic Periods}
Finally, we compare the forecasting accuracies during the normal times, the GFC period and the COVID-19 pandemic period in more detail.
\begin{table}[H]
    \scriptsize
%    \small
    \centering
    \setlength{\tabcolsep}{3.5pt}
    \begin{tabular}{rl rr rr rr}
        \hline
         & & \multicolumn{2}{c}{Normal} & \multicolumn{2}{c}{GFC} & \multicolumn{2}{c}{COVID-19} \\
        \cmidrule(lr){3-4} \cmidrule(lr){5-6} \cmidrule(lr){7-8}
        \# & Variable & DFSVL & DFSVML & DFSVL & DFSVML & DFSVL & DFSVML \\
        \hline
        1 & GDPC1 & 6.64 & 6.29 & \textcolor{blue}{-15.64} & \textcolor{blue}{-13.80} & \textcolor{blue}{-11.60} & \textcolor{blue}{-11.61} \\
        2 & PCECTPI & -7.07 & -7.53 & 6.43 & 5.73 & \textcolor{red}{30.92} & \textcolor{red}{30.13} \\
        3 & FEDFUNDS & \textcolor{blue}{-31.64} & \textcolor{blue}{-36.08} & \textcolor{red}{56.82} & \textcolor{red}{52.12} & \textcolor{red}{45.98} & \textcolor{red}{46.12} \\
        4 & PCECC96 & 6.32 & 8.49 & \textcolor{red}{19.26} & \textcolor{red}{16.83} & \textcolor{blue}{-13.93} & \textcolor{blue}{-15.55} \\
        5 & CMRMTSPLx & \textcolor{red}{11.03} & \textcolor{red}{10.21} & \textcolor{red}{20.30} & \textcolor{red}{20.73} & \textcolor{blue}{-54.54} & \textcolor{blue}{-53.03} \\
        6 & INDPRO & \textcolor{blue}{-22.63} & \textcolor{blue}{-23.11} & -6.00 & -0.08 & \textcolor{blue}{-25.36} & \textcolor{blue}{-24.55} \\
        7 & CUMFNS & 5.11 & -0.36 & \textcolor{blue}{-11.02} & -7.32 & \textcolor{blue}{-20.66} & \textcolor{blue}{-19.91} \\
        8 & UNRATE & \textcolor{blue}{-10.57} & \textcolor{blue}{-11.90} & -7.32 & -5.17 & \textcolor{blue}{-15.36} & \textcolor{blue}{-13.87} \\
        9 & PAYEMS & 7.64 & 4.20 & 0.68 & 3.65 & \textcolor{blue}{-10.57} & \textcolor{blue}{-11.96} \\
        10 & CES0600000007 & 1.52 & 3.80 & \textcolor{blue}{-11.36} & \textcolor{blue}{-12.88} & 7.75 & 6.68 \\
        11 & CES0600000008 & 4.83 & 3.78 & \textcolor{red}{43.00} & \textcolor{red}{38.86} & \textcolor{red}{22.32} & \textcolor{red}{19.22} \\
        12 & WPSFD49207 & -5.44 & -5.86 & \textcolor{red}{12.10} & \textcolor{red}{13.65} & \textcolor{red}{10.74} & \textcolor{red}{10.21} \\
        13 & PPIACO & 5.84 & 5.20 & 0.95 & -1.66 & \textcolor{red}{17.36} & \textcolor{red}{17.47} \\
        14 & AMDMNOx & \textcolor{red}{12.93} & \textcolor{red}{11.54} & 4.66 & 7.36 & \textcolor{blue}{-45.61} & \textcolor{blue}{-44.76} \\
        15 & HOUST & 6.36 & 5.57 & \textcolor{red}{36.26} & \textcolor{red}{34.04} & \textcolor{blue}{-32.03} & \textcolor{blue}{-29.10} \\
        16 & S\&P 500 & \textcolor{red}{12.23} & \textcolor{red}{11.87} & \textcolor{red}{14.26} & \textcolor{red}{10.11} & \textcolor{blue}{-47.57} & \textcolor{blue}{-37.57} \\
        17 & EXUSUKx & 6.71 & 6.83 & 7.58 & 8.94 & -7.62 & -6.64 \\
        18 & TB3SMFFM & \textcolor{red}{35.35} & \textcolor{red}{36.34} & \textcolor{red}{67.28} & \textcolor{red}{69.91} & \textcolor{red}{82.91} & \textcolor{red}{81.50} \\
        19 & T5YFFM & \textcolor{red}{12.09} & \textcolor{red}{16.76} & \textcolor{red}{76.40} & \textcolor{red}{75.21} & \textcolor{red}{68.82} & \textcolor{red}{63.57} \\
        20 & AAAFFM & \textcolor{red}{10.81} & \textcolor{red}{14.13} & \textcolor{red}{54.70} & \textcolor{red}{46.35} & \textcolor{red}{69.37} & \textcolor{red}{66.45} \\
        \hline
    \end{tabular}
    \caption{Percentage gains of DFSVL and DFSVML models in forecast accuracy (1-step ahead) relative to the LSVVAR benchmark.  Red: values $\geq 10$; blue: values $\leq -10$.}
    \label{tab:crisis_gain_comparison_dfsv_1step}
\end{table}
\noindent
Table \ref{tab:crisis_gain_comparison_dfsv_1step} summarizes the percentage gain in 1-step-ahead forecast accuracies of DFSVL and DFSVML models relative to the benchmark LSVVAR model across three distinct periods. The gain is defined as $100 \times \left(1 - \text{CSFE}_{\text{model}}/\text{CSFE}_{\text{LSVVAR}}\right)$, where a positive value indicates an improvement in predictive accuracy. `Normal' refers to the normal period since 2000Q1, excluding the `GFC' period (2008Q1--2009Q4) and the `COVID-19' pandemic period (2020Q1--2021Q4).
During the GFC period, the DFSVL and DFSVML models notably improve the forecasting accuracies for most variables (e.g. \#3--\#5, \#11, \#12, \#15, \#16, \#18--\#20). On the other hand, during the COVID-19 pandemic period, they perform poorly, especially for those variables related to the first factor or the real economic activity and growth (e.g., \#1, \#4--\#9, \#14--\#16), while they outperform the benchmark model for variables related to the second factor (e.g., \#2, \#12, \#13), the third factor (e.g., \#3, \#18--\#20), and \#11. 
During normal times, the DFSVL and DFSVML models outperform, especially for variables 
such as \#5, \#14, \#16, \#18--\#20. We note that our proposed models always perform well with respect to \#18--\#20 for all periods. 

Table \ref{tab:crisis_gain_comparison_fsv_1step} summarizes the percentage gain in 1-step-ahead forecast accuracies of static factor models (FSVL and FSVML) relative to the benchmark LSVVAR model.  
They are found to improve the forecasting accuracies during the COVID-19 pandemic period, while their performances are not so good as dynamic factor models during the normal times and the GFC period. This indicates the performance of the econometric models is strongly affected by the social and economic conditions, and the sudden and unforeseen structural break may have occurred during the COVID-19 pandemic period.

In summary, a comprehensive comparison with the LSVVAR benchmark reveals a nuanced picture across different economic regimes. While the LSVVAR model performed robustly during relatively stable periods (as evidenced by the results for `Normal' times in Table \ref{tab:crisis_gain_comparison_dfsv_1step}), our proposed factor models demonstrate clear advantages during the two major economic downturns: the GFC and the COVID-19 pandemic periods.
This crucial divergence indicates that while the benchmark may capture the conditional mean adequately, our proposed factor SV models provide a more accurate characterization of the entire predictive density. By better capturing time-varying uncertainty and tail risk, our models offer more reliable density forecasts, which is a significant advantage for risk management and economic policy analysis in times of crisis.
\begin{table}[H]
    \scriptsize
%    \small
    \centering
    \setlength{\tabcolsep}{3.5pt}
    \begin{tabular}{rl  rr !{\hspace{3pt}} rr !{\hspace{3pt}} rr}
        \hline
         & & \multicolumn{2}{c}{Normal} & \multicolumn{2}{c}{GFC} & \multicolumn{2}{c}{COVID-19} \\
        \cmidrule(lr){3-4} \cmidrule(lr){5-6} \cmidrule(lr){7-8}
        \# & Variable & FSVL & FSVML & FSVL & FSVML & FSVL & FSVML \\
        \hline
        1 & GDPC1 & -7.53 & -6.25 & \textcolor{blue}{-122.17} & \textcolor{blue}{-118.90} & 5.77 & 5.81 \\
        2 & PCECTPI & 6.63 & 7.23 & 6.28 & 7.56 & \textcolor{red}{15.71} & \textcolor{red}{14.33} \\
        3 & FEDFUNDS & \textcolor{blue}{-47.79} & \textcolor{blue}{-54.03} & \textcolor{red}{48.76} & \textcolor{red}{46.41} & \textcolor{red}{58.24} & \textcolor{red}{58.35} \\
        4 & PCECC96 & -4.59 & -3.97 & \textcolor{blue}{-35.14} & \textcolor{blue}{-35.83} & -9.09 & \textcolor{blue}{-10.63} \\
        5 & CMRMTSPLx & -3.36 & -6.56 & \textcolor{blue}{-25.85} & \textcolor{blue}{-30.30} & -3.76 & -0.46 \\
        6 & INDPRO & \textcolor{blue}{-61.12} & \textcolor{blue}{-61.74} & \textcolor{blue}{-88.38} & \textcolor{blue}{-88.10} & 7.27 & 8.54 \\
        7 & CUMFNS & \textcolor{blue}{-28.21} & \textcolor{blue}{-33.83} & \textcolor{blue}{-112.50} & \textcolor{blue}{-111.89} & \textcolor{red}{14.92} & \textcolor{red}{16.67} \\
        8 & UNRATE & \textcolor{blue}{-10.89} & \textcolor{blue}{-12.29} & \textcolor{blue}{-94.31} & \textcolor{blue}{-97.03} & \textcolor{blue}{-13.99} & \textcolor{blue}{-13.65} \\
        9 & PAYEMS & \textcolor{blue}{-15.41} & \textcolor{blue}{-18.10} & \textcolor{blue}{-112.71} & \textcolor{blue}{-115.47} & -9.23 & \textcolor{blue}{-11.05} \\
        10 & CES0600000007 & -2.00 & -2.15 & \textcolor{blue}{-130.68} & \textcolor{blue}{-148.56} & \textcolor{red}{29.53} & \textcolor{red}{30.63} \\
        11 & CES0600000008 & 3.88 & 4.87 & -5.89 & -4.82 & \textcolor{red}{13.76} & \textcolor{red}{10.58} \\
        12 & WPSFD49207 & -4.51 & -4.11 & 8.95 & 9.29 & -5.13 & -5.75 \\
        13 & PPIACO & -1.48 & -1.81 & -9.63 & \textcolor{blue}{-11.17} & 5.01 & 5.75 \\
        14 & AMDMNOx & -3.29 & -4.33 & \textcolor{blue}{-53.44} & \textcolor{blue}{-57.04} & 3.72 & 5.35 \\
        15 & HOUST & \textcolor{red}{10.06} & 6.84 & \textcolor{red}{19.19} & \textcolor{red}{15.65} & 1.31 & 4.55 \\
        16 & S\&P 500 & \textcolor{red}{15.08} & \textcolor{red}{13.09} & \textcolor{red}{12.26} & \textcolor{red}{10.54} & \textcolor{blue}{-23.83} & \textcolor{blue}{-14.53} \\
        17 & EXUSUKx & 9.01 & 8.05 & 9.06 & 9.66 & -0.63 & 0.66 \\
        18 & TB3SMFFM & \textcolor{red}{30.96} & \textcolor{red}{30.17} & \textcolor{red}{58.49} & \textcolor{red}{56.94} & \textcolor{red}{86.42} & \textcolor{red}{86.47} \\
        19 & T5YFFM & 6.73 & 6.42 & \textcolor{red}{62.70} & \textcolor{red}{60.69} & \textcolor{red}{75.10} & \textcolor{red}{73.06} \\
        20 & AAAFFM & 1.53 & 3.14 & \textcolor{red}{34.52} & \textcolor{red}{36.14} & \textcolor{red}{77.02} & \textcolor{red}{75.70} \\
        \hline
    \end{tabular}
    \caption{Percentage gains of FSVL and FSVML models in forecast accuracy (1-step ahead) relative to the LSVVAR benchmark. Red: values $\geq 10$; blue: values $\leq -10$.}
    \label{tab:crisis_gain_comparison_fsv_1step}
\end{table}
%
%\noindent

\section{Conclusion}
\label{sec:conclusion}

This paper has proposed a novel dynamic factor stochastic volatility-in-mean (DFSVM) model, providing a scalable Bayesian framework for analyzing large-scale macroeconomic datasets where time-varying uncertainty can directly influence economic outcomes. By integrating a factor structure with both stochastic volatility and in-mean components, our model is designed to capture key empirical features of macroeconomic data, including leverage effects and risk premium, within a computationally efficient MCMC estimation scheme.

Our empirical application to a large panel of U.S. macroeconomic data yields several key insights. 
The in-sample analysis successfully extracts three economically interpretable latent factors—real economic activity, price levels/inflation, and financial conditions—and finds the strong evidence of negative in-mean effect associated with the real economic activity variables. This confirms the presence of a time-varying risk premium linked to real-side uncertainty. 

The out-of-sample forecasting exercise, splitting the evaluation into three distinct regimes (Normal, GFC, and COVID-19 pandemic), reveals a nuanced picture of model performance. In normal periods, the competitive LSVVAR benchmark remains highly effective for certain level variables, most notably the effective federal funds rate (\#3). However, our factor specifications yield superior or comparable performance for a majority of variables even during these stable intervals, particularly for financial spreads (\#18, \#19, and \#20). A striking reversal is observed during the GFC period, where our models demonstrate superior adaptability in capturing sharp monetary policy adjustments, delivering a gain of over 50\% for the federal funds rate and robustly outperforming the benchmark.

During the turbulent COVID-19 pandemic period, the relative performance between model variants underscores the importance of structural flexibility. Initially, more adaptive specifications without strong factor dynamics, such as the FSVL and FSVML models, demonstrated superior short-term forecasting ability, especially in density prediction. However, as the period progressed into 2021, the results for price-related indicators suggested a return of persistence that was better captured by explicit factor dynamics (DFSVL and DFSVML). These advantages for key financial indicators are qualitatively robust across forecast horizons, although the predictive gains for certain level variables can diminish at longer horizons due to idiosyncratic constraints such as the zero lower bound.

The implications of these findings are twofold. First, the superior forecasting performance during crisis regimes (dynamic models for the GFC period and static models for the COVID-19 pandemic period) highlights our model's strength in capturing heightened uncertainty and tail risks more accurately than traditional large VARs, a crucial advantage for policy analysis and risk management. This advantage was confirmed not only for real economic indicators but also for key financial variables, such as the policy rate, during periods of such extreme stress.

Second, the shifting relative performance between model variants across all three regimes underscores the importance of model flexibility in adapting to different economic conditions. Our investigation of an alternative specification further suggests that the in-mean effect is better modeled as an idiosyncratic, variable-specific phenomenon rather than a common factor-driven one. This allows for a more granular analysis of how common uncertainty propagates into specific macroeconomic outcomes, reinforcing the appropriateness of the proposed DFSVM structure.

This research opens several avenues for future work. While we have focused on forecasting, the proposed framework could be extended to structural analysis by computing generalized impulse response functions to trace the effects of uncertainty shocks. Furthermore, allowing the in-mean coefficients ($\bm{\beta}$) to be time-varying could offer deeper insights into how risk premium evolve in response to changing monetary policy regimes or macroeconomic conditions.
\small
\bibliography{ref_DFSVM_paper}	
\normalsize
%\clearpage
%
\newpage
\appendix
\setcounter{page}{1}
\setcounter{footnote}{0}
\def\thefootnote{\fnsymbol{footnote}}
%\section*{Supplementary Material for ``Dynamic Factor Stochastic Volatility-in-Mean VAR for Large Macroeconomic Panels"} 
\begin{center}
{\LARGE
\textbf{Supplementary Material for \\ 
``Dynamic Factor Stochastic Volatility-in-Mean VAR for Large Macroeconomic Panels"}
}

\vspace{0.5cm}

{\large Daichi Hiraki\footnote{Graduate School of Economics, University of Tokyo, Tokyo 113-0033, Japan}, 
Siddhartha Chib\footnote{Olin School of Business, Washington University, St Louis, USA} 
and Yasuhiro Omori\footnote{Faculty of Economics, University of Tokyo, Tokyo 113-0033, Japan}}
\end{center}

\vspace{0.5cm}

\section{MCMC algorithm}
\label{sec:MCMC algorithm}

For notation simplicity, we first define 
\begin{eqnarray}
    &&\mu_{it} = \exp(h_{it}/2) \rho_i \sigma_i^{-1} \{ h_{i,t+1} - \mu_i - \phi_i (h_{it} - \mu_i) \} I(t<n), \notag \\
    &&\sigma_{it}^2 = \exp(h_{it}) \{ 1-\rho_i^2 I(t<n) \}, \notag \\
    &&\bm{\mu}_{1t} = (\mu_{1t}, \ldots, \mu_{pt})', \quad \bm{\mu}_{2t} =  (\mu_{p+1,t}, \ldots, \mu_{p+q,t})' \label{mu} \\
    &&\mathbf{\Omega}_{1t} = \mbox{diag}(\sigma_{1t}^2, \ldots, \sigma_{pt}^2), \quad \mathbf{\Omega}_{2t} = \mbox{diag}(\sigma_{p+1,t}^2, \ldots, \sigma_{p+q,t}^2), \label{Omega}
\end{eqnarray}
and denote $\beta_j$ be $j$-th component of $\bm{\beta}$.

\subsection*{Generation of $\bm{\beta}$}
\label{Generation beta}
The conditional posterior distribution of $\bm{\beta}$ is $N_p(\hat{\bm{m}}_{\bm{\beta}}, \hat{\mathbf{S}}_{\bm{\beta}})$ where
\begin{eqnarray*}
    &\hat{\bm{m}}_{\bm{\beta}} &= \hat{\mathbf{S}}_{\bm{\beta}} \left\{ \sum_{t=1}^n \mathbf{\Lambda}_{t} \mathbf{\Omega}_{1t}^{-1} \{ \bm{y}_{t} - \sum_{\ell=1}^{L}\mathbf{B}_{\ell}\,\bm{y}_{t-\ell}
    - \mathbf{B}\bm{f}_{t} - \bm{\mu}_{1t} \} + \mathbf{S}_{\bm{\beta}}^{-1} \bm{m}_{\bm{\beta}} \right\} \\
    &\hat{\mathbf{S}}_{\bm{\beta}} &= \left\{ (n-1) \mbox{diag} ( (1-\rho_{1}^2)^{-1}, \ldots, (1-\rho_{p}^2)^{-1} ) + \mathbf{I}_p + \mathbf{S}_{\bm{\beta}}^{-1} \right\}^{-1}.
\end{eqnarray*}
Especially, if we assume independent prior distribution $\beta_i \sim N(m_{\beta}, v_{\beta}^2)$ for $i=1,\ldots,p$, the posterior distribution is $N(\hat{m}_{\beta}, \hat{v}_\beta^2)$ where
\begin{eqnarray*}
    &\hat{m}_{\beta} &= \hat{v}_{\beta}^2 \left\{ \sum_{t=1}^n \exp(\lambda_i h_{it}) \sigma_{it}^{-2} \{ y_{it} - \sum_{\ell=1}^L ( \mathbf{B}_\ell )_{i\cdot} \bm{y}_{t-\ell}
    - \mathbf{B}_{i\cdot} \bm{f}_{t} - \mu_{it} \} + v_{\beta}^{-2} m_{\beta} \right\} \\
    &\hat{v}_{\beta}^2 &= \left\{ (n-1) (1-\rho_{i}^2)^{-1} + 1 + v_{\beta}^{-2} \right\}^{-1}.
\end{eqnarray*}
Here, $( \mathbf{B}_\ell )_{i\cdot}$ is $i$-the row of $\mathbf{B}_\ell$.

\subsection*{Generation of $(\bm{h}, \bm{\alpha})$}
\label{Generation h alpha}
Define
\begin{eqnarray*}
    w_{it} =
    \begin{cases}
        y_{it} - \sum_{\ell=1}^L ( \mathbf{B}_\ell )_{i\cdot} \bm{y}_{t-\ell} - \mathbf{B}_{i \cdot} \bm{f}_t, & i=1, \ldots, p, \\
        f_{i-p,t} - \gamma_{i-p} - \psi_{i-p}(f_{i-p,t-1} - \gamma_{i-p}), & i=p+1, \ldots, p+q.
    \end{cases}
\end{eqnarray*}
where $( \mathbf{B}_\ell )_{i\cdot}$ is $i$-the row of $\mathbf{B}_\ell$.
Conditioned on other parameters, the original DFSVML model reduced to
\begin{eqnarray*}
    &&w_{it} = \begin{cases}
         \beta_{i} \exp(h_{it}/2) +\exp(h_{it}/2) \epsilon_{it}, & i = 1,\ldots, p, \\
        \exp(h_{it}/2) \epsilon_{it}, & i = p+1,\ldots, p+q, 
    \end{cases} \\
    &&h_{i,t+1} = \mu_{i} + \phi_i (h_{it} - \mu_{i}) + \eta_{it}, \\
    &&(\epsilon_{it}, \eta_{it})' \sim N_2(\bm{0}_2, \mathbf{\Sigma}_i).
\end{eqnarray*}
This is the univariate stochastic volatility in mean and standard stochastic volatility model. Thus, we use the generalized mixture sampler, which is a highly efficient sampling method introduced in \cite{HirakiChibOmori(25)}.

\subsection*{Generation of $(\mathbf{B}, \mathbf{\bar{B}})$}
\label{Generation barB} 
Let $\mathbf{Y}_{t_1:t_2} = (\bm{y}_{t_1}, \ldots, \bm{y}_{t_2})'$, $\mathbf{X}_1 = (\bm{f}_1,\ldots,\bm{f}_n)'$, and $\mathbf{X}_2 = (\mathbf{Y}_{0:n-1}, \ldots, \mathbf{Y}_{-L+1:n-L})$. Then, given other parameters and latent volatility, $(\mathbf{B}, \mathbf{\bar{B}})$ is the regression coefficient when regressing $\mathbf{Y}_{1:n}$ on $(\mathbf{X}_1, \mathbf{X}_2)$. Its update can be performed using a straightforward Gibbs sampler.

\subsection*{Generation of $\bm{\psi}$}
\label{Generation psi}
Given $\bm{h}$ and $\bm{\theta}_{\backslash \bm{\psi}}$, we consider the linear model
\begin{equation*}
    \bm{f}_{t} = \bm{\gamma} + \mathbf{\Psi} (\bm{f}_{t-1}-\bm{\gamma}) + \bm{\mu}_{2t} + \bm{\epsilon}_{2t|\bm{h}}, \quad \bm{\epsilon}_{2t|\bm{h}}\sim N_q(\bm{0}_q, \mathbf{\Omega}_{2t}).
\end{equation*}
Then, we define
\begin{eqnarray*}
    \hat{m}_{\psi_j} &=& \hat{v}_{\psi_j}^2 \left\{ \sum_{t=1}^n \sigma_{p+j,t}^{-2} (f_{j,t-1}-\gamma_j) \{ f_{j,t}-\gamma_j - \mu_{p+j,t} \} \right\}, \\
    \hat{v}_{\psi_j}^2 &=& \left( \sum_{t=1}^n \sigma_{p+j,t}^{-2} (f_{j,t-1}-\gamma_j)^2 \right)^{-1}.
\end{eqnarray*}
Given the current value $\psi_j$, we generate a candidate $\psi_j^\dag$ for $j=1, \ldots, q$ from $TN_{(-1,1)}(\hat{m}_{\psi_j}, \hat{v}_{\psi_j}^2)$ and accept it with Metropolis-Hastings probability \citep{ChibGreenberg(95)}
\begin{equation*}
    \min \left\{ 1, \frac{(1+\psi_j^\dag)^{a_\psi-1} (1-\psi_j^\dag)^{b_\psi-1}}{(1+\psi_j)^{a_\psi-1} (1-\psi_j)^{b_\psi-1}} \right\}.
\end{equation*}

\subsection*{Generation of $\bm{\gamma}$}
\label{Generation gamma}
The conditional posterior distribution of $\bm{\gamma}$ is $N_q(\hat{\bm{m}}_{\bm{\gamma}}, \hat{\mathbf{S}}_{\bm{\gamma}})$ where
\begin{eqnarray*}
    &\hat{\bm{m}}_{\bm{\gamma}} &= \hat{\mathbf{S}}_{\bm{\gamma}} \left\{ \sum_{t=1}^n (\mathbf{I}_q-\mathbf{\Psi})' \mathbf{\Omega}_{2t}^{-1} \{ \bm{f}_{t} - \mathbf{\Psi} \bm{f}_{t-1} -\bm{\mu}_{2t} \} + \mathbf{S}_{\bm{\gamma}}^{-1} \bm{m}_{\bm{\gamma}} \right\} \\  &\hat{\mathbf{S}}_{\bm{\gamma}} &= \left( \sum_{t=1}^n (\mathbf{I}_q-\mathbf{\Psi})' \mathbf{\Omega}_{2t}^{-1} (\mathbf{I}_q-\mathbf{\Psi}) + \mathbf{S}_{\bm{\gamma}}^{-1} \right)^{-1}.
%    = \left( \sum_{t=1}^n \mbox{diag} ( \{ \sigma_{p+j,t}^{-2} (1-\psi_j)^2 \}_{j=1,\ldots, q} ) + \mathbf{S}_{\bm{\gamma}}^{-1} \right)^{-1}
\end{eqnarray*}

\subsection*{Generation of $\bm{f}$}
\label{Generation f}
Let us define $\bm{z}_t$ to be $\bm{z}_t = \bm{y}_t - \sum_{\ell=1}^L \mathbf{B}_\ell \bm{y}_{t-\ell}$. Given $\bm{h}$ and parameters, the DFSVML model is reduced to the linear Gaussian state space model:
\begin{eqnarray*}
    \bm{z}_{t} &=& \mathbf{X}_t \bm{\delta} + \mathbf{B}\bm{f}_t + \mathbf{G}_t \bm{u}_{t}, \quad t=1, \ldots, n, \\
    \bm{f}_{t+1} &=& \mathbf{W}_t \bm{\delta} + \mathbf{\Psi} \bm{f}_t + \mathbf{H}_t \bm{u}_{t}, \quad t=0, \ldots, n-1, \\
    \bm{f}_0 &\equiv& \bm{0}_q, \quad \bm{u}_t \overset{\text{\it i.i.d.}}{\sim} N_{p+q}(\bm{0}_{p+q}, \mathbf{I}_{p+q}),
\end{eqnarray*}
where $\bm{\delta} = \bm{d} + \mathbf{N} \bm{\gamma} = (\bm{1}_{p+q}', \bm{\gamma}')'$ with $\bm{d} = (\bm{1}_{p+q}', \bm{0}_{q}')'$ and $\mathbf{N} = [\mathbf{O}_{(p+q) \times q}', \mathbf{I}_q']'$,
\begin{eqnarray*}    
    \mathbf{X}_{t} &=& \left[ \mbox{diag}(\mathbf{\Lambda}_{t}\bm{\beta} + \bm{\mu}_{1t}), \mathbf{O}_{p \times 2q} \right], \quad \mathbf{G}_{t} = \left[ \mathbf{\Omega}_{1t}^{1/2}, \mathbf{O}_{p \times q} \right], \\
    \mathbf{W}_{t} &=& \left[ \mathbf{O}_{q \times p}, \mbox{diag}( \bm{\mu}_{2,t+1}), \mathbf{I}_{q} - \mathbf{\Psi} \right], \quad \mathbf{H}_{t} = \left[ \mathbf{O}_{q \times p}, \mathbf{\Omega}_{2,t+1}^{1/2} \right], \\
    \mathbf{W}_{0} &=& \left[ \mathbf{O}_{q \times p}, \mbox{diag}( \bm{\mu}_{2,1}), \mathbf{I}_{q} \right].
\end{eqnarray*}
We generate $\bm{f}$ using a simulation smoother introduced by \cite{DeShephard(95)} and \cite{DurbinKoopman(02)}.
\\

\noindent
{\it Remark}. Generation of $(\bm{f}, \bm{\gamma})$ is more efficient if we use the augmented Kalman filter and the simulation smoother (see \cite{De(91)}; \cite{DeShephard(95)}; \cite{KimShephardChib(98)}).
The state equation above is defined over $t=1, \ldots, n-1$, but over $t=1,\ldots,n$ in \cite{DeShephard(95)}. $\mathbf{W}_n$ and $\bm{\mathbf{H}_n}$ (and of course $\mathbf{\Psi}$ at time $t=n$) is arbitrary. For the sake of calculation, we defined $\bm{f}_{n+1} = \mathbf{\Psi} \bm{f}_n$. That is, we set $\mathbf{W}_n = \mathbf{O}_{q \times (p+2q) }$ and $\mathbf{H}_n = \mathbf{O}_{q \times (p+q)}$.

\newpage
\section{Estimation Results}
\label{sec:estimation_results}
\subsection{Estimation Results for $\phi_i$ and $\rho_i$}
\begin{table}[H]
%    \scriptsize
    \footnotesize
%    \small
    \centering
    \begin{tabular}{rrcrrrcrr}
         \hline
         & $\phi_i$ &  &  &  & $\rho_i$ &  &  &  \\
         \hline
        $i$ & Mean & 95\% interval & IF  & Pr(+) & Mean & 95\% interval & IF  & Pr(+) \\
        \hline
        1 & 0.878 & ( 0.704,  0.986) & 15  & 1.000 & -0.117 & (-0.524, 0.283) & 24  & 0.287 \\
        2 & 0.934 & ( 0.779,  0.996) & 23  & 1.000 & 0.177 & (-0.351, 0.621) & 33  & 0.775 \\
        3 & 0.963 & ( 0.904,  0.997) & 11  & 1.000 & 0.130 & (-0.232, 0.492) & 19  & 0.757 \\
        4 & 0.873 & ( 0.727,  0.971) & 10  & 1.000 & -0.122 & (-0.448, 0.193) & 15  & 0.235 \\
        5 & 0.863 & ( 0.679,  0.984) & 29  & 1.000 & -0.034 & (-0.478, 0.445) & 32  & 0.430 \\
        6 & 0.785 & ( 0.564,  0.948) & 89  & 1.000 & -0.538 & (-0.842, -0.139) & 65  & 0.005 \\
        7 & 0.880 & ( 0.643,  0.997) & 194  & 1.000 & -0.565 & (-0.918, 0.761) & 222  & 0.083 \\
        8 & 0.760 & ( 0.552,  0.903) & 10  & 1.000 & 0.091 & (-0.182, 0.356) & 10  & 0.744 \\
        9 & 0.782 & ( 0.620,  0.903) & 10  & 1.000 & 0.191 & (-0.067, 0.432) & 12  & 0.929 \\
        10 & 0.830 & ( 0.481,  0.991) & 29  & 1.000 & -0.257 & (-0.773, 0.308) & 42  & 0.174 \\
        11 & 0.922 & ( 0.720,  0.997) & 33  & 1.000 & 0.411 & (-0.158, 0.829) & 41  & 0.929 \\
        12 & 0.888 & ( 0.566,  0.998) & 34  & 1.000 & -0.276 & (-0.902, 0.741) & 138  & 0.292 \\
        13 & 0.962 & ( 0.809,  0.999) & 54  & 1.000 & 0.507 & (-0.373, 0.906) & 110  & 0.905 \\
        14 & 0.824 & ( 0.549,  0.989) & 35  & 1.000 & 0.066 & (-0.352, 0.564) & 27  & 0.600 \\
        15 & 0.906 & ( 0.710,  0.996) & 23  & 1.000 & -0.333 & (-0.749, 0.102) & 29  & 0.072 \\
        16 & 0.682 & ( 0.427,  0.872) & 11  & 1.000 & -0.459 & (-0.709, -0.179) & 12  & 0.001 \\
        17 & 0.972 & ( 0.913,  0.999) & 6  & 1.000 & 0.085 & (-0.251, 0.400) & 11  & 0.696 \\
        18 & 0.934 & ( 0.857,  0.988) & 5  & 1.000 & -0.282 & (-0.597, 0.055) & 17  & 0.049 \\
        19 & 0.888 & ( 0.657,  0.996) & 40  & 1.000 & -0.388 & (-0.859, 0.159) & 60  & 0.084 \\
        20 & 0.980 & ( 0.918,  0.999) & 61  & 1.000 & -0.175 & (-0.781, 0.543) & 72  & 0.308 \\
        21 & 0.962 & ( 0.884,  0.995) & 123  & 1.000 & -0.377 & (-0.609, -0.114) & 14  & 0.003 \\
        22 & 0.985 & ( 0.956,  0.998) & 29  & 1.000 & 0.254 & (-0.063, 0.542) & 15  & 0.944 \\
        23 & 0.991 & ( 0.975,  0.999) & 8  & 1.000 & 0.115 & (-0.201, 0.431) & 20  & 0.754 \\
        \hline
    \end{tabular}
    \caption{Estimation result of $\phi_i$ and $\rho_i$ with $i=21, 22, 23$ for the first, second and third factors.}
    \label{tab:result_phi_rho}
    \normalsize
\end{table}
\subsection{Estimation Results for $\mu_i$ and $\sigma_i$}
\begin{table}[H]
    \footnotesize
%    \small
    \centering
    \begin{tabular}{rrcrrrcrr}
        \hline
         & $\mu_i$ &  &  &  & $\sigma_i$ &  &  &  \\
         \hline
        $i$ & Mean & 95\% interval & IF  & Pr(+) & Mean & 95\% interval & IF  & Pr(+) \\
        \hline
        1 & -1.638 & (-2.357, -0.890) & 3 & 0.005 & 0.443 & ( 0.166, 0.775) & 25 & 1.000 \\
        2 & -2.690 & (-3.795, -1.649) & 2 & 0.004 & 0.296 & ( 0.101, 0.648) & 31 & 1.000 \\
        3 & -6.173 & (-9.513, -2.080) & 2 & 0.011 & 0.608 & ( 0.356, 0.949) & 17 & 1.000 \\
        4 & -1.739 & (-2.594, -0.908) & 2 & 0.004 & 0.621 & ( 0.375, 0.938) & 14 & 1.000 \\
        5 & -1.809 & (-2.461, -1.069) & 6 & 0.003 & 0.476 & ( 0.154, 0.863) & 46 & 1.000 \\
        6 & -4.394 & (-5.390, -3.663) & 48 & 0.000 & 0.847 & ( 0.414, 1.434) & 93 & 1.000 \\
        7 & -8.709 & (-13.783, -5.534) & 65 & 0.007 & 1.032 & ( 0.324, 2.177) & 214 & 1.000 \\
        8 & -4.985 & (-5.624, -4.326) & 5 & 0.000 & 0.935 & ( 0.613, 1.352) & 14 & 1.000 \\
        9 & -4.288 & (-5.084, -3.473) & 4 & 0.000 & 1.145 & ( 0.803, 1.574) & 14 & 1.000 \\
        10 & -2.980 & (-3.423, -2.517) & 4 & 0.001 & 0.280 & ( 0.050, 0.707) & 43 & 1.000 \\
        11 & -1.500 & (-2.240, -0.687) & 2 & 0.007 & 0.248 & ( 0.056, 0.633) & 41 & 1.000 \\
        12 & -2.870 & (-3.681, -1.929) & 8 & 0.003 & 0.174 & ( 0.037, 0.563) & 83 & 1.000 \\
        13 & -2.943 & (-5.543, -0.212) & 6 & 0.022 & 0.294 & ( 0.073, 0.814) & 86 & 1.000 \\
        14 & -1.447 & (-1.984, -0.836) & 4 & 0.004 & 0.447 & ( 0.102, 0.885) & 44 & 1.000 \\
        15 & -3.011 & (-3.711, -2.327) & 2 & 0.002 & 0.268 & ( 0.070, 0.622) & 32 & 1.000 \\
        16 & -0.539 & (-0.962, -0.117) & 4 & 0.008 & 0.673 & ( 0.332, 1.080) & 16 & 1.000 \\
        17 & -1.721 & (-7.451, 3.245) & 1 & 0.145 & 0.605 & ( 0.377, 0.906) & 10 & 1.000 \\
        18 & -3.175 & (-4.964, -1.329) & 1 & 0.007 & 0.678 & ( 0.434, 0.990) & 9 & 1.000 \\
        19 & -3.632 & (-4.604, -2.813) & 2 & 0.002 & 0.399 & ( 0.083, 0.901) & 52 & 1.000 \\
        20 & -7.158 & (-13.229, 0.255) & 15 & 0.027 & 0.518 & ( 0.167, 1.217) & 113 & 1.000 \\
        21 & 0.000 & - & - & - & 0.687 & ( 0.435, 1.026) & 15 & 1.000 \\
        22 & 0.000 & - & - & - & 0.506 & ( 0.268, 0.805) & 14 & 1.000 \\
        23 & 0.000 & - & - & - & 0.614 & ( 0.383, 0.924) & 13 & 1.000 \\
        \hline
    \end{tabular}
    \caption{Estimation result of $\mu_i$ and $\sigma_i$ with $i=21, 22, 23$ for the first, second and third factors..}
    \label{tab:result_mu_sigma}
    \normalsize
\end{table}
\begin{sidewaystable}
\subsection{Estimation Results for Factor Loading Matrix}
    \footnotesize
%    \scriptsize
    \centering
    \begin{tabular}{rl rcrr rcrr rcrr}
        \hline
        & & \multicolumn{4}{c}{$\mathbf{B}_{i1}$} & \multicolumn{4}{c}{$\mathbf{B}_{i2}$} & \multicolumn{4}{c}{$\mathbf{B}_{i3}$} \\
        \cline{3-6} \cline{7-10} \cline{11-14}
        \# & Variable & Mean & 95\% interval & IF & Pr(+) & Mean & 95\% interval & IF & Pr(+) & Mean & 95\% interval & IF & Pr(+) \\
        \hline
        1 & GDPC1 & 1.333 & ( 0.683,  2.051) & 249 & 1.000 & -0.056 & (-0.364,  0.229) & 100 & 0.350 & -0.156 & (-0.705,  0.397) & 110 & 0.291 \\
        2 & PCECTPI & 0.036 & (-0.117,  0.195) & 61 & 0.696 & 1.368 & ( 0.820,  2.002) & 167 & 1.000 & 0.284 & (-0.058,  0.663) & 29 & 0.949 \\
        3 & FEDFUNDS & 0.058 & ( 0.023,  0.101) & 185 & 1.000 & 0.008 & (-0.016,  0.034) & 58 & 0.750 & 1.063 & ( 0.713,  1.417) & 252 & 1.000 \\
        4 & PCECC96 & 0.757 & ( 0.370,  1.204) & 218 & 1.000 & -0.165 & (-0.412,  0.032) & 72 & 0.053 & -0.174 & (-0.639,  0.287) & 46 & 0.225 \\
        5 & CMRMTSPLx & 1.604 & ( 0.835,  2.442) & 262 & 1.000 & -0.106 & (-0.437,  0.188) & 131 & 0.237 & -0.206 & (-0.810,  0.398) & 124 & 0.256 \\
        6 & INDPRO & 1.644 & ( 0.870,  2.485) & 272 & 1.000 & 0.176 & (-0.063,  0.484) & 217 & 0.930 & -0.017 & (-0.495,  0.512) & 212 & 0.473 \\
        7 & CUMFNS & 0.526 & ( 0.276,  0.793) & 273 & 1.000 & 0.069 & ( 0.001,  0.167) & 233 & 0.978 & 0.002 & (-0.147,  0.166) & 219 & 0.515 \\
        8 & UNRATE & -0.219 & (-0.341, -0.111) & 243 & 0.000 & -0.007 & (-0.059,  0.043) & 96 & 0.391 & -0.081 & (-0.192,  0.022) & 123 & 0.064 \\
        9 & PAYEMS & 0.522 & ( 0.272,  0.791) & 263 & 1.000 & 0.053 & (-0.036,  0.161) & 165 & 0.887 & 0.021 & (-0.170,  0.219) & 157 & 0.582 \\
        10 & CES0600000007 & 0.535 & ( 0.276,  0.820) & 259 & 1.000 & 0.051 & (-0.064,  0.185) & 117 & 0.811 & 0.041 & (-0.199,  0.289) & 107 & 0.625 \\
        11 & CES0600000008 & 0.164 & ( 0.026,  0.351) & 78 & 0.992 & 0.161 & (-0.014,  0.375) & 30 & 0.964 & -0.330 & (-0.790,  0.085) & 19 & 0.061 \\
        12 & WPSFD49207 & 0.014 & (-0.231,  0.251) & 78 & 0.557 & 2.368 & ( 1.420,  3.455) & 171 & 1.000 & 0.273 & (-0.149,  0.752) & 32 & 0.894 \\
        13 & PPIACO & 0.101 & (-0.108,  0.336) & 76 & 0.834 & 2.319 & ( 1.400,  3.373) & 171 & 1.000 & 0.289 & (-0.103,  0.750) & 37 & 0.919 \\
        14 & AMDMNOx & 1.464 & ( 0.757,  2.218) & 258 & 1.000 & 0.396 & ( 0.066,  0.838) & 114 & 0.992 & 0.244 & (-0.378,  0.899) & 116 & 0.773 \\
        15 & HOUST & 0.237 & ( 0.113,  0.385) & 215 & 1.000 & 0.031 & (-0.064,  0.133) & 39 & 0.744 & -0.273 & (-0.510, -0.067) & 51 & 0.004 \\
        16 & S\&P 500 & 0.202 & (-0.058,  0.498) & 40 & 0.938 & -0.049 & (-0.427,  0.301) & 5 & 0.395 & -0.128 & (-0.739,  0.451) & 6 & 0.337 \\
        17 & EXUSUKx & 0.056 & (-0.012,  0.139) & 41 & 0.948 & 0.123 & (-0.096,  0.391) & 12 & 0.857 & 0.082 & (-0.297,  0.450) & 7 & 0.682 \\
        18 & TB3SMFFM & -0.032 & (-0.095,  0.030) & 45 & 0.134 & 0.015 & (-0.041,  0.073) & 19 & 0.708 & -1.319 & (-1.861, -0.857) & 174 & 0.000 \\
        19 & T5YFFM & -0.041 & (-0.131,  0.043) & 102 & 0.156 & 0.206 & ( 0.093,  0.363) & 120 & 1.000 & -2.473 & (-3.279, -1.682) & 249 & 0.000 \\
        20 & AAAFFM & -0.091 & (-0.168, -0.031) & 198 & 0.001 & 0.125 & ( 0.053,  0.225) & 148 & 1.000 & -1.979 & (-2.596, -1.332) & 259 & 0.000 \\
        \hline
    \end{tabular}
    \caption{Estimation results of the factor loading matrix, $\mathbf{B}$.}
    \label{tab:result_B}
\end{sidewaystable}
\newpage
\section{Posterior densities of $\bm{\beta}$ and $\mathbf{B}$}
\label{sec:posterior_summary}
\subsection{Posterior densities of $\bm{\beta}$}
\begin{figure}[H]
    \centering
    \includegraphics[width=0.9\linewidth]{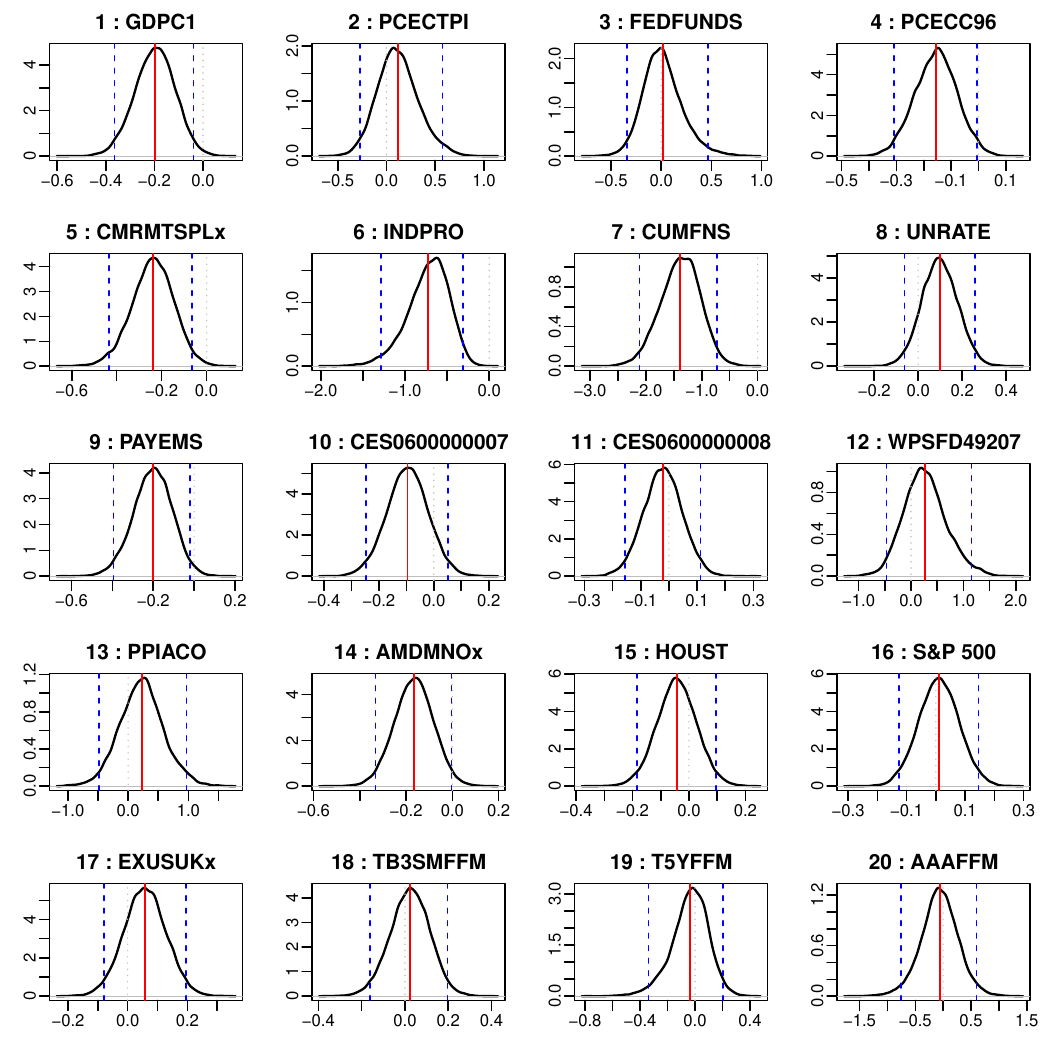}
    \caption{Posterior densities of the risk premium coefficients $\bm{\beta}$ with  posterior means (solid red), and 95\% credible intervals (dashed blue).}
    \label{fig:beta_density}
\end{figure}

To check the possible label switching problem for
$\bm{\beta}$, Figure \ref{fig:beta_density} shows the posterior densities of $\bm{\beta}$ for the pre-pandemic period. All posterior densities are unimodal, which implies no label switching for these parameters. The results are similar for other periods and hence omitted.

\subsection{Posterior densities of ($\mathbf{B}_{15,1}, \mathbf{B}_{15,2}, \mathbf{B}_{15,3}$)}
\begin{figure}[H]
    \centering
    \includegraphics[width=0.9\linewidth]{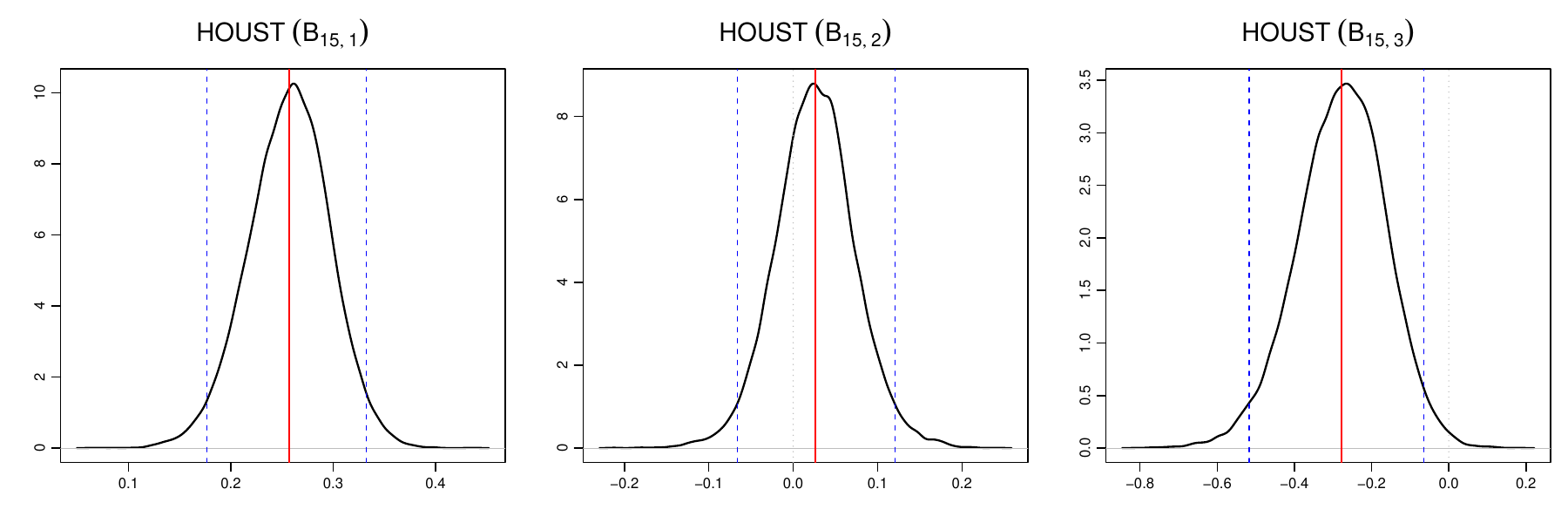}
    \caption{Posterior densities of the selected factor loading elements for AMDMNOx ($\mathbf{B}_{15,1}, \mathbf{B}_{15,2}, \mathbf{B}_{15,3}$). The unimodal posterior distributions confirm the stability of factor identification.}
    \label{fig:B_density}
\end{figure}

Figure \ref{fig:B_density} focuses on the factor loadings for \# 15 HOUST as an example. This variable is particularly suitable for diagnosing the potential label switching because its posterior means across the three factors are notably distinct (0.24, 0.03, and -0.27). If label switching or sign-invariance issues had occurred during the MCMC iterations, these densities would exhibit clear multimodality. However, the estimated densities are unimodal, confirming that our factor identification remains stable throughout the sampling process without the need for artificial constraints.

\newpage
\section{Prediction Procedure}
\label{sec:prediction}

As pointed out in \cite{Kastner(19)}, approximating the posterior predictive distribution by sampling future values of the latent factors can be numerically unstable. To circumvent this issue, we cast the model into a VAR(1) representation to compute the predictive moments analytically, conditional on a future path of volatilities.
The prediction procedure for each MCMC draw is as follows. First, we simulate a path of future log-volatilities, $\{h_{t+1}, h_{t+2}, \ldots, h_{t+k}\}$, from their autoregressive processes. Second, conditional on the model parameters and this simulated path, the model becomes a linear VAR process. This allows for the direct computation of the predictive distribution's moments.

We define a $(pL+q) \times 1$ vector $\bm{z}_t = (\bm{y}'_t, \bm{y}'_{t-1}, \ldots, \bm{y}'_{t-L+1}, \bm{f}'_{t+1})'$. The evolution of this vector can be described by the following VAR(1) system:
\begin{equation}
    \bm{z}_t
    =
    \begin{pmatrix}
        \mathbf{\Lambda}_{t+1} \bm{\beta} \\
        \bm{0}\\
        \vdots \\
        \bm{0} \\
        (\mathbf{I}_q - \mathbf{\Psi})\,\bm{\gamma}
    \end{pmatrix}
    +
    \begin{pmatrix}
        \mathbf{B}_{1} & \mathbf{B}_{2} & \ldots & \mathbf{B}_{L-1} & \mathbf{B}_{L} & \mathbf{B} \\
        \mathbf{I}_{p} & \mathbf{O} & \ldots & \mathbf{O} & \mathbf{O} & \mathbf{O} & \\
         &  & \vdots &  &  \\
        \mathbf{O} & \mathbf{O} & \ldots & \mathbf{I}_{p} & \mathbf{O} & \mathbf{O} \\
        \mathbf{O} & \mathbf{O} & \ldots & \mathbf{O} & \mathbf{O} & \mathbf{\Psi} 
    \end{pmatrix}
    \bm{z}_{t-1}
    +
    \bm{\zeta}_{t},
\end{equation}
where the error term $\bm{\zeta}_{t}$ is normally distributed, $\bm{\zeta}_{t} \sim N(\bm{0}, \mathbf{\Sigma}_{t})$, with a block-diagonal covariance matrix:
\begin{equation}
    \mathbf{\Sigma}_{t} = \mathrm{diag}(\mathbf{V}_{1t}, \mathbf{O}_{p(L-1)}, \mathbf{V}_{2,t+1}).
\end{equation}
Using this VAR representation, we compute the conditional mean $E[\bm{y}_{t+k} | \mathcal{D}_t, \bm{\theta}, \bm{h}_{t+1:t+k}]$ and the associated variance, where $\mathcal{D}_t$ denotes the information set up to time $t$.
The final posterior predictive mean, $E[\bm{y}_{t+k} | \mathcal{D}_t]$, and the log predictive likelihood are then obtained by averaging these conditional moments over all MCMC draws of the parameters and the corresponding simulated volatility paths.

\section{Details of Predictive Performance Comparison}
\label{sec:details-predictive-performance}
\subsection{1-step Ahead Forecast for All Variables}
\begin{figure}[H]
    \centering
    \includegraphics[width=0.85\linewidth]{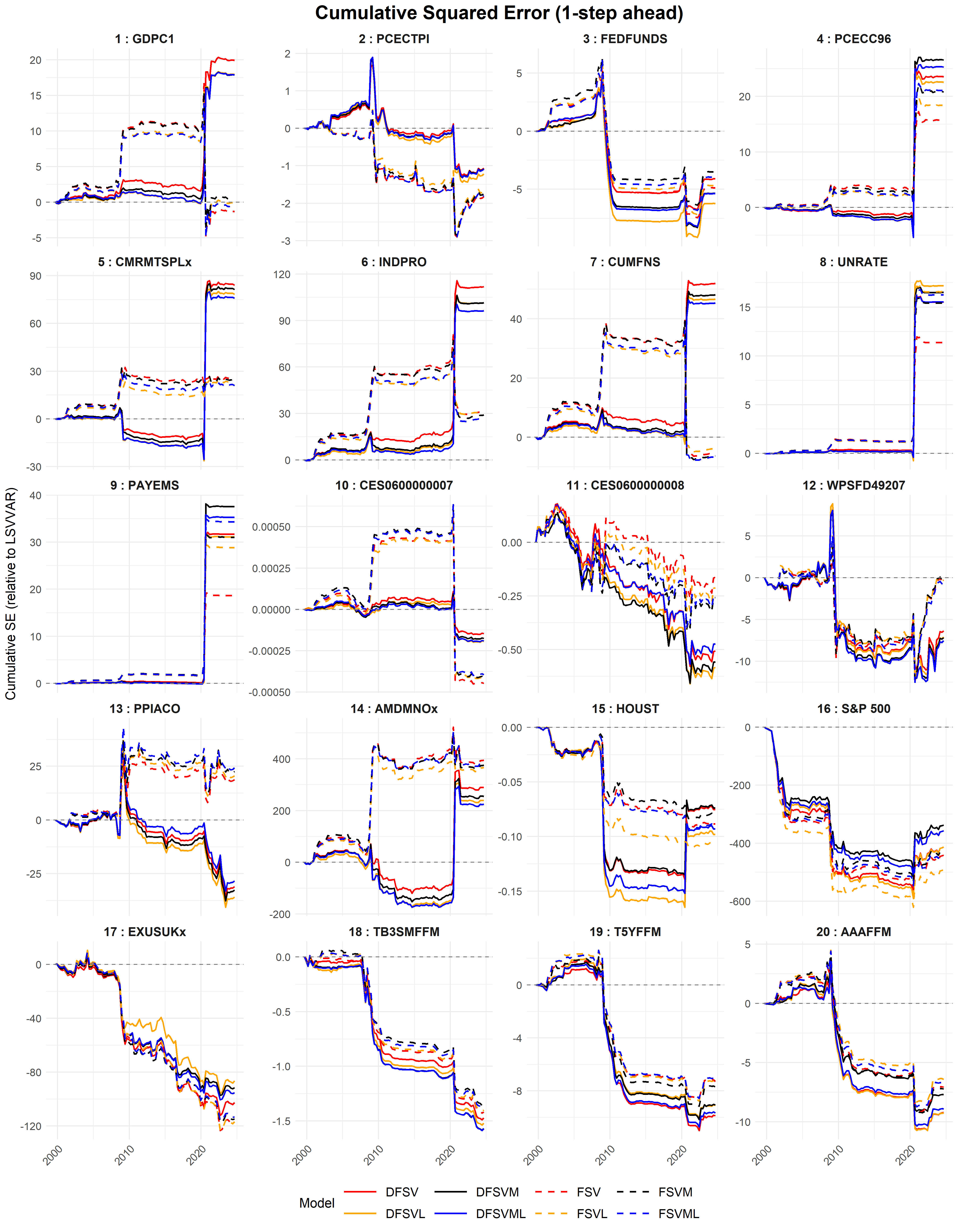}
    \caption{Cumulative Squared Forecast Error (CSFE) relative to the LSVVAR benchmark (zero-line). 2000Q1-2024Q3.}
    \label{fig:result_csfe_1step}
\end{figure}

\begin{figure}[H]
    \centering
    \includegraphics[width=0.85\linewidth]{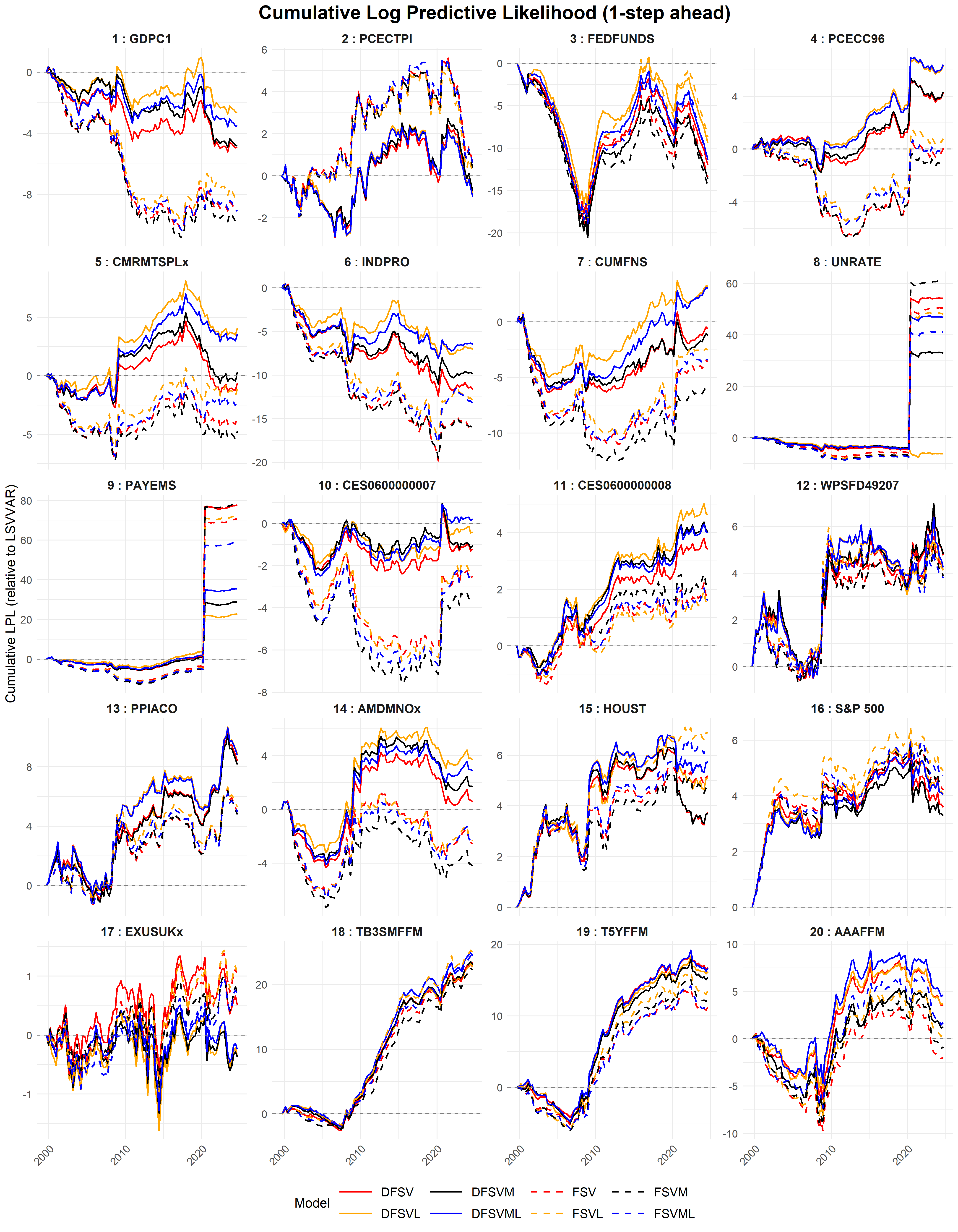}
    \caption{Cumulative log predictive likelihood (LPL) relative to the LSVVAR benchmark (zero-line). 2000Q1-2024Q3.}
    \label{fig:result_clpl_1step}
\end{figure}

\subsection{4-step Ahead Forecast for All Variables}
\begin{figure}[H]
    \centering
    \includegraphics[width=0.85\linewidth]{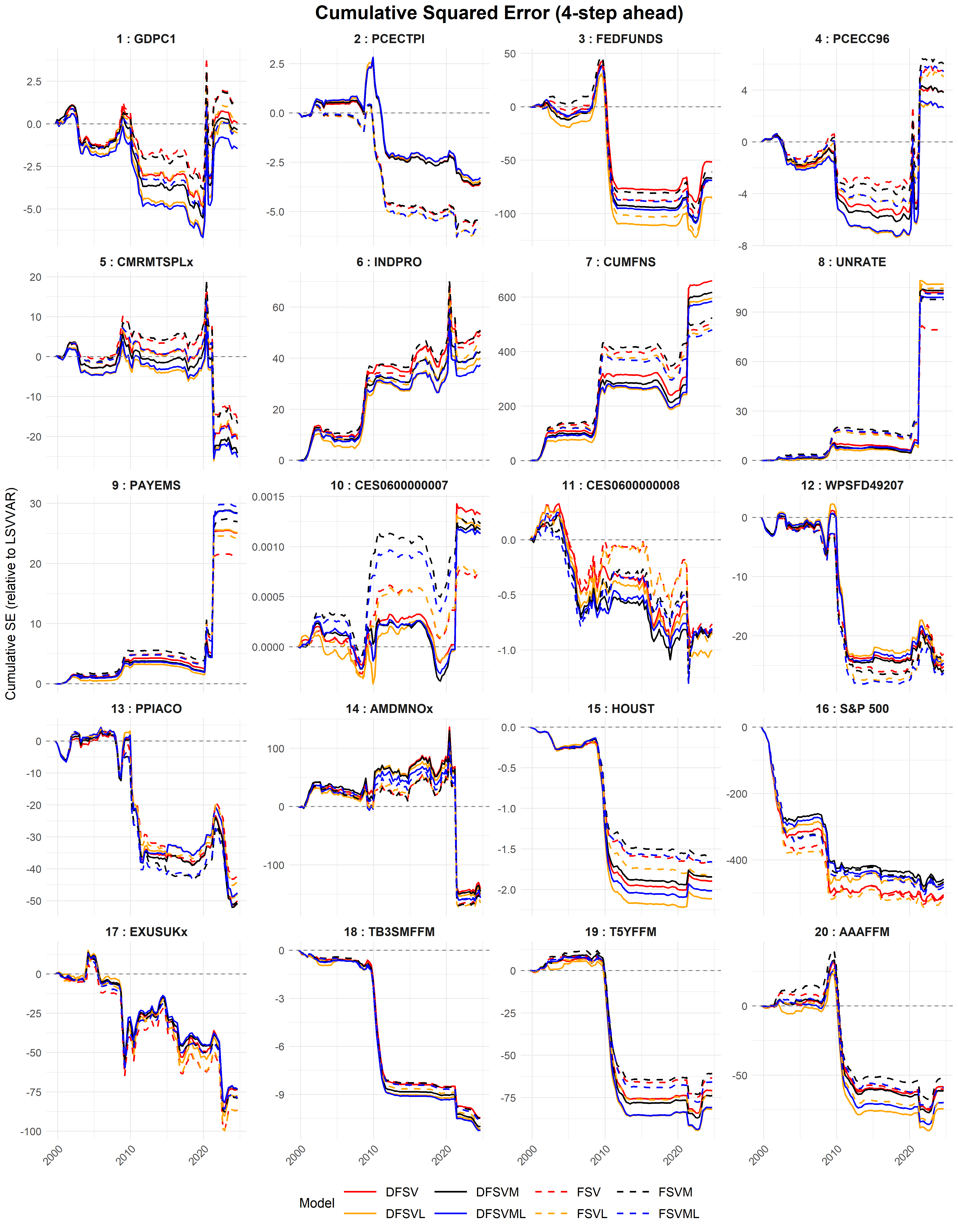}
    \caption{Cumulative Squared Forecast Error (CSFE) relative to the LSVVAR benchmark (zero-line). 2000Q1-2024Q3.}
    \label{fig:result_csfe_4step}
\end{figure}

\begin{figure}[H]
    \centering
    \includegraphics[width=0.85\linewidth]{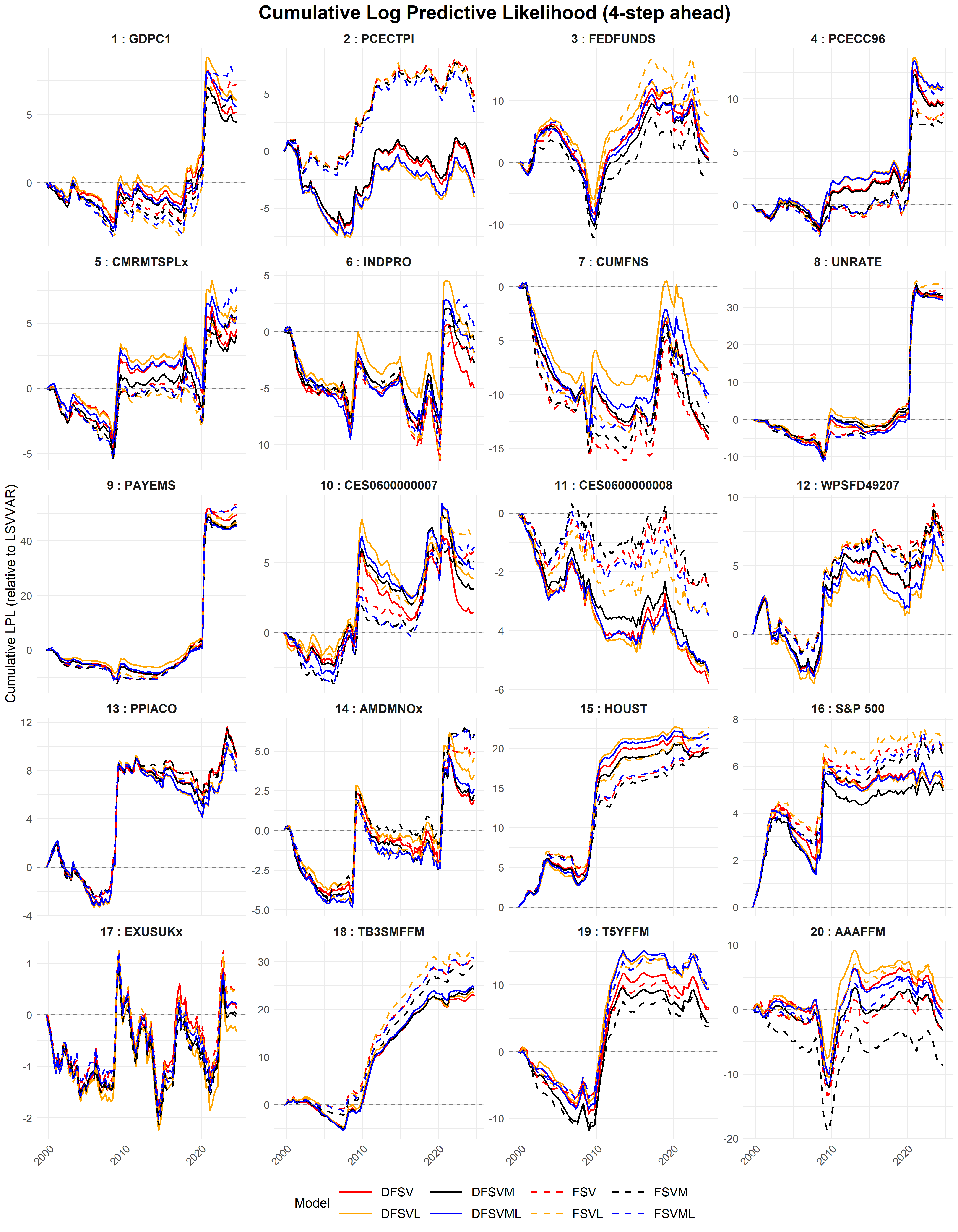}
    \caption{Cumulative log predictive likelihood (LPL) relative to the LSVVAR benchmark (zero-line). 2000Q1-2024Q3.}
    \label{fig:result_clpl_4step}
\end{figure}

\subsection{4-step Ahead Forecast during GFC Period}
\begin{figure}[H]
    \centering
    \includegraphics[width=0.9\linewidth]{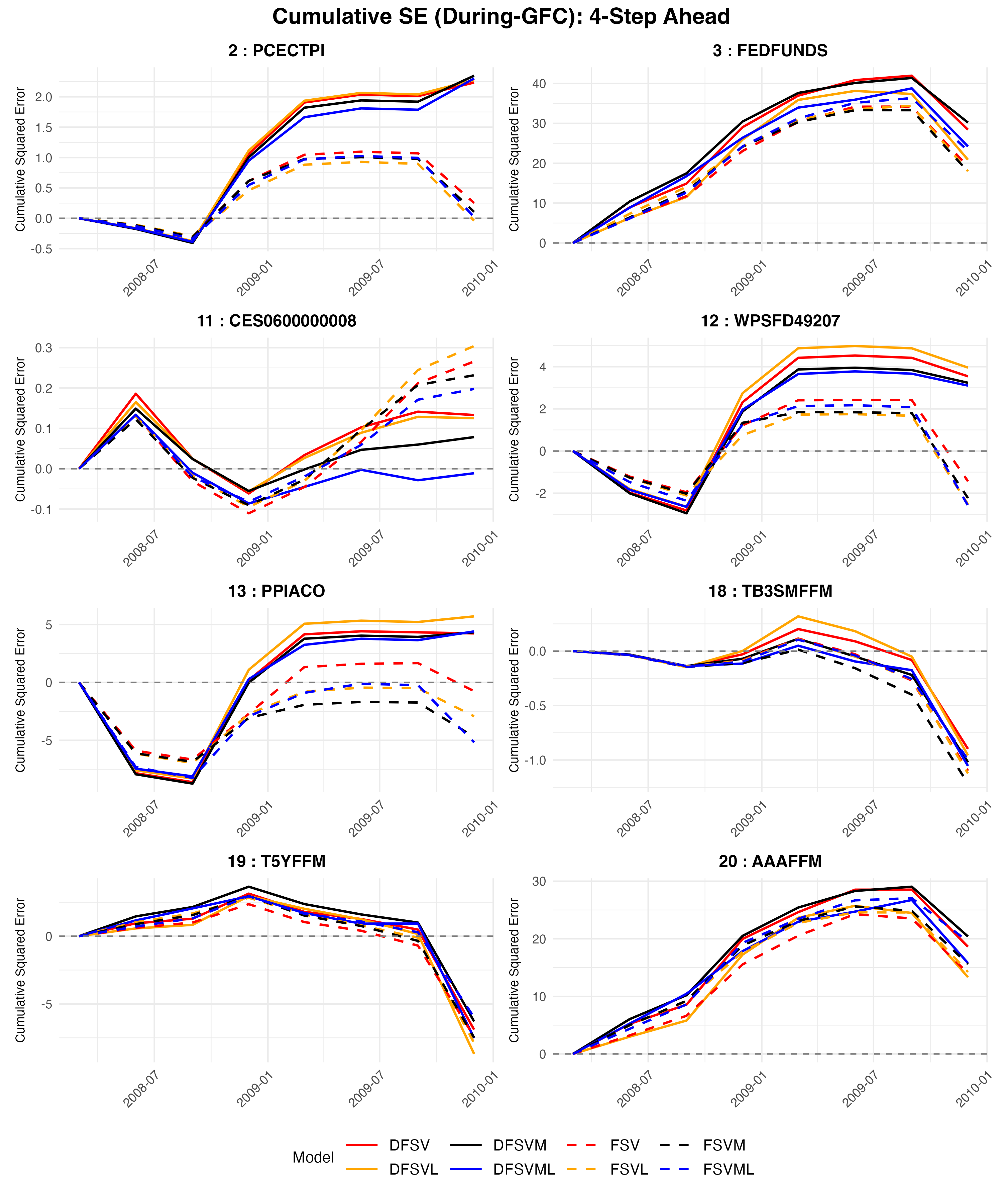}
    \caption{Cumulative squared forecast error (CSFE) relative to the LSVVAR benchmark (zero-line). 2008Q1-2009Q4.}
    \label{fig:result_csfe_gfc_4step}
\end{figure}
\begin{figure}[H]
    \centering
    \includegraphics[width=0.9\linewidth]{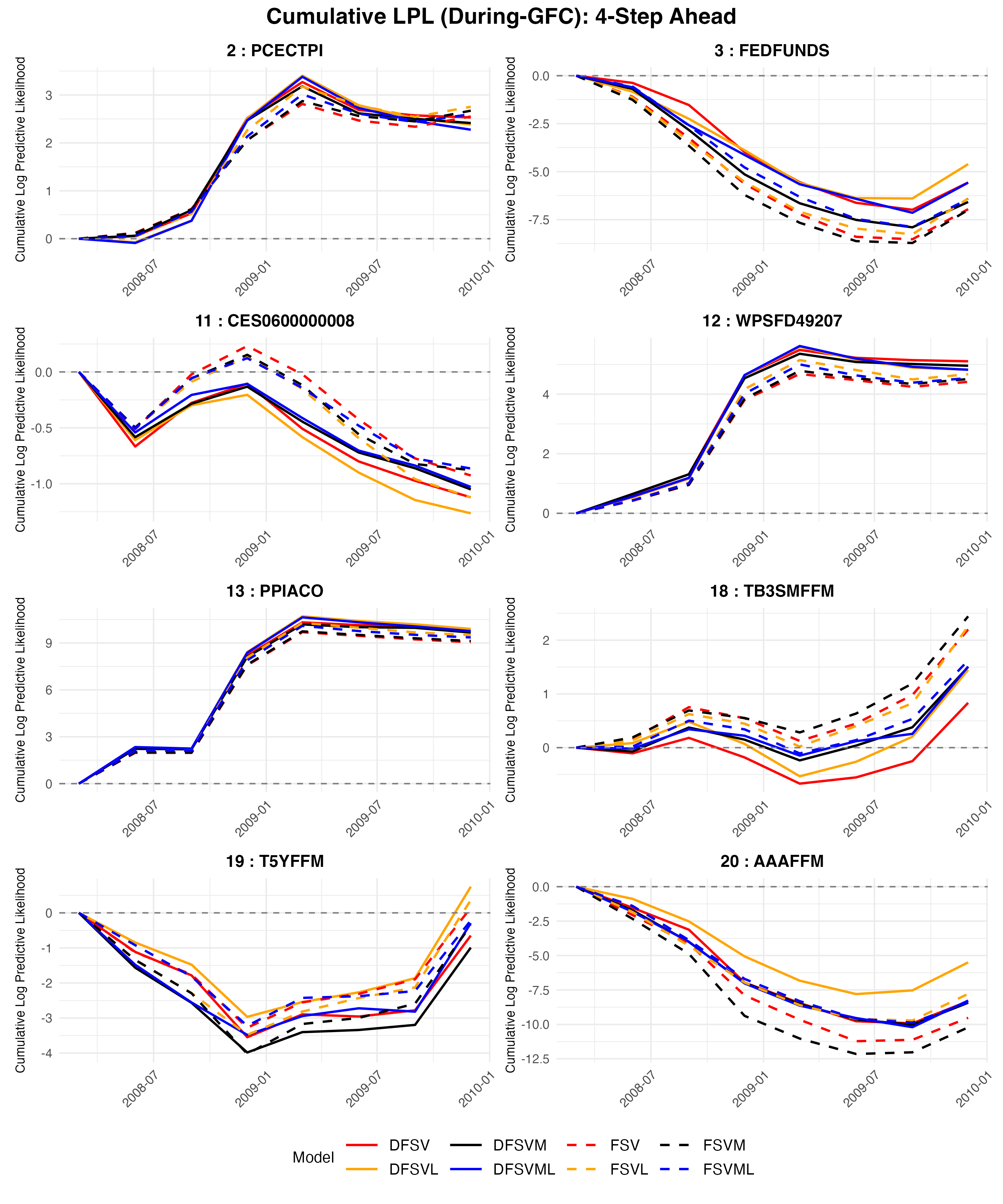}
    \caption{Cumulative log predictive likelihood (LPL) relative to the LSVVAR benchmark (zero-line). 2008Q1-2009Q4.}
    \label{fig:result_clpl_gfc_4step}
\end{figure}

\subsection{4-step Ahead Forecast during COVID-19 Pandemic Period}
\begin{figure}[H]
    \centering
    \includegraphics[width=0.9\linewidth]{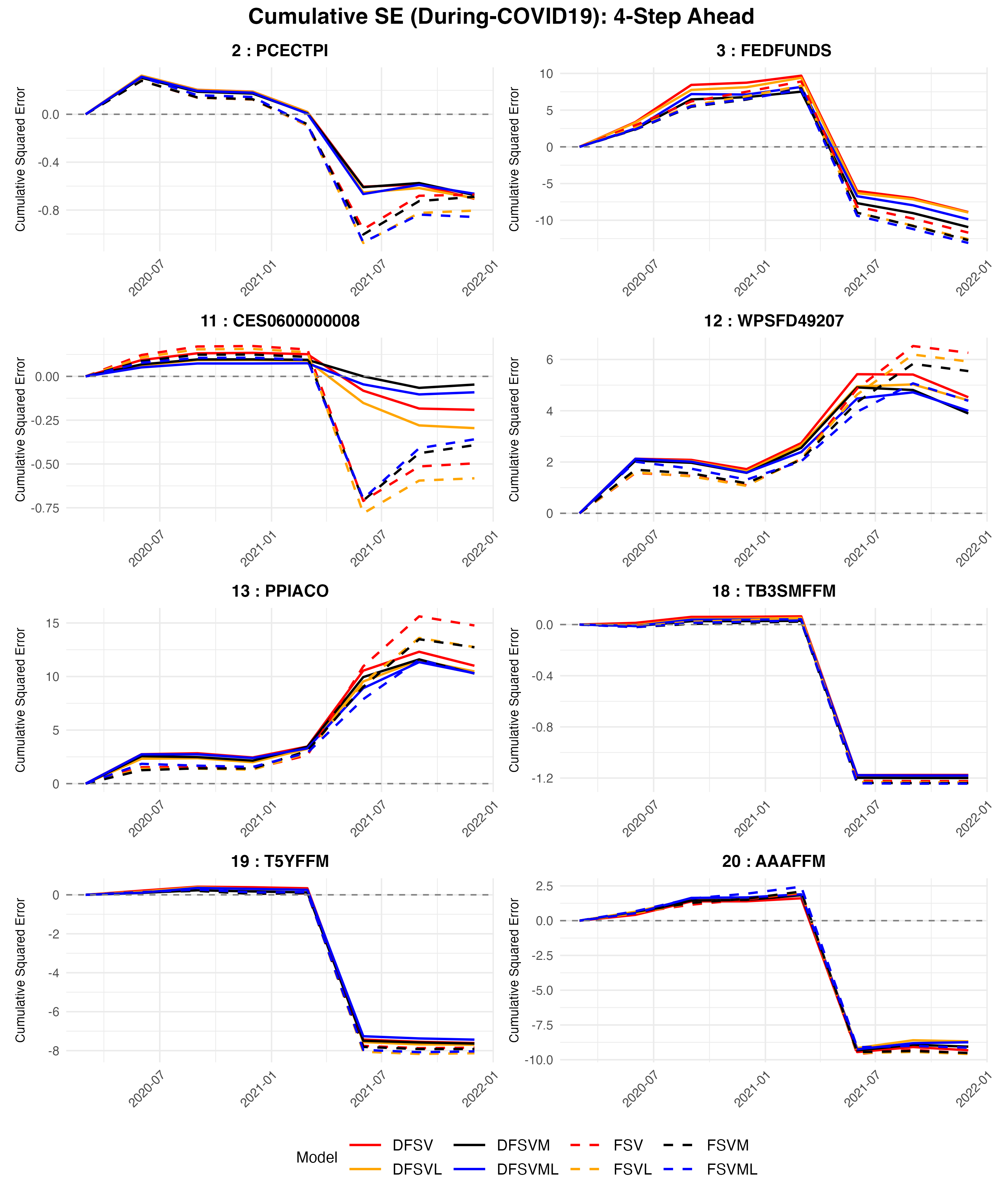}
    \caption{Cumulative squared forecast error (CSFE) relative to the LSVVAR benchmark (zero-line). 2020Q1-2021Q4.}
    \label{fig:result_csfe_covid19_4step}
\end{figure}
\noindent
\begin{figure}[H]
    \centering
    \includegraphics[width=0.9\linewidth]{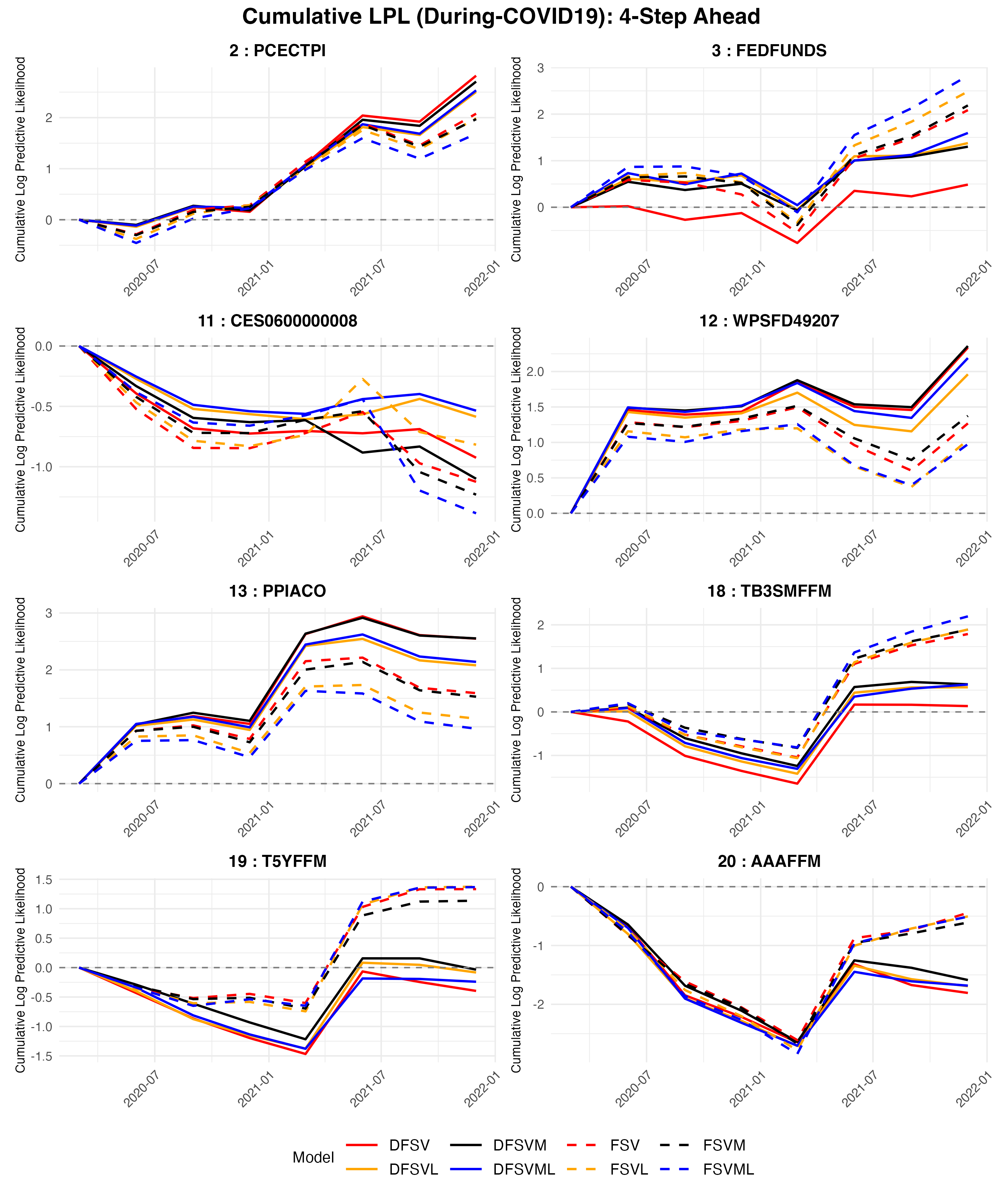}
    \caption{Cumulative log predictive likelihood (LPL) relative to the LSVVAR benchmark (zero-line). 2020Q1-2021Q4.}
    \label{fig:result_clpl_covid19_4step}
\end{figure}

\section{Predictive Performance Comparison with Alternative Specification of Stochastic volatility in Mean}
\label{sec:sv-in-mean-comparison}
\subsection{During Global Financial Crisis Period}
\begin{figure}[H]
    \centering
    \includegraphics[width=0.9\linewidth]{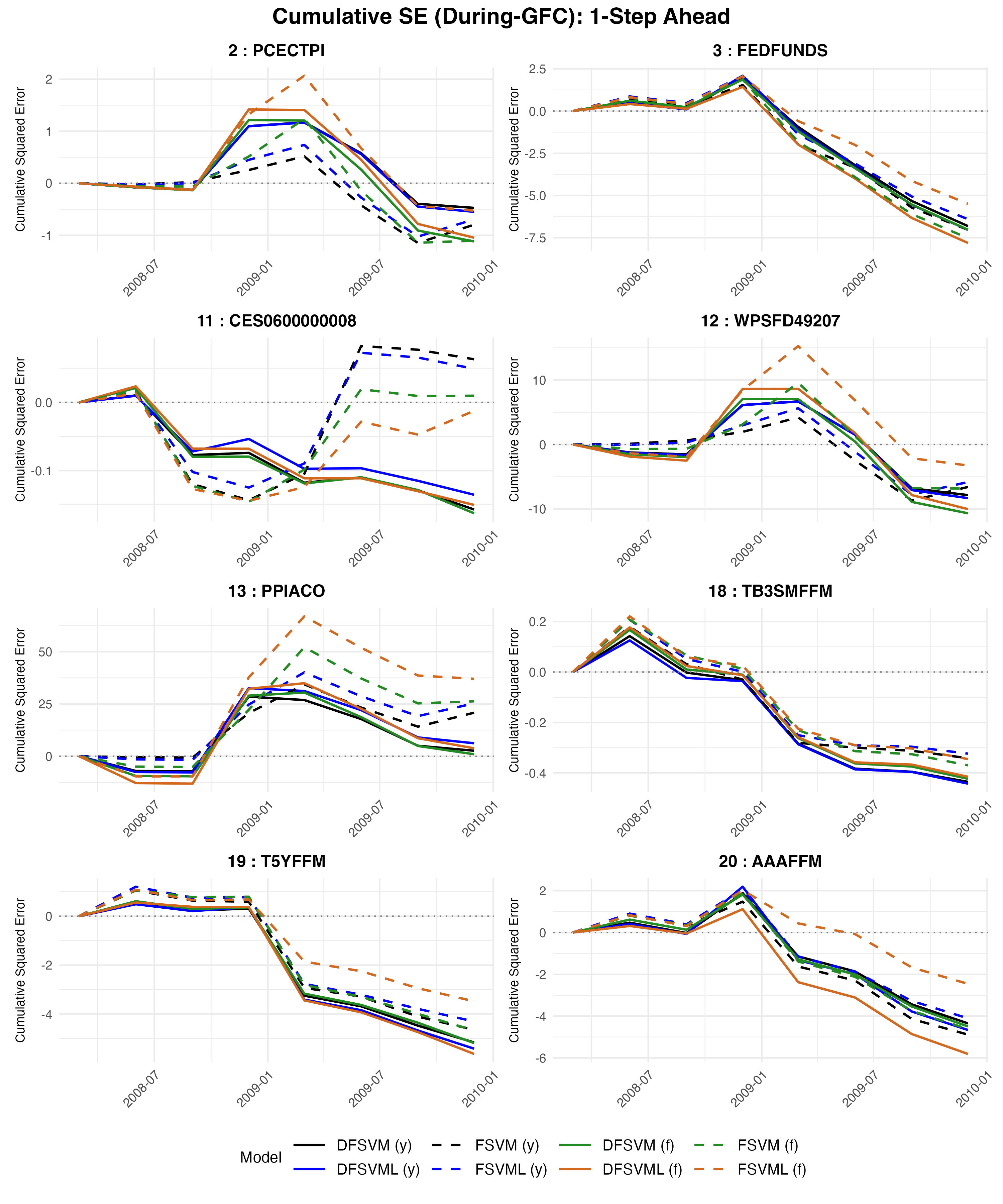}
    \caption{Cumulative squared forecast error (CSFE) relative to the LSVVAR benchmark (zero-line). 2008Q1-2009Q4. (y): stochastic volatility in mean of the $\bm{y}_t$. (f): stochastic volatility in mean of the $\bm{f}_t$. }
    \label{fig:result_csfe_gfc_y_vs_f_1step}
\end{figure}

\begin{figure}[H]
    \centering
    \includegraphics[width=0.9\linewidth]{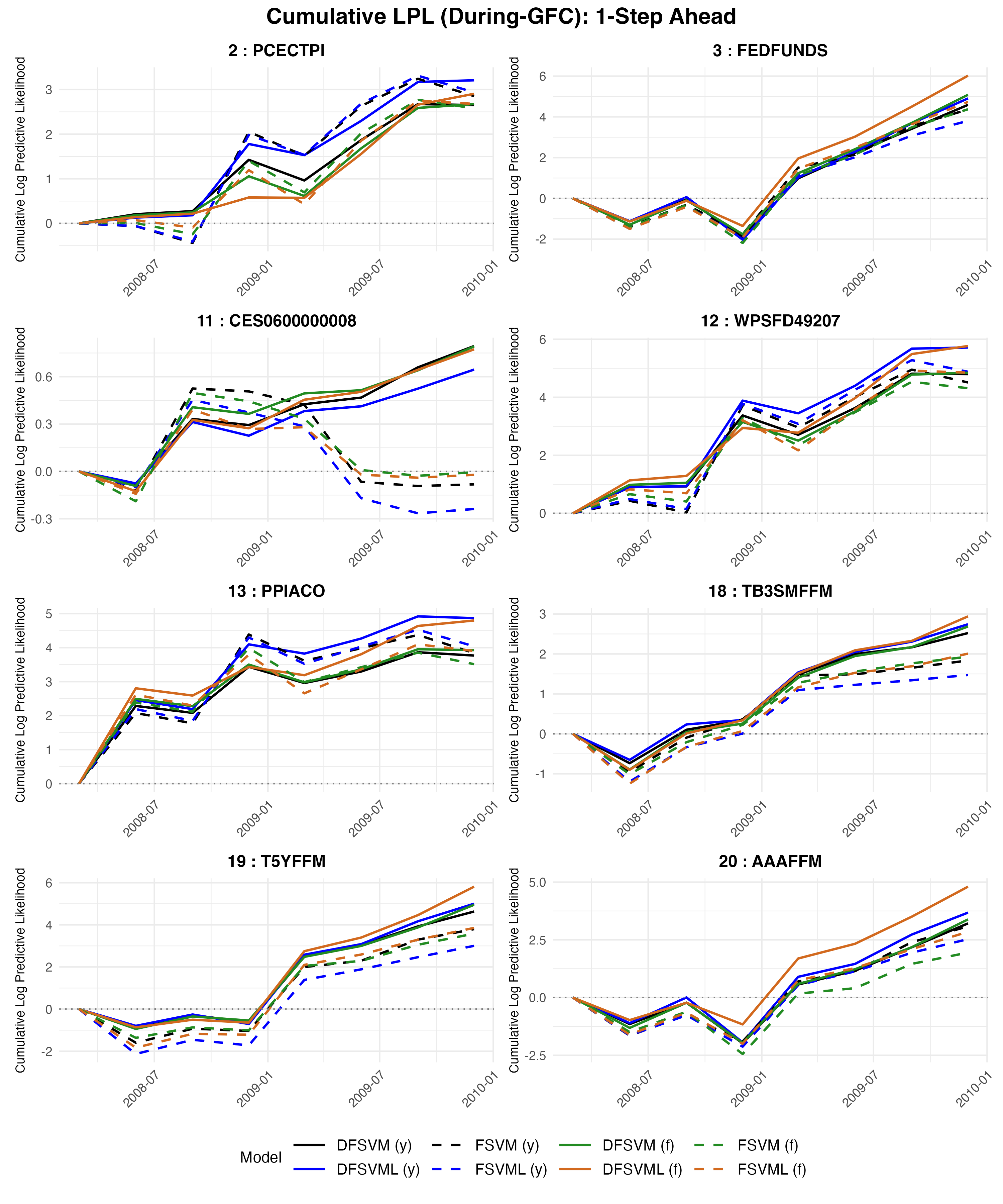}
    \caption{Cumulative log predictive likelihood (LPL) relative to the LSVVAR benchmark (zero-line). 2008Q1-2009Q4. (y): stochastic volatility in mean of the $\bm{y}_t$. (f): stochastic volatility in mean of the $\bm{f}_t$.}
    \label{fig:result_clpl_gfc_y_vs_f_1step}
\end{figure}

\begin{figure}[H]
    \centering
    \includegraphics[width=0.9\linewidth]{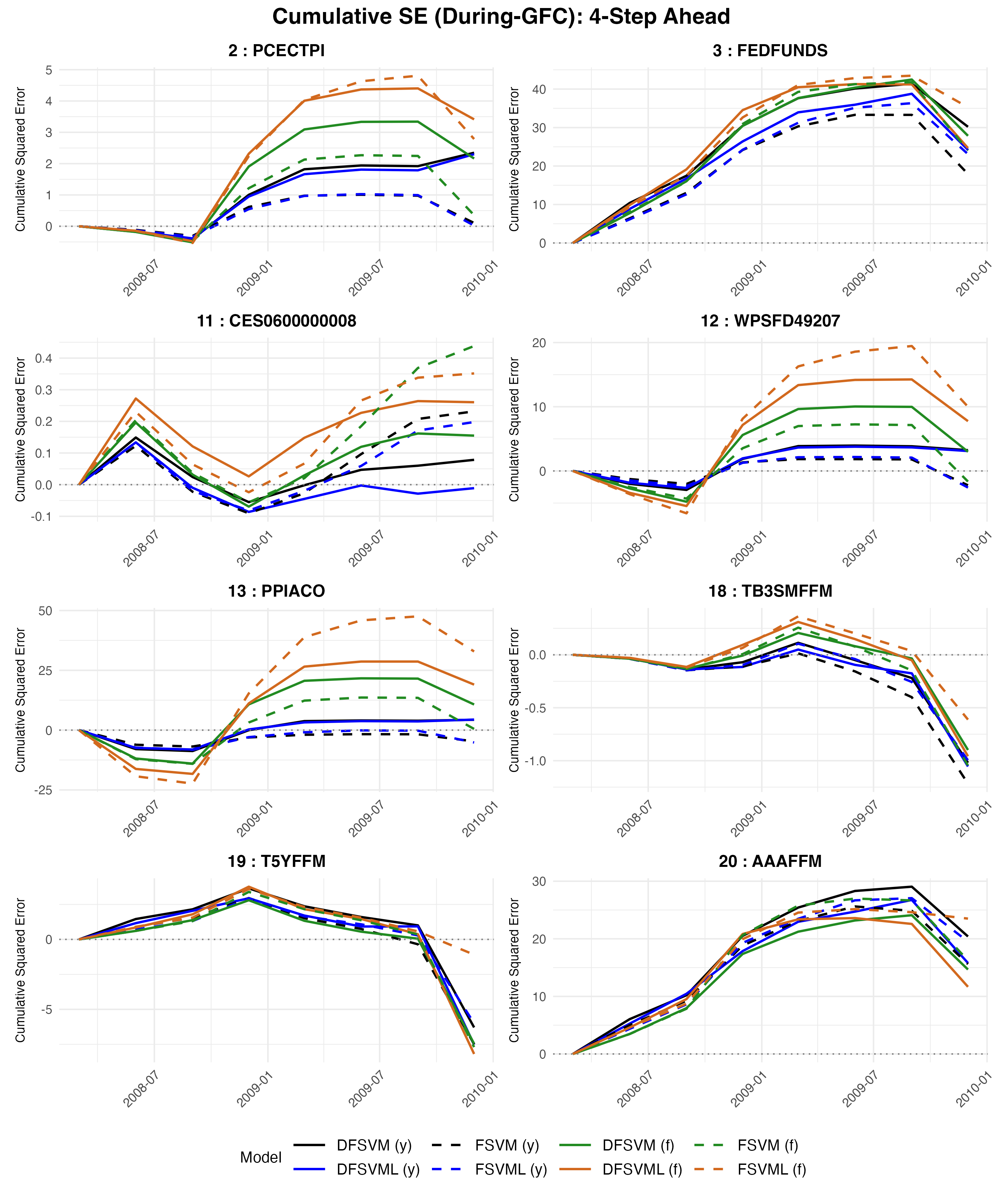}
    \caption{Cumulative squared forecast error (CSFE) relative to the LSVVAR benchmark (zero-line). 2008Q1-2009Q4. (y): stochastic volatility in mean of the $\bm{y}_t$. (f): stochastic volatility in mean of the $\bm{f}_t$. }
    \label{fig:result_csfe_gfc_y_vs_f_4step}
\end{figure}

\begin{figure}[H]
    \centering
    \includegraphics[width=0.9\linewidth]{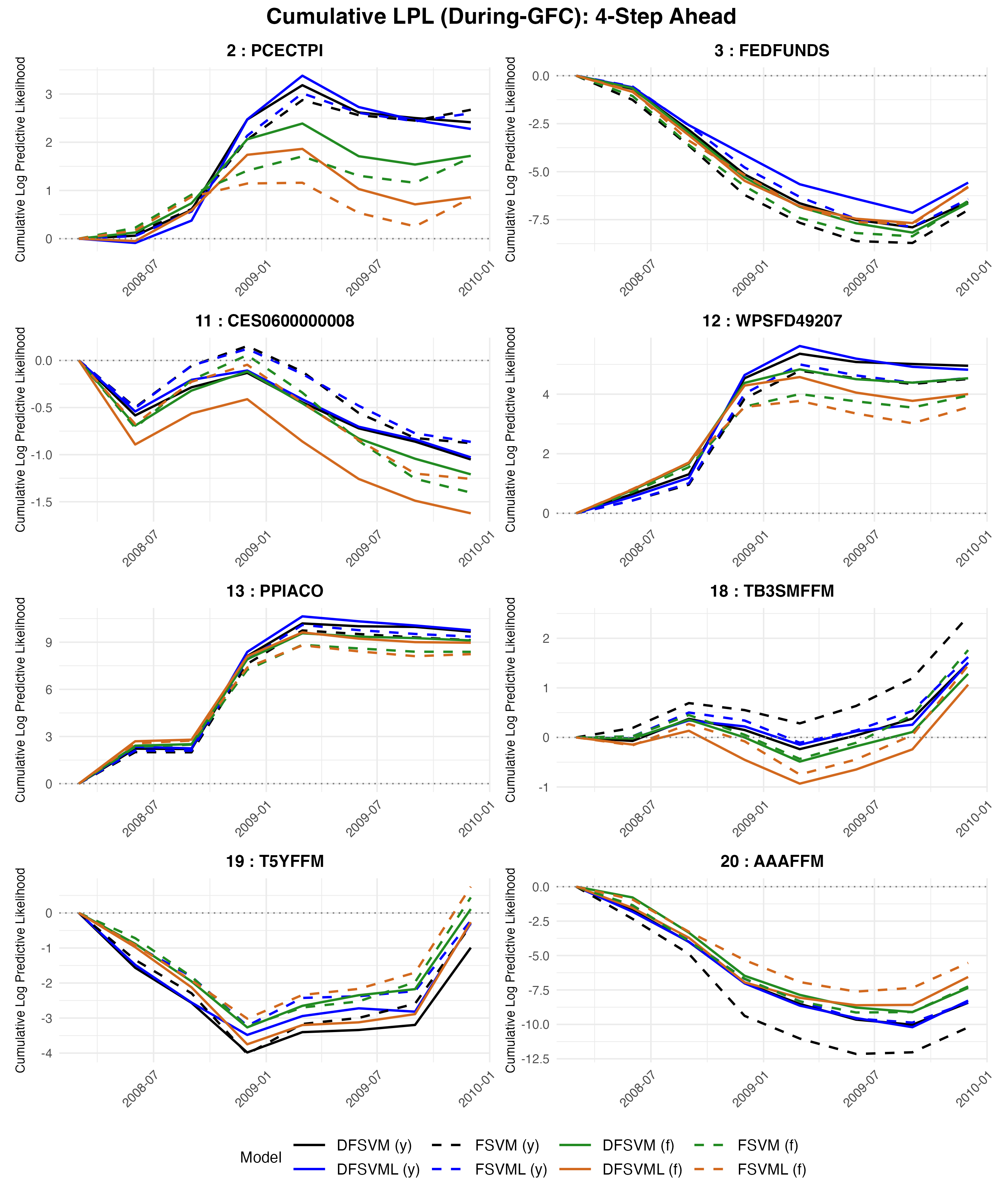}
    \caption{Cumulative log predictive likelihood (LPL) relative to the LSVVAR benchmark (zero-line). 2008Q1-2009Q4. (y): stochastic volatility in mean of the $\bm{y}_t$. (f): stochastic volatility in mean of the $\bm{f}_t$.}
    \label{fig:result_clpl_gfc_y_vs_f_4step}
\end{figure}

\subsection{During COVID-19 Pandemic Period}

\begin{figure}[H]
    \centering
    \includegraphics[width=0.9\linewidth]{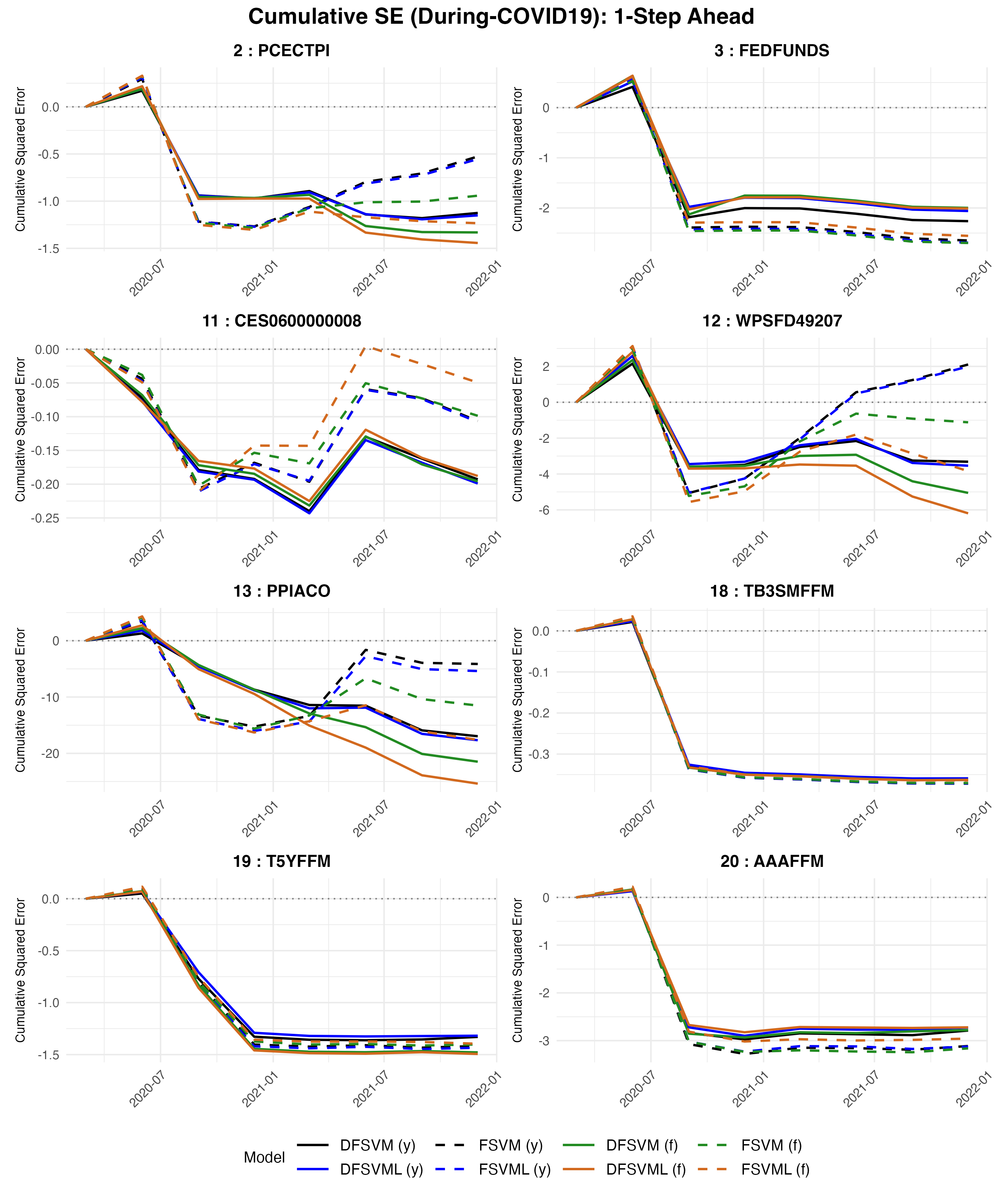}
    \caption{Cumulative squared forecast error (CSFE) relative to the LSVVAR benchmark (zero-line). 2020Q1-2021Q4. (y): stochastic volatility in mean of the $\bm{y}_t$. (f): stochastic volatility in mean of the $\bm{f}_t$.}
    \label{fig:result_csfe_covid19_y_vs_f_1step}
\end{figure}

\begin{figure}[H]
    \centering
    \includegraphics[width=0.9\linewidth]{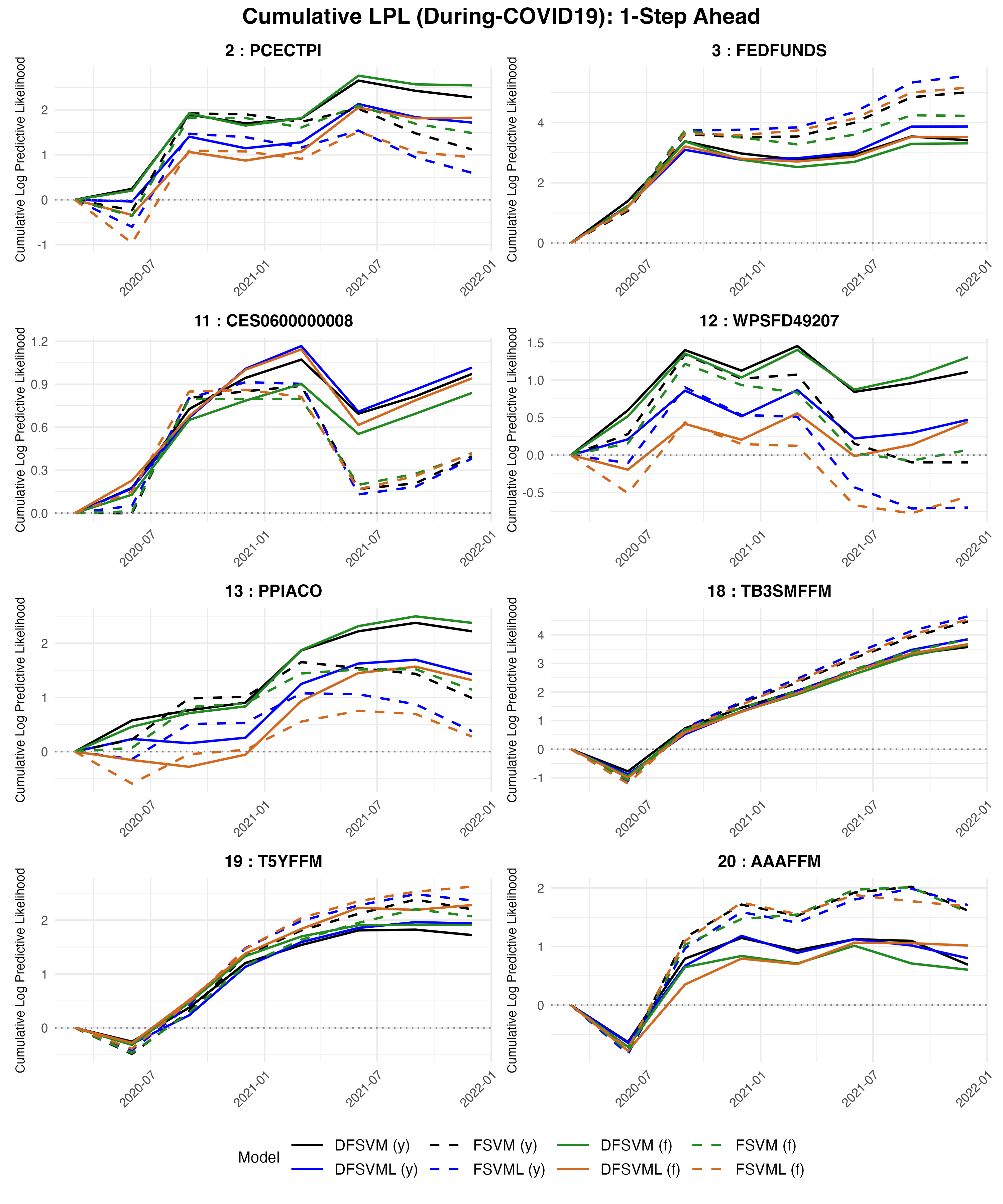}
    \caption{Cumulative log predictive likelihood (LPL) relative to the LSVVAR benchmark (zero-line). 2020Q1-2021Q4. (y): stochastic volatility in mean of the $\bm{y}_t$. (f): stochastic volatility in mean of the $\bm{f}_t$.}
    \label{fig:result_clpl_covid19_y_vs_f_1step}
\end{figure}

\begin{figure}[H]
    \centering
    \includegraphics[width=0.9\linewidth]{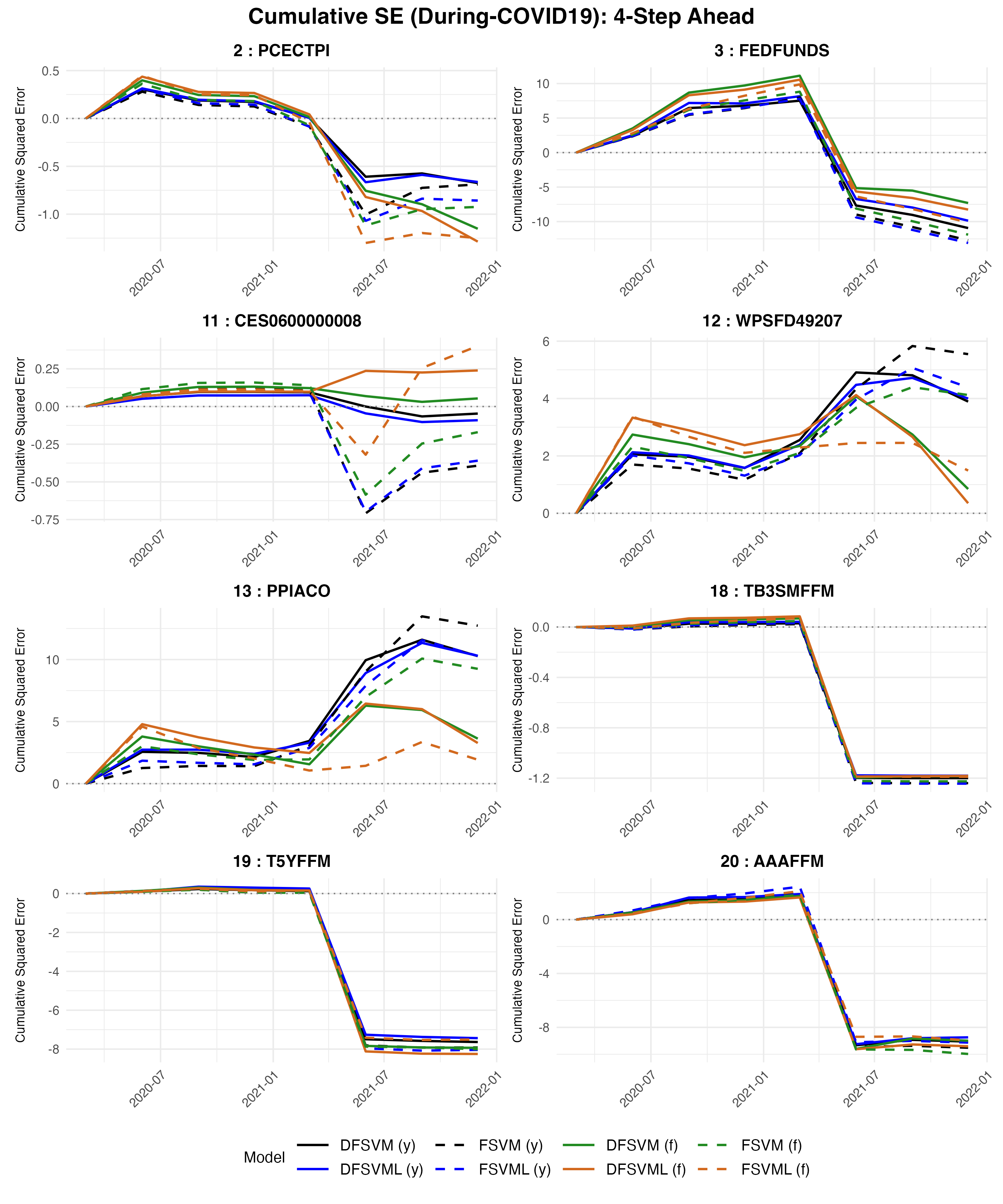}
    \caption{Cumulative squared forecast error (CSFE) relative to the LSVVAR benchmark (zero-line). 2020Q1-2021Q4. (y): stochastic volatility in mean of the $\bm{y}_t$. (f): stochastic volatility in mean of the $\bm{f}_t$.}
    \label{fig:result_csfe_covid19_y_vs_f_4step}
\end{figure}

\begin{figure}[H]
    \centering
    \includegraphics[width=0.9\linewidth]{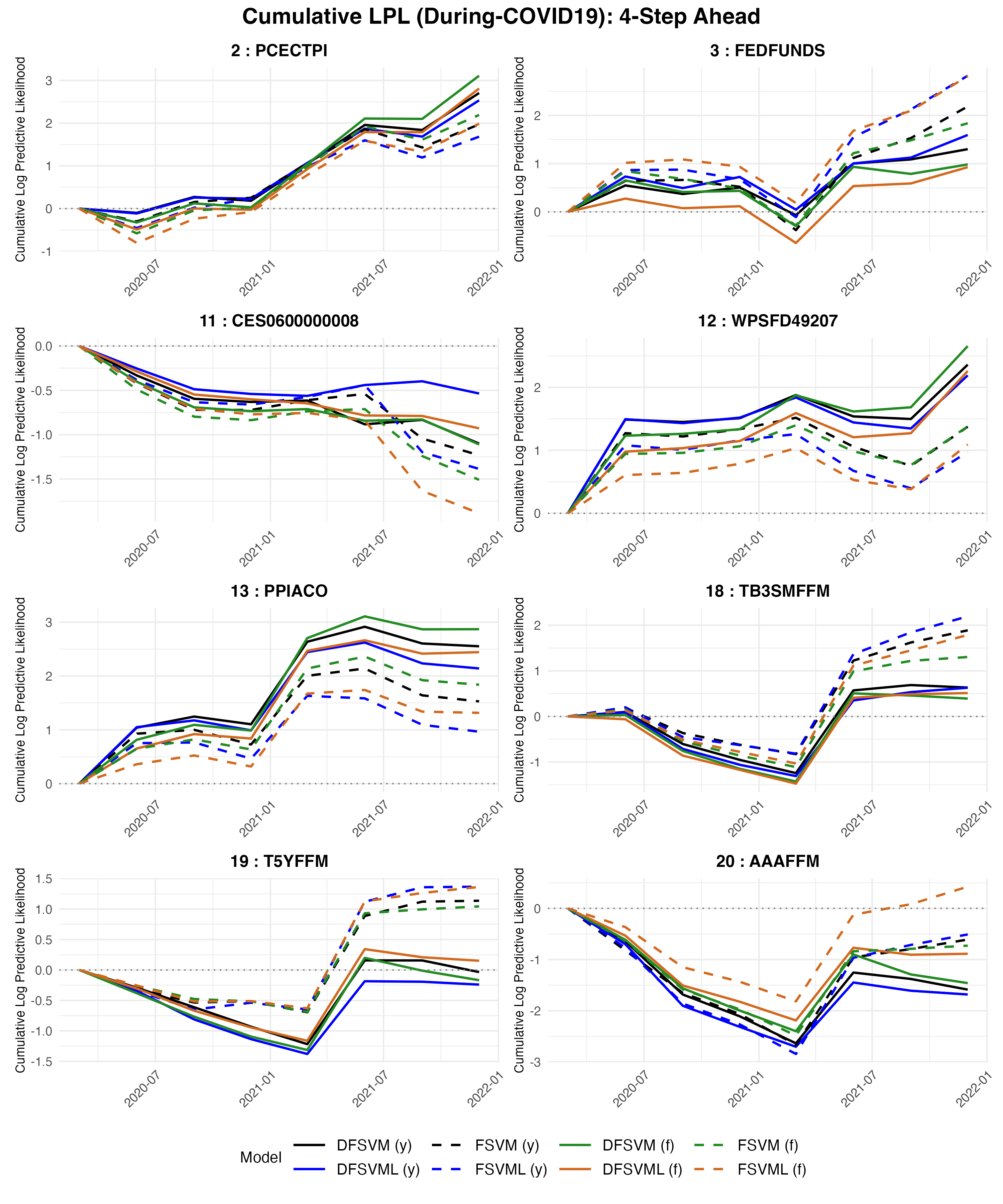}
    \caption{Cumulative log predictive likelihood (LPL) relative to the LSVVAR benchmark (zero-line). 2020Q1-2021Q4. (y): stochastic volatility in mean of the $\bm{y}_t$. (f): stochastic volatility in mean of the $\bm{f}_t$.}
    \label{fig:result_clpl_covid19_y_vs_f_4step}
\end{figure}

\section{Comparison of Normal, Global Financial Crisis, and COVID-19 Pandemic Periods (4-step Ahead)}
\label{sec:out-of-sample result 4step}
\subsection{Percentage gains of models in forecast accuracy (4-step ahead)}
\begin{table}[H]
    \footnotesize
    \centering
    \setlength{\tabcolsep}{3.5pt}
    \begin{tabular}{rl rr rr rr}
        \hline
         & & \multicolumn{2}{c}{Normal} & \multicolumn{2}{c}{GFC} & \multicolumn{2}{c}{COVID-19} \\
        \cmidrule(lr){3-4} \cmidrule(lr){5-6} \cmidrule(lr){7-8}
        \# & Variable & DFSVL & DFSVML & DFSVL & DFSVML & DFSVL & DFSVML \\
        \hline
        1 & GDPC1 & \textcolor{red}{28.24} & \textcolor{red}{29.65} & \textcolor{blue}{-15.64} & -3.98 & -4.51 & -4.29 \\
        2 & PCECTPI & \textcolor{red}{28.91} & \textcolor{red}{28.50} & \textcolor{blue}{-23.42} & \textcolor{blue}{-23.71} & 10.50 & 9.97 \\
        3 & FEDFUNDS & \textcolor{red}{44.65} & \textcolor{red}{38.55} & \textcolor{blue}{-85.49} & \textcolor{blue}{-96.95} & \textcolor{red}{21.75} & \textcolor{red}{25.63} \\
        4 & PCECC96 & \textcolor{red}{29.11} & \textcolor{red}{32.03} & \textcolor{red}{21.19} & \textcolor{red}{19.27} & -5.97 & -6.15 \\
        5 & CMRMTSPLx & 8.39 & 10.29 & -1.30 & -2.71 & 11.30 & 11.90 \\
        6 & INDPRO & \textcolor{blue}{-27.99} & \textcolor{blue}{-28.71} & \textcolor{blue}{-20.86} & \textcolor{blue}{-19.92} & -1.05 & -0.28 \\
        7 & CUMFNS & -13.70 & \textcolor{blue}{-22.23} & \textcolor{blue}{-55.49} & \textcolor{blue}{-51.49} & \textcolor{blue}{-226.78} & \textcolor{blue}{-213.78} \\
        8 & UNRATE & 11.53 & -0.35 & \textcolor{blue}{-18.34} & \textcolor{blue}{-18.95} & \textcolor{blue}{-60.80} & \textcolor{blue}{-54.77} \\
        9 & PAYEMS & 14.15 & 2.47 & \textcolor{blue}{-26.93} & \textcolor{blue}{-26.05} & -12.59 & -13.92 \\
        10 & CES0600000007 & 1.66 & 4.73 & 3.13 & -0.68 & \textcolor{blue}{-65.33} & \textcolor{blue}{-64.33} \\
        11 & CES0600000008 & 13.63 & 11.44 & -11.17 & 10.10 & 11.19 & 4.05 \\
        12 & WPSFD49207 & \textcolor{red}{21.24} & \textcolor{red}{20.78} & -4.42 & -3.01 & -9.95 & -9.05 \\
        13 & PPIACO & \textcolor{red}{15.96} & \textcolor{red}{15.54} & -0.55 & -0.11 & -9.20 & -9.09 \\
        14 & AMDMNOx & -11.44 & -10.95 & 1.50 & 2.20 & \textcolor{red}{19.21} & \textcolor{red}{18.96} \\
        15 & HOUST & \textcolor{red}{44.14} & \textcolor{red}{42.05} & \textcolor{red}{28.80} & \textcolor{red}{26.83} & \textcolor{blue}{-81.79} & \textcolor{blue}{-64.91} \\
        16 & S\&P 500 & 10.68 & 10.76 & 8.87 & 8.29 & 10.00 & \textcolor{red}{15.42} \\
        17 & EXUSUKx & 6.20 & 6.36 & 4.85 & 3.46 & -13.88 & -7.75 \\
        18 & TB3SMFFM & \textcolor{red}{74.07} & \textcolor{red}{74.48} & \textcolor{red}{48.79} & \textcolor{red}{50.68} & \textcolor{red}{86.38} & \textcolor{red}{86.04} \\
        19 & T5YFFM & \textcolor{red}{47.18} & \textcolor{red}{47.64} & \textcolor{red}{69.44} & \textcolor{red}{61.51} & \textcolor{red}{86.09} & \textcolor{red}{83.16} \\
        20 & AAAFFM & \textcolor{red}{46.25} & \textcolor{red}{45.46} & \textcolor{blue}{-90.06} & \textcolor{blue}{-107.64} & \textcolor{red}{71.50} & \textcolor{red}{71.99} \\
        \hline
    \end{tabular}
    \caption{Percentage gains of DFSVL and DFSVML models in forecast accuracy (4-step ahead) relative to the LSVVAR benchmark.  Red: values $\geq 15$; blue: values $\leq -15$.}
    \label{tab:crisis_gain_comparison_dfsv_4step}
\end{table} 
\begin{table}[H]
    \footnotesize
    \centering
    \setlength{\tabcolsep}{3.5pt}
    \begin{tabular}{rl  rr !{\hspace{3pt}} rr !{\hspace{3pt}} rr}
        \hline
         & & \multicolumn{2}{c}{Normal} & \multicolumn{2}{c}{GFC} & \multicolumn{2}{c}{COVID-19} \\
        \cmidrule(lr){3-4} \cmidrule(lr){5-6} \cmidrule(lr){7-8}
        \# & Variable & FSVL & FSVML & FSVL & FSVML & FSVL & FSVML \\
        \hline
        1 & GDPC1 & \textcolor{red}{23.50} & \textcolor{red}{26.49} & -6.65 & -6.21 & -4.07 & -3.92 \\
        2 & PCECTPI & \textcolor{red}{29.33} & \textcolor{red}{28.64} & 1.64 & 1.04 & 11.84 & 12.60 \\
        3 & FEDFUNDS & \textcolor{red}{41.38} & \textcolor{red}{37.02} & \textcolor{blue}{-70.24} & \textcolor{blue}{-92.14} & \textcolor{red}{34.53} & \textcolor{red}{36.50} \\
        4 & PCECC96 & \textcolor{red}{18.62} & \textcolor{red}{22.05} & 13.03 & 13.36 & -5.80 & -6.39 \\
        5 & CMRMTSPLx & 2.53 & 5.05 & -4.32 & -5.65 & 12.10 & 12.68 \\
        6 & INDPRO & \textcolor{blue}{-39.04} & \textcolor{blue}{-34.35} & \textcolor{blue}{-21.60} & \textcolor{blue}{-22.58} & -0.32 & -0.27 \\
        7 & CUMFNS & \textcolor{blue}{-23.52} & \textcolor{blue}{-23.40} & \textcolor{blue}{-88.52} & \textcolor{blue}{-87.62} & \textcolor{blue}{-85.29} & \textcolor{blue}{-81.92} \\
        8 & UNRATE & \textcolor{red}{20.44} & 11.69 & \textcolor{blue}{-49.58} & \textcolor{blue}{-51.90} & \textcolor{blue}{-54.95} & \textcolor{blue}{-51.70} \\
        9 & PAYEMS & 5.10 & 3.57 & \textcolor{blue}{-42.08} & \textcolor{blue}{-42.77} & -11.23 & -13.92 \\
        10 & CES0600000007 & 8.13 & 7.71 & -13.45 & \textcolor{blue}{-22.45} & \textcolor{blue}{-30.77} & \textcolor{blue}{-33.37} \\
        11 & CES0600000008 & 9.38 & 10.52 & \textcolor{blue}{-39.34} & \textcolor{blue}{-23.12} & \textcolor{red}{21.01} & 13.33 \\
        12 & WPSFD49207 & \textcolor{red}{18.56} & \textcolor{red}{18.46} & 6.11 & 6.38 & -13.49 & -10.39 \\
        13 & PPIACO & 13.59 & 14.04 & 2.41 & 3.34 & -11.17 & -9.26 \\
        14 & AMDMNOx & -9.75 & -9.83 & 2.17 & 1.35 & \textcolor{red}{18.92} & \textcolor{red}{18.64} \\
        15 & HOUST & \textcolor{red}{37.88} & \textcolor{red}{33.42} & \textcolor{red}{22.19} & \textcolor{red}{20.37} & -9.52 & -0.26 \\
        16 & S\&P 500 & 14.74 & 13.04 & 6.66 & 5.99 & 9.22 & 14.75 \\
        17 & EXUSUKx & 7.39 & 6.10 & 4.59 & 4.58 & -12.09 & -5.65 \\
        18 & TB3SMFFM & \textcolor{red}{69.91} & \textcolor{red}{68.77} & \textcolor{red}{52.01} & \textcolor{red}{49.61} & \textcolor{red}{89.63} & \textcolor{red}{90.39} \\
        19 & T5YFFM & \textcolor{red}{40.63} & \textcolor{red}{37.20} & \textcolor{red}{63.11} & \textcolor{red}{51.02} & \textcolor{red}{90.43} & \textcolor{red}{89.33} \\
        20 & AAAFFM & \textcolor{red}{42.83} & \textcolor{red}{41.52} & \textcolor{blue}{-95.55} & \textcolor{blue}{-133.79} & \textcolor{red}{78.71} & \textcolor{red}{75.10} \\
        \hline
    \end{tabular}
    \caption{Percentage gains of FSVL and FSVML models in forecast accuracy (4-step ahead) relative to the LSVVAR benchmark. Red: values $\geq 15$; blue: values $\leq -15$.}
    \label{tab:crisis_gain_comparison_fsv_4step}
\end{table}
\subsection{Percentage gains of DFSVL, DFSVM and DSVML models in forecast accuracy}
Table \ref{tab:crisis_gain_comparison_vs_dfsv_1step} shows the percentage gain of  DFSVL, DFSVM and DSVML models in forecast accuracy (1-step ahead) relative to the DFSV model. During the GFC period, the synergistic effect between leverage and in-mean components is most evident in real economic activity variables (Factor 1). For indicators such as \#1, \#5--\#9 and \#18, the DFSVML model outperforms both the DFSVL and DFSVM models. This provides compelling empirical evidence that incorporating the risk-premium effect and the leverage effect is essential for capturing the contraction of the real economy during periods of severe uncertainty.

In contrast, for variables such as \#3, \#4, and \#20, the predictive gains are primarily driven by the leverage component, as reflected in the high performance of the DFSVL model. While the models show limited or negative gains during the COVID-19 pandemic and normal times for certain variables, the substantial improvements observed during the GFC period demonstrate the necessity of the proposed structure in modeling extreme economic distress.

The 4-step-ahead forecast results, detailed in Supplementary Material \ref{sec:out-of-sample result 4step}, present several distinct characteristics compared to the one-step horizon. For indicators such as \#1--\#4, \#12, \#13, \#15, \#18--\#20, the proposed dynamic models demonstrate superior predictive accuracy during normal times. However, some of those substantial gains tend to diminish during the GFC or COVID-19 pandemic periods. Overall, while the models maintain competitive performance during stable intervals, the relative advantage over the benchmark often becomes less pronounced during these crisis regimes as the forecast horizon increases. Despite these variations, the forecasting gains for the third factor variables (\#18, \#19, and \#20) remain robust across both horizons, with the notable exception of \#20 during the GFC period.
\begin{sidewaystable}
%    \footnotesize
    \scriptsize
    \centering
    \setlength{\tabcolsep}{3pt}
    \begin{tabular}{rl rrr rrr rrr}
        \hline
        & & \multicolumn{3}{c}{Normal} & \multicolumn{3}{c}{GFC} & \multicolumn{3}{c}{COVID-19} \\
         \cmidrule(lr){3-5} \cmidrule(lr){6-8}  \cmidrule(lr){9-11}
        \# & Variable & DFSVL & DFSVM & DFSVML & DFSVL & DFSVM & DFSVML & DFSVL & DFSVM & DFSVML \\
        \hline
        1 & GDPC1 & 5.04 & 2.35 & 4.67 & \textcolor{red}{13.03} & 9.75 & \textcolor{red}{14.41} & 0.07 & 0.44 & 0.06 \\
        2 & PCECTPI & -1.20 & -0.61 & -1.64 & 2.72 & 1.19 & 1.99 & 0.49 & -1.51 & -0.65 \\
        3 & FEDFUNDS & 1.39 & 1.02 & -1.94 & \textcolor{red}{29.85} & \textcolor{red}{15.53} & \textcolor{red}{22.22} & -4.27 & 3.80 & -4.00 \\
        4 & PCECC96 & 0.92 & 0.24 & 3.21 & \textcolor{red}{16.87} & \textcolor{red}{11.25} & \textcolor{red}{14.36} & -0.04 & -1.88 & -1.46 \\
        5 & CMRMTSPLx & 2.63 & -2.40 & 1.74 & \textcolor{red}{10.54} & 7.93 & \textcolor{red}{11.03} & -0.25 & -0.10 & 0.73 \\
        6 & INDPRO & 9.47 & 3.40 & 9.12 & \textcolor{red}{10.08} & \textcolor{red}{12.72} & \textcolor{red}{15.11} & 0.26 & 0.59 & 0.90 \\
        7 & CUMFNS & 8.78 & -1.56 & 3.52 & 9.54 & \textcolor{red}{12.05} & \textcolor{red}{12.55} & 0.35 & 0.34 & 0.97 \\
        8 & UNRATE & -2.46 & 1.31 & -3.70 & \textcolor{red}{10.50} & 8.63 & \textcolor{red}{12.29} & -1.39 & -0.89 & -0.08 \\
        9 & PAYEMS & 3.85 & -0.61 & 0.27 & \textcolor{red}{14.48} & \textcolor{red}{13.88} & \textcolor{red}{17.04} & 0.09 & -1.85 & -1.17 \\
        10 & CES0600000007 & -1.23 & 2.23 & 1.11 & 9.54 & 1.35 & 8.30 & 1.14 & -0.29 & -0.00 \\
        11 & CES0600000008 & 1.71 & 2.39 & 0.63 & -3.38 & -2.53 & \textcolor{blue}{-10.89} & 0.66 & -4.00 & -3.31 \\
        12 & WPSFD49207 & 0.09 & -0.01 & -0.31 & 0.58 & 1.46 & 2.35 & 1.02 & 0.06 & 0.44 \\
        13 & PPIACO & -0.17 & -0.12 & -0.84 & 1.95 & 0.74 & -0.63 & 0.83 & 0.22 & 0.96 \\
        14 & AMDMNOx & 2.66 & -0.15 & 1.10 & 6.20 & 5.56 & 8.86 & -1.14 & -0.42 & -0.55 \\
        15 & HOUST & 0.50 & -0.63 & -0.34 & 8.61 & -0.18 & 5.43 & -0.74 & 0.24 & 1.49 \\
        16 & S\&P 500 & -0.75 & -2.20 & -1.16 & 1.66 & -2.97 & -3.10 & -3.11 & 3.31 & 3.88 \\
        17 & EXUSUKx & -0.99 & -1.13 & -0.86 & -1.86 & -0.63 & -0.36 & 0.09 & 0.74 & 0.99 \\
        18 & TB3SMFFM & 1.39 & 3.51 & 3.45 & 7.38 & \textcolor{red}{12.08} & \textcolor{red}{14.84} & -0.40 & -3.29 & -8.68 \\
        19 & T5YFFM & -7.47 & -4.67 & -1.76 & 9.41 & -7.57 & 4.84 & -0.27 & \textcolor{blue}{-14.16} & \textcolor{blue}{-17.15} \\
        20 & AAAFFM & -3.53 & -5.32 & 0.33 & \textcolor{red}{11.15} & \textcolor{blue}{-17.71} & -5.23 & -2.06 & -6.66 & \textcolor{blue}{-11.77} \\
        \hline
    \end{tabular}
    \vspace{5pt}
    \caption{\footnotesize Percentage gain of DFSVL, DFSVM and DFSVML models in forecast accuracy (1-step ahead) relative to the DFSV model. Red: values $\geq 10$; blue: values $\leq -10$.}   \label{tab:crisis_gain_comparison_vs_dfsv_1step}
\end{sidewaystable}
\begin{sidewaystable}
%    \footnotesize
    \scriptsize
    \centering
    \setlength{\tabcolsep}{3pt}
    \begin{tabular}{rl rrr rrr rrr}
        \hline
        & & \multicolumn{3}{c}{Normal} & \multicolumn{3}{c}{GFC} & \multicolumn{3}{c}{COVID-19} \\
         \cmidrule(lr){3-5} \cmidrule(lr){6-8}  \cmidrule(lr){9-11}
        \# & Variable & DFSVL & DFSVM & DFSVML & DFSVL & DFSVM & DFSVML & DFSVL & DFSVM & DFSVML \\
        \hline
        1 & GDPC1 & 4.09 & 3.02 & 5.97 & 3.09 & 2.02 & 3.73 & -0.47 & -0.32 & -0.26 \\
        2 & PCECTPI & -0.87 & 0.88 & -1.44 & -0.38 & -0.99 & -0.61 & -0.11 & -0.51 & -0.70 \\
        3 & FEDFUNDS & \textcolor{red}{17.40} & 9.82 & 8.29 & 10.93 & -3.62 & 5.43 & -0.35 & 8.64 & 4.62 \\
        4 & PCECC96 & 7.08 & 3.69 & 10.90 & 8.43 & 4.02 & 6.19 & -0.32 & -0.50 & -0.49 \\
        5 & CMRMTSPLx & 4.15 & 2.66 & 6.14 & 3.93 & 1.73 & 2.58 & -1.33 & 0.36 & -0.65 \\
        6 & INDPRO & 8.84 & 3.89 & 8.32 & 2.86 & 2.31 & 3.62 & -0.16 & 0.66 & 0.60 \\
        7 & CUMFNS & 10.11 & 2.91 & 3.36 & 5.21 & 3.23 & 7.65 & 0.66 & 2.86 & 4.61 \\
        8 & UNRATE & -5.26 & -1.29 & \textcolor{blue}{-19.40} & 7.08 & 3.00 & 6.60 & -2.61 & -0.85 & 1.24 \\
        9 & PAYEMS & 10.22 & 2.80 & -1.98 & 5.24 & 2.34 & 5.89 & -0.56 & -1.74 & -1.75 \\
        10 & CES0600000007 & -1.80 & 6.28 & 1.38 & 3.22 & -5.39 & -0.59 & 2.84 & 1.10 & 3.44 \\
        11 & CES0600000008 & 1.30 & 1.78 & -1.21 & 1.09 & 7.61 & \textcolor{red}{20.01} & 4.00 & -5.39 & -3.72 \\
        12 & WPSFD49207 & -0.63 & -0.24 & -1.21 & -0.69 & 0.57 & 0.67 & 0.13 & 1.35 & 0.94 \\
        13 & PPIACO & -0.40 & 0.00 & -0.91 & -0.56 & 0.01 & -0.12 & 0.37 & 0.59 & 0.47 \\
        14 & AMDMNOx & -0.15 & -0.80 & 0.29 & 1.89 & 1.62 & 2.59 & -0.65 & -0.20 & -0.96 \\
        15 & HOUST & 6.22 & -1.31 & 2.72 & 4.46 & -1.83 & 1.81 & -1.38 & 6.52 & 8.04 \\
        16 & S\&P 500 & -1.18 & -1.88 & -1.08 & -0.85 & -1.34 & -1.49 & 0.19 & 6.43 & 6.21 \\
        17 & EXUSUKx & -0.27 & -0.53 & -0.10 & 0.86 & 0.03 & -0.59 & 1.41 & 6.72 & 6.71 \\
        18 & TB3SMFFM & \textcolor{red}{15.68} & 11.89 & \textcolor{red}{16.98} & 9.94 & 8.69 & 13.27 & 4.34 & 10.96 & 1.89 \\
        19 & T5YFFM & 11.14 & 4.28 & 11.93 & \textcolor{red}{31.89} & -6.89 & 14.21 & 7.59 & 3.48 & -11.93 \\
        20 & AAAFFM & 11.44 & 5.65 & 10.15 & 12.96 & -6.82 & 4.91 & \textcolor{blue}{-21.36} & -8.69 & \textcolor{blue}{-19.30} \\
        \hline
        \hline
    \end{tabular}
    \vspace{5pt}
    \caption{\footnotesize Percentage gain of DFSVL, DFSVM and DFSVML models in forecast accuracy (4-step ahead) relative to the DFSV model. Red: values $\geq 10$; blue: values $\leq -10$.}
    \label{tab:crisis_gain_comparison_vs_dfsv_4step}
\end{sidewaystable}

\newpage
\section{Diebold-Mariano test and Model Confidence Set}
\label{sec:appendix_prediction_tests}
To rigorously evaluate the forecasting performance of the proposed models, we employ two statistical approaches. First, we conduct the Diebold-Mariano (DM) test to perform pairwise comparisons between each factor model and the LSVVAR benchmark. Second, we utilize the Model Confidence Set (MCS) procedure proposed by \cite{HansenLundeNason(11)} to simultaneously evaluate the entire set of candidate models without specifying a benchmark \textit{a priori}, thereby accounting for data snooping bias inherent in multiple model comparisons.

\subsection{Diebold-Mariano Test}
\label{subsec:dm_test}
We report the $p$-values for the one-sided DM test where the null hypothesis is equal predictive accuracy, and the alternative hypothesis is that the factor model provides more accurate forecasts than the benchmark ($H_a$: Factor Model $<$ LSVVAR). A $p$-value below 0.05 indicates that the factor model significantly outperforms the LSVVAR benchmark at the 5\% level.

Tables \ref{tab:dm_test_less_4var_first} and \ref{tab:dm_test_less_4var_second} present the DM test results across the three defined periods for the eight focal variables. The results provide statistical evidence supporting the superiority of the factor specifications, although the performance varies across variables and horizons. For example, horizons), all factor models significantly outperform the LSVVAR benchmark during the normal period for variables such as \#12 (4-step horizon) and \#18 (1-step and 4-step). 

During the GFC period, a notable finding is observed for the wage index (\#11); the dynamic factor specifications (DFSV, DFSVL, and DFSVM) significantly outperform the benchmark in the 1-step forecast, with $p$-values as low as 0.037. This underscores the effectiveness of the proposed dynamic structure in capturing labor market adjustments under extreme financial stress. During the COVID-19 pandemic period, while many $p$-values are higher, the DFSVM-type models maintain a significant edge for \#13 (PPIACO) at the 1-step horizon.

For a broader perspective, Tables \ref{tab:dm_test_full_period_1step_less} and \ref{tab:dm_test_full_period_4step_less} present the results for the full 20-variable panel over the entire evaluation period. These comprehensive tests confirm that the predictive gains of the factor models extend beyond the core indicators discussed above. For instance, the factor specifications demonstrate significant outperformance for several financial series and housing-related variables (e.g., \#16--\#19 for 1-step ahead forecast, and \#15, \#16, \#18 for 4-step ahead forecast). While LSVVAR remains robust for many real activity variables, the DM tests highlight that our proposed framework provides a statistically significant advantage for variables driven by systemic financial factors and long-term price dynamics.
\begin{table}[H]
    \centering
    \scriptsize
    \resizebox{\textwidth}{!}{
    \begin{tabular}{llcccccccc}
        \toprule
        & & \multicolumn{8}{c}{Model vs. LSVVAR ($H_a$: Factor Model $<$ LSVVAR)} \\
        \cmidrule(lr){3-10}
        Variable & Horizon & DFSV & DFSVL & DFSVM & DFSVML & FSV & FSVL & FSVM & FSVML \\
        \midrule
        \multicolumn{10}{c}{\textbf{Normal}} \\
        2 : PCECTPI & 1-step & 0.748 & 0.769 & 0.768 & 0.792 & 0.278 & 0.270 & 0.291 & 0.253 \\
         & 4-step & 0.062 & 0.070 & 0.062 & 0.077 & \textbf{0.045} & \textbf{0.046} & 0.050 & 0.058 \\
        3 : FEDFUNDS & 1-step & 0.853 & 0.823 & 0.834 & 0.867 & 0.933 & 0.918 & 0.978 & 0.946 \\
         & 4-step & 0.194 & 0.149 & 0.179 & 0.183 & 0.173 & 0.152 & 0.221 & 0.172 \\
        11 : CES0600000008 & 1-step & 0.324 & 0.238 & 0.195 & 0.276 & 0.337 & 0.286 & 0.205 & 0.229 \\
         & 4-step & 0.137 & 0.116 & 0.100 & 0.164 & 0.213 & 0.197 & 0.161 & 0.171 \\
        12 : WPSFD49207 & 1-step & 0.763 & 0.764 & 0.761 & 0.778 & 0.815 & 0.787 & 0.826 & 0.771 \\
         & 4-step & \textbf{0.029} & \textbf{0.041} & \textbf{0.033} & \textbf{0.043} & \textbf{0.028} & \textbf{0.031} & \textbf{0.032} & \textbf{0.035} \\
        \midrule
        \multicolumn{10}{c}{\textbf{GFC}} \\
        2 : PCECTPI & 1-step & 0.418 & 0.374 & 0.403 & 0.390 & 0.275 & 0.349 & 0.286 & 0.327 \\
         & 4-step & 0.913 & 0.917 & 0.932 & 0.941 & 0.537 & 0.458 & 0.498 & 0.475 \\
        3 : FEDFUNDS & 1-step & 0.195 & 0.123 & 0.154 & 0.144 & 0.132 & 0.147 & 0.126 & 0.162 \\
         & 4-step & 0.858 & 0.777 & 0.884 & 0.841 & 0.794 & 0.759 & 0.766 & 0.844 \\
        11 : CES0600000008 & 1-step & \textbf{0.042} & \textbf{0.042} & \textbf{0.037} & 0.051 & 0.593 & 0.541 & 0.547 & 0.538 \\
         & 4-step & 0.677 & 0.672 & 0.570 & 0.306 & 0.745 & 0.778 & 0.750 & 0.727 \\
        12 : WPSFD49207 & 1-step & 0.295 & 0.303 & 0.278 & 0.270 & 0.258 & 0.305 & 0.268 & 0.304 \\
         & 4-step & 0.635 & 0.657 & 0.620 & 0.624 & 0.334 & 0.272 & 0.288 & 0.290 \\
        \midrule
        \multicolumn{10}{c}{\textbf{COVID-19}} \\
        2 : PCECTPI & 1-step & 0.178 & 0.178 & 0.185 & 0.184 & 0.383 & 0.370 & 0.385 & 0.380 \\
         & 4-step & 0.185 & 0.192 & 0.185 & 0.196 & 0.218 & 0.200 & 0.221 & 0.194 \\
        3 : FEDFUNDS & 1-step & 0.235 & 0.248 & 0.219 & 0.232 & 0.221 & 0.222 & 0.223 & 0.218 \\
         & 4-step & 0.361 & 0.359 & 0.309 & 0.328 & 0.309 & 0.288 & 0.281 & 0.276 \\
        11 : CES0600000008 & 1-step & 0.116 & 0.110 & 0.157 & 0.150 & 0.259 & 0.268 & 0.331 & 0.328 \\
         & 4-step & 0.247 & 0.170 & 0.331 & 0.239 & 0.231 & 0.208 & 0.249 & 0.255 \\
        12 : WPSFD49207 & 1-step & 0.308 & 0.302 & 0.309 & 0.315 & 0.599 & 0.571 & 0.582 & 0.577 \\
         & 4-step & 0.983 & 0.992 & 0.974 & 0.990 & 0.988 & 0.991 & 0.993 & 0.995 \\
        \bottomrule
    \end{tabular}
    }
    \caption{$p$-values of Diebold-Mariano test where the alternative hypothesis is that the factor model's mean squared error is less than the LSVVAR benchmark's. The values less than 0.05 are in bold.}
    \label{tab:dm_test_less_4var_first}
\end{table}

\begin{table}[H]
    \centering
    \scriptsize
    \resizebox{\textwidth}{!}{
    \begin{tabular}{llcccccccc}
        \toprule
        & & \multicolumn{8}{c}{Model vs. LSVVAR ($H_a$: Factor Model $<$ LSVVAR)} \\
        \cmidrule(lr){3-10}
        Variable & Horizon & DFSV & DFSVL & DFSVM & DFSVML & FSV & FSVL & FSVM & FSVML \\
        \midrule
        \multicolumn{10}{c}{\textbf{Normal}} \\
        13 : PPIACO & 1-step & 0.169 & 0.171 & 0.176 & 0.201 & 0.644 & 0.606 & 0.648 & 0.629 \\
         & 4-step & \textbf{0.027} & \textbf{0.040} & \textbf{0.030} & \textbf{0.044} & \textbf{0.021} & \textbf{0.024} & \textbf{0.020} & \textbf{0.029} \\
        18 : TB3SMFFM & 1-step & \textbf{0.006} & \textbf{0.005} & \textbf{0.006} & \textbf{0.004} & \textbf{0.012} & \textbf{0.025} & \textbf{0.036} & \textbf{0.024} \\
         & 4-step & \textbf{0.035} & \textbf{0.035} & \textbf{0.034} & \textbf{0.033} & \textbf{0.033} & \textbf{0.032} & \textbf{0.035} & \textbf{0.030} \\
        19 : T5YFFM & 1-step & 0.089 & 0.190 & 0.156 & 0.125 & 0.279 & 0.324 & 0.286 & 0.329 \\
         & 4-step & 0.131 & 0.112 & 0.130 & 0.112 & 0.136 & 0.119 & 0.160 & 0.124 \\
        20 : AAAFFM & 1-step & 0.164 & 0.245 & 0.273 & 0.176 & 0.496 & 0.463 & 0.509 & 0.421 \\
         & 4-step & 0.124 & 0.102 & 0.120 & 0.104 & 0.128 & 0.106 & 0.161 & 0.110 \\
        \midrule
        \multicolumn{10}{c}{\textbf{GFC}} \\
        13 : PPIACO & 1-step & 0.522 & 0.480 & 0.506 & 0.531 & 0.735 & 0.744 & 0.747 & 0.753 \\
         & 4-step & 0.500 & 0.539 & 0.499 & 0.509 & 0.354 & 0.270 & 0.203 & 0.243 \\
        18 : TB3SMFFM & 1-step & 0.081 & 0.082 & 0.058 & 0.053 & 0.110 & 0.112 & 0.104 & 0.122 \\
         & 4-step & 0.151 & 0.161 & 0.116 & 0.104 & 0.107 & 0.117 & 0.089 & 0.100 \\
        19 : T5YFFM & 1-step & 0.076 & 0.097 & 0.083 & 0.085 & 0.117 & 0.125 & 0.111 & 0.132 \\
         & 4-step & 0.203 & 0.189 & 0.218 & 0.190 & 0.158 & 0.182 & 0.177 & 0.189 \\
        20 : AAAFFM & 1-step & 0.176 & 0.157 & 0.223 & 0.217 & 0.193 & 0.247 & 0.173 & 0.250 \\
         & 4-step & 0.829 & 0.750 & 0.870 & 0.819 & 0.808 & 0.785 & 0.813 & 0.886 \\
        \midrule
        \multicolumn{10}{c}{\textbf{COVID-19}} \\
        13 : PPIACO & 1-step & \textbf{0.022} & \textbf{0.023} & \textbf{0.022} & \textbf{0.029} & 0.435 & 0.412 & 0.420 & 0.402 \\
         & 4-step & 0.984 & 0.984 & 0.983 & 0.991 & 0.959 & 0.967 & 0.969 & 0.981 \\
        18 : TB3SMFFM & 1-step & 0.176 & 0.175 & 0.174 & 0.175 & 0.176 & 0.176 & 0.173 & 0.173 \\
         & 4-step & 0.128 & 0.123 & 0.118 & 0.121 & 0.119 & 0.120 & 0.115 & 0.117 \\
        19 : T5YFFM & 1-step & 0.088 & 0.088 & 0.092 & 0.090 & 0.092 & 0.088 & 0.089 & 0.086 \\
         & 4-step & 0.115 & 0.112 & 0.105 & 0.111 & 0.106 & 0.108 & 0.105 & 0.109 \\
        20 : AAAFFM & 1-step & 0.181 & 0.185 & 0.192 & 0.187 & 0.193 & 0.185 & 0.191 & 0.185 \\
         & 4-step & 0.154 & 0.168 & 0.161 & 0.170 & 0.157 & 0.158 & 0.160 & 0.173 \\
        \bottomrule
    \end{tabular}
    }
    \caption{$p$-values of Diebold-Mariano test where the alternative hypothesis is that the factor model's mean squared error is less than the LSVVAR benchmark's. The values less than 0.05 are in bold.}
    \label{tab:dm_test_less_4var_second}
\end{table}

\begin{table}[H]
    \scriptsize
    \centering
    \setlength{\tabcolsep}{4pt} % Adjust column spacing
    \begin{tabular}{rlcccccccc}
        \toprule
        & & \multicolumn{8}{c}{Model vs. LSVVAR ($H_a$: Factor Model $<$ LSVVAR)} \\
        \cmidrule(lr){3-10}
        \# & Variable & DFSV & DFSVL & DFSVM & DFSVML & FSV & FSVL & FSVM & FSVML \\
        \midrule
        1 & GDPC1 & 0.930 & 0.901 & 0.910 & 0.896 & 0.479 & 0.499 & 0.511 & 0.491 \\
        2 & PCECTPI & 0.300 & 0.287 & 0.306 & 0.308 & 0.201 & 0.228 & 0.213 & 0.217 \\
        3 & FEDFUNDS & 0.251 & 0.171 & 0.201 & 0.205 & 0.223 & 0.234 & 0.293 & 0.269 \\
        4 & PCECC96 & 0.791 & 0.778 & 0.803 & 0.794 & 0.942 & 0.920 & 0.924 & 0.918 \\
        5 & CMRMTSPLx & 0.800 & 0.776 & 0.791 & 0.773 & 0.918 & 0.858 & 0.891 & 0.866 \\
        6 & INDPRO & 0.912 & 0.883 & 0.888 & 0.871 & 0.725 & 0.760 & 0.723 & 0.726 \\
        7 & CUMFNS & 0.881 & 0.849 & 0.875 & 0.863 & 0.457 & 0.465 & 0.450 & 0.443 \\
        8 & UNRATE & 0.838 & 0.827 & 0.842 & 0.845 & 0.915 & 0.894 & 0.906 & 0.904 \\
        9 & PAYEMS & 0.850 & 0.843 & 0.843 & 0.841 & 0.896 & 0.878 & 0.874 & 0.870 \\
        10 & CES0600000007 & 0.277 & 0.216 & 0.261 & 0.229 & 0.336 & 0.343 & 0.352 & 0.356 \\
        11 & CES0600000008 & 0.087 & 0.054 & 0.057 & 0.090 & 0.363 & 0.315 & 0.282 & 0.287 \\
        12 & WPSFD49207 & 0.325 & 0.318 & 0.308 & 0.304 & 0.477 & 0.478 & 0.481 & 0.469 \\
        13 & PPIACO & 0.225 & 0.190 & 0.218 & 0.267 & 0.687 & 0.685 & 0.703 & 0.701 \\
        14 & AMDMNOx & 0.742 & 0.692 & 0.713 & 0.686 & 0.885 & 0.888 & 0.882 & 0.889 \\
        15 & HOUST & 0.247 & 0.209 & 0.246 & 0.203 & 0.059 & 0.050 & 0.082 & 0.060 \\
        16 & S\&P 500 & 0.055 & 0.060 & 0.079 & 0.071 & \textbf{0.018} & \textbf{0.018} & \textbf{0.014} & \textbf{0.013} \\
        17 & EXUSUKx & \textbf{0.009} & \textbf{0.016} & \textbf{0.007} & \textbf{0.007} & \textbf{0.004} & \textbf{0.003} & \textbf{0.002} & \textbf{0.002} \\
        18 & TB3SMFFM & \textbf{0.005} & \textbf{0.005} & \textbf{0.003} & \textbf{0.003} & \textbf{0.008} & \textbf{0.010} & \textbf{0.011} & \textbf{0.011} \\
        19 & T5YFFM & \textbf{0.013} & \textbf{0.031} & \textbf{0.022} & \textbf{0.020} & \textbf{0.047} & 0.056 & \textbf{0.046} & 0.061 \\
        20 & AAAFFM & \textbf{0.049} & 0.055 & 0.086 & 0.068 & 0.114 & 0.133 & 0.105 & 0.124 \\
        \bottomrule
    \end{tabular}
    \caption{Diebold-Mariano test ($p$-values, 1-step ahead) for the full period. The alternative hypothesis is that the factor model's mean squared error is less than the LSVVAR benchmark's. $p$-values less than 0.05 are in bold, indicating that the factor model significantly outperforms the LSVVAR benchmark at the 5\% level.}
        \label{tab:dm_test_full_period_1step_less}
\end{table}

\begin{table}[H]

    \centering
    \scriptsize
    \setlength{\tabcolsep}{4pt} % Adjust column spacing
    \begin{tabular}{rlcccccccc}
        \toprule
        & & \multicolumn{8}{c}{Model vs. LSVVAR ($H_a$: Factor Model $<$ LSVVAR)} \\
        \cmidrule(lr){3-10}
        \# & Variable & DFSV & DFSVL & DFSVM & DFSVML & FSV & FSVL & FSVM & FSVML \\
        \midrule
        1 & GDPC1 & 0.507 & 0.475 & 0.480 & 0.403 & 0.593 & 0.530 & 0.591 & 0.466 \\
        2 & PCECTPI & 0.174 & 0.187 & 0.174 & 0.188 & 0.058 & \textbf{0.047} & 0.059 & 0.059 \\
        3 & FEDFUNDS & 0.307 & 0.232 & 0.276 & 0.270 & 0.248 & 0.213 & 0.283 & 0.250 \\
        4 & PCECC96 & 0.665 & 0.610 & 0.654 & 0.608 & 0.762 & 0.723 & 0.756 & 0.722 \\
        5 & CMRMTSPLx & 0.228 & 0.177 & 0.195 & 0.171 & 0.288 & 0.239 & 0.280 & 0.226 \\
        6 & INDPRO & 0.954 & 0.932 & 0.940 & 0.922 & 0.935 & 0.926 & 0.943 & 0.931 \\
        7 & CUMFNS & 0.959 & 0.945 & 0.956 & 0.952 & 0.975 & 0.971 & 0.974 & 0.971 \\
        8 & UNRATE & 0.870 & 0.860 & 0.868 & 0.871 & 0.913 & 0.895 & 0.910 & 0.905 \\
        9 & PAYEMS & 0.899 & 0.890 & 0.891 & 0.888 & 0.935 & 0.927 & 0.928 & 0.916 \\
        10 & CES0600000007 & 0.812 & 0.799 & 0.796 & 0.793 & 0.808 & 0.816 & 0.866 & 0.865 \\
        11 & CES0600000008 & 0.114 & 0.087 & 0.083 & 0.102 & 0.198 & 0.186 & 0.170 & 0.168 \\
        12 & WPSFD49207 & 0.081 & 0.099 & 0.074 & 0.086 & 0.078 & 0.070 & 0.071 & 0.065 \\
        13 & PPIACO & 0.073 & 0.092 & 0.072 & 0.089 & 0.078 & 0.063 & \textbf{0.047} & 0.054 \\
        14 & AMDMNOx & 0.279 & 0.270 & 0.273 & 0.255 & 0.243 & 0.240 & 0.238 & 0.251 \\
        15 & HOUST & \textbf{0.037} & \textbf{0.041} & \textbf{0.043} & \textbf{0.042} & \textbf{0.022} & \textbf{0.022} & \textbf{0.027} & \textbf{0.027} \\
        16 & S\&P 500 & \textbf{0.011} & \textbf{0.015} & \textbf{0.007} & \textbf{0.006} & \textbf{0.009} & \textbf{0.009} & \textbf{0.005} & \textbf{0.005} \\
        17 & EXUSUKx & 0.137 & 0.132 & 0.112 & 0.110 & 0.094 & 0.104 & 0.096 & 0.090 \\
        18 & TB3SMFFM & \textbf{0.019} & \textbf{0.019} & \textbf{0.018} & \textbf{0.016} & \textbf{0.015} & \textbf{0.016} & \textbf{0.016} & \textbf{0.014} \\
        19 & T5YFFM & 0.088 & 0.076 & 0.091 & 0.077 & 0.083 & 0.076 & 0.105 & 0.080 \\
        20 & AAAFFM & 0.198 & 0.155 & 0.196 & 0.165 & 0.185 & 0.161 & 0.231 & 0.191 \\
        \bottomrule
    \end{tabular}
    \caption{Diebold-Mariano test ($p$-values, 4-step ahead) for the full period. The alternative hypothesis is that the factor model's mean squared error is less than the LSVVAR benchmark's. $p$-values less than 0.05 are in bold, indicating that the factor model significantly outperforms the LSVVAR benchmark at the 5\% level.}
    \label{tab:dm_test_full_period_4step_less}
\end{table}

\newpage
\subsection{Model Confidence Set}
\label{subsec:mcs}
While the DM test focuses on pairwise comparisons against a specific benchmark, the MCS procedure identifies a set of models, denoted as $\widehat{\mathcal{M}}^*_{1-\alpha}$, that contains the best model(s) with a given confidence level $1-\alpha$. We set the significance level to $\alpha=0.1$, corresponding to a 90\% confidence level.
A model is included in the MCS if its MCS $p$-value is greater than $\alpha$. We calculate the $p$-values using the $T_{max}$ statistic with 5,000 block bootstrap replications. 
%\subsubsection{Summary of MCS Results}
%

Table \ref{tab:mcs_summary} summarizes the frequencies with which each model is included in the 90\% MCS across all 20 variables. The results indicate that the factor stochastic volatility models without factor dynamics in the observation equation (FSV and FSVL models) are slightly more robust, achieving high inclusion rates across all horizons for both the squared errors (SE) and the negative log predictive likelihood (LPL).
%
% --- Summary Table ---
\begin{table}[H]
\footnotesize
\centering
\begin{tabular}{lcccccccr}
\toprule
 Loss & \multicolumn{3}{c}{SE} & \multicolumn{3}{c}{LPL} \\
\cmidrule(lr){2-4} \cmidrule(lr){5-7}
Model & 1-step & 4-step & Subtotal & 1-step  & 4-step & Subtotal & Total & Rank\\
\midrule
DFSV & 20 & 19 & 39 & 19 & 18  & 37 & 76 & 6 \\
DFSVL & 20 & 20 & 40 & 20 & 17 & 37 & 77 & 3\\
DFSVM & 19 & 20 & 39 & 19 & 18 & 37 & 76 & 6 \\
DFSVML & 20 & 20 & 40 & 20 & 17 & 37 & 77 & 3\\
FSV & 20 & 20 & 40 & 19 & 20 & 39 & 79 & 2\\
FSVL & 20 & 20 & 40 & 20 & 20 & 40 & 80 & 1\\
FSVM & 19 & 19 & 38 & 15 & 20 & 35 & 73 & 8\\
FSVML & 20 & 19 & 39 & 18 & 20 & 38 & 77 & 3\\
LSVVAR & 17 & 18 & 35 & 19 & 18 & 37 & 72 & 9\\
\bottomrule
\end{tabular}
\caption{Frequency of Inclusion in the 90\% Model Confidence Set. Numbers indicate how many variables (out of 20) the model was included in the MCS.}
\label{tab:mcs_summary}
\end{table}
\noindent
Further, Tables \ref{tab:mcs_se_1} through \ref{tab:mcs_lpl_4} report the detailed MCS $p$-values\footnote{ 
In the tables, entries with ``$-$'' indicate that the model has been eliminated from the MCS ($p$-value $< 0.10$), signifying it is statistically inferior to the surviving models.}.
Those MCS results reinforce and extend the findings from the DM tests as follows:
\begin{itemize}
    \item Real activity variables such as \#1 and \#6, the LSVVAR benchmark typically remains in the MCS alongside the factor models (often with $p$-values of 1.000) with respect to the SE and the negative LPL. This confirms that for standard macroeconomic aggregates, the predictive accuracy of the factor models is statistically indistinguishable from the LSVVAR; they perform equally well.
    
    \item For financial market variables and the spread variables,  the MCS analysis highlights the superiority of factor models in financial sectors more clearly than the DM test as shown in Table \ref{tab:mcs_se_1} (1-step ahead SE). The LSVVAR model is explicitly excluded from the MCS (indicated by ``$-$'') for \#16, \#17 and \#18, while the FSV and DFSV models remain. This exclusion implies that LSVVAR is not just ``not significantly beaten'' (as a DM test might obscure if variance is high), but is statistically inferior to the best set of models.
    
    \item Regarding horizon differences, the simpler FSV models seems to be more robust overall at the 1-step ahead SE, whereas the DFSV models regain competitiveness at the 4-step ahead horizon SE (though the DFSV models are found to be more robust at the 1-step ahead negative LPL). This suggests that while simple factor structures handle short-term volatility well (especially post-structural breaks), factor dynamics become increasingly relevant for medium-term forecasting with respect to the SE. 
\end{itemize}
\noindent
These findings are visually supported in more detail by the time series plots of Cumulative Squared Forecast Error (CSFE) in Figures \ref{fig:result_csfe_1step} and \ref{fig:result_csfe_4step}, and Cumulative Log Predictive Likelihood (CLPL) in Figures \ref{fig:result_clpl_1step} and \ref{fig:result_clpl_4step}, where the superior performance of factor models in specific sectors is discernible against the LSVVAR baseline.

\clearpage
% --- Detailed Tables SE ---
\begin{table}[H]
\small
\centering
\setlength{\tabcolsep}{3pt}
\begin{tabular}{rlccccccccc}
\toprule
\# & Variable & DFSV & DFSVL & DFSVM & DFSVML & FSV & FSVL & FSVM & FSVML & LSVVAR \\
\midrule
1 & GDPC1 & 0.624 & 0.677 & 0.667 & 0.677 & \textbf{1.000} & \textbf{1.000} & \textbf{1.000} & \textbf{1.000} & \textbf{1.000} \\
2 & PCECTPI & 0.972 & \textbf{1.000} & 0.975 & 0.982 & \textbf{1.000} & \textbf{1.000} & \textbf{1.000} & \textbf{1.000} & 0.505 \\
3 & FEDFUNDS & \textbf{1.000} & \textbf{1.000} & \textbf{1.000} & \textbf{1.000} & \textbf{1.000} & \textbf{1.000} & 0.955 & 0.998 & 0.895 \\
4 & PCECC96 & 0.759 & 0.810 & 0.690 & 0.723 & \textbf{1.000} & \textbf{1.000} & 0.830 & 0.817 & \textbf{1.000} \\
5 & CMRMTSPLx & 0.561 & 0.579 & 0.572 & 0.584 & \textbf{1.000} & \textbf{1.000} & \textbf{1.000} & \textbf{1.000} & \textbf{1.000} \\
6 & INDPRO & 0.543 & 0.593 & 0.592 & 0.610 & \textbf{1.000} & \textbf{1.000} & \textbf{1.000} & \textbf{1.000} & \textbf{1.000} \\
7 & CUMFNS & 0.603 & 0.646 & 0.636 & 0.646 & \textbf{1.000} & \textbf{1.000} & \textbf{1.000} & \textbf{1.000} & \textbf{1.000} \\
8 & UNRATE & 0.670 & 0.655 & 0.651 & 0.666 & \textbf{1.000} & 0.156 & 0.139 & 0.131 & \textbf{1.000} \\
9 & PAYEMS & 0.613 & 0.627 & 0.608 & 0.612 & \textbf{1.000} & 0.253 & $-$ & 0.131 & \textbf{1.000} \\
10 & CES0600000007 & 0.722 & 0.835 & 0.800 & 0.842 & \textbf{1.000} & \textbf{1.000} & \textbf{1.000} & \textbf{1.000} & 0.641 \\
11 & CES0600000008 & \textbf{1.000} & \textbf{1.000} & \textbf{1.000} & \textbf{1.000} & 0.583 & 0.635 & 0.900 & 0.799 & 0.450 \\
12 & WPSFD49207 & \textbf{1.000} & \textbf{1.000} & \textbf{1.000} & \textbf{1.000} & 0.814 & 0.703 & 0.760 & 0.763 & 0.900 \\
13 & PPIACO & \textbf{1.000} & \textbf{1.000} & \textbf{1.000} & \textbf{1.000} & 0.254 & 0.106 & 0.116 & 0.102 & 0.984 \\
14 & AMDMNOx & \textbf{1.000} & \textbf{1.000} & \textbf{1.000} & \textbf{1.000} & 0.859 & 0.879 & 0.879 & 0.860 & \textbf{1.000} \\
15 & HOUST & 0.997 & \textbf{1.000} & 0.991 & \textbf{1.000} & \textbf{1.000} & \textbf{1.000} & \textbf{1.000} & \textbf{1.000} & 0.457 \\
16 & S\&P 500 & \textbf{1.000} & \textbf{1.000} & 0.362 & 0.653 & \textbf{1.000} & \textbf{1.000} & \textbf{1.000} & \textbf{1.000} & $-$ \\
17 & EXUSUKx & 0.878 & 0.105 & $-$ & 0.114 & \textbf{1.000} & \textbf{1.000} & \textbf{1.000} & \textbf{1.000} & $-$ \\
18 & TB3SMFFM & \textbf{1.000} & \textbf{1.000} & \textbf{1.000} & \textbf{1.000} & 0.599 & 0.699 & 0.359 & 0.417 & $-$ \\
19 & T5YFFM & \textbf{1.000} & \textbf{1.000} & \textbf{1.000} & \textbf{1.000} & 0.986 & \textbf{1.000} & \textbf{1.000} & 0.969 & 0.388 \\
20 & AAAFFM & \textbf{1.000} & \textbf{1.000} & \textbf{1.000} & \textbf{1.000} & \textbf{1.000} & 0.947 & \textbf{1.000} & \textbf{1.000} & 0.488 \\
\bottomrule
\end{tabular}
\caption{MCS $p$-values for squared error (SE) based on $T_{max}$ (1-step ahead). Entries with `$-$' indicate exclusion from $\widehat{\mathcal{M}}^*_{90\%}$.}
\label{tab:mcs_se_1}
\end{table}

\clearpage
% --- Detailed Tables LPL ---
\begin{table}[H]
\small
\centering
\setlength{\tabcolsep}{3pt}
\begin{tabular}{rlccccccccc}
\toprule
\# & Variable & DFSV & DFSVL & DFSVM & DFSVML & FSV & FSVL & FSVM & FSVML & LSVVAR \\
\midrule
1 & GDPC1 & 0.995 & \textbf{1.000} & \textbf{1.000} & \textbf{1.000} & 0.143 & 0.231 & $-$ & $-$ & \textbf{1.000} \\
2 & PCECTPI & 0.901 & 0.923 & 0.974 & 0.929 & \textbf{1.000} & \textbf{1.000} & \textbf{1.000} & \textbf{1.000} & \textbf{1.000} \\
3 & FEDFUNDS & 0.912 & \textbf{1.000} & 0.417 & 0.967 & 0.416 & \textbf{1.000} & 0.236 & 0.994 & \textbf{1.000} \\
4 & PCECC96 & \textbf{1.000} & \textbf{1.000} & \textbf{1.000} & \textbf{1.000} & 0.245 & 0.254 & $-$ & $-$ & 0.686 \\
5 & CMRMTSPLx & 0.935 & \textbf{1.000} & \textbf{1.000} & \textbf{1.000} & 0.106 & 0.819 & $-$ & 0.612 & \textbf{1.000} \\
6 & INDPRO & $-$ & \textbf{1.000} & 0.645 & \textbf{1.000} & $-$ & 0.160 & $-$ & 0.188 & \textbf{1.000} \\
7 & CUMFNS & \textbf{1.000} & \textbf{1.000} & \textbf{1.000} & \textbf{1.000} & 0.664 & 0.946 & 0.185 & 0.843 & \textbf{1.000} \\
8 & NRATE & \textbf{1.000} & 0.549 & 0.585 & \textbf{1.000} & \textbf{1.000} & \textbf{1.000} & \textbf{1.000} & \textbf{1.000} & 0.585 \\
9 & PAYEMS & \textbf{1.000} & 0.556 & 0.553 & 0.556 & \textbf{1.000} & \textbf{1.000} & \textbf{1.000} & \textbf{1.000} & 0.488 \\
10 & CES0600000007 & \textbf{1.000} & \textbf{1.000} & \textbf{1.000} & \textbf{1.000} & 0.839 & 0.975 & 0.692 & 0.972 & \textbf{1.000} \\
11 & CES0600000008 & \textbf{1.000} & \textbf{1.000} & \textbf{1.000} & \textbf{1.000} & 0.495 & 0.179 & 0.805 & 0.298 & 0.237 \\
12 & WPSFD49207 & \textbf{1.000} & \textbf{1.000} & \textbf{1.000} & \textbf{1.000} & \textbf{1.000} & \textbf{1.000} & \textbf{1.000} & \textbf{1.000} & 0.518 \\
13 & PPIACO & \textbf{1.000} & \textbf{1.000} & \textbf{1.000} & \textbf{1.000} & 0.751 & 0.874 & 0.665 & 0.751 & 0.315 \\
14 & AMDMNOx & \textbf{1.000} & \textbf{1.000} & \textbf{1.000} & \textbf{1.000} & 0.390 & 0.463 & $-$ & 0.472 & \textbf{1.000} \\
15 & HOUST & 0.872 & \textbf{1.000} & 0.907 & \textbf{1.000} & \textbf{1.000} & \textbf{1.000} & \textbf{1.000} & \textbf{1.000} & 0.376 \\
16 & S\&P 500 & \textbf{1.000} & \textbf{1.000} & 0.976 & \textbf{1.000} & \textbf{1.000} & \textbf{1.000} & \textbf{1.000} & \textbf{1.000} & 0.282 \\
17 & EXUSUKx & \textbf{1.000} & 0.400 & $-$ & 0.356 & \textbf{1.000} & \textbf{1.000} & \textbf{1.000} & \textbf{1.000} & 0.985 \\
18 & TB3SMFFM & 0.768 & 0.960 & 0.907 & \textbf{1.000} & \textbf{1.000} & \textbf{1.000} & 0.886 & \textbf{1.000} & $-$ \\
19 & T5YFFM & \textbf{1.000} & \textbf{1.000} & \textbf{1.000} & \textbf{1.000} & 0.697 & \textbf{1.000} & \textbf{1.000} & 0.895 & 0.150 \\
20 & AAAFFM & \textbf{1.000} & \textbf{1.000} & \textbf{1.000} & \textbf{1.000} & 0.329 & 0.985 & 0.655 & \textbf{1.000} & \textbf{1.000} \\
\bottomrule
\end{tabular}
\caption{MCS $p$-values for the negative log predictive likelihood (LPL) based on $T_{max}$ (1-step ahead). Entries with `$-$' indicate exclusion from $\widehat{\mathcal{M}}^*_{90\%}$.}
\label{tab:mcs_lpl_1}
\end{table}

\begin{table}[H]
\small
\centering
\setlength{\tabcolsep}{3pt}
\begin{tabular}{rlccccccccc}
\toprule
\# & Variable & DFSV & DFSVL & DFSVM & DFSVML & FSV & FSVL & FSVM & FSVML & LSVVAR \\
\midrule
1 & GDPC1 & \textbf{1.000} & \textbf{1.000} & \textbf{1.000} & \textbf{1.000} & 0.536 & 0.984 & 0.355 & \textbf{1.000} & \textbf{1.000} \\
2 & PCECTPI & 0.933 & 0.852 & 0.906 & 0.745 & \textbf{1.000} & \textbf{1.000} & \textbf{1.000} & \textbf{1.000} & 0.372 \\
3 & FEDFUNDS & 0.648 & \textbf{1.000} & \textbf{1.000} & \textbf{1.000} & \textbf{1.000} & \textbf{1.000} & \textbf{1.000} & \textbf{1.000} & 0.773 \\
4 & PCECC96 & \textbf{1.000} & \textbf{1.000} & \textbf{1.000} & \textbf{1.000} & 0.408 & 0.497 & 0.153 & 0.538 & \textbf{1.000} \\
5 & CMRMTSPLx & \textbf{1.000} & \textbf{1.000} & \textbf{1.000} & \textbf{1.000} & 0.820 & \textbf{1.000} & 0.951 & \textbf{1.000} & 0.716 \\
6 & INDPRO & $-$ & 0.986 & 0.402 & \textbf{1.000} & 0.254 & 0.479 & 0.127 & 0.379 & \textbf{1.000} \\
7 & CUMFNS & 0.403 & 0.695 & 0.567 & 0.690 & \textbf{1.000} & \textbf{1.000} & 0.942 & \textbf{1.000} & \textbf{1.000} \\
8 & UNRATE & 0.629 & 0.647 & 0.631 & 0.643 & \textbf{1.000} & 0.116 & 0.110 & $-$ & \textbf{1.000} \\
9 & PAYEMS & 0.653 & 0.701 & 0.653 & 0.670 & \textbf{1.000} & 0.710 & $-$ & 0.140 & \textbf{1.000} \\
10 & CES0600000007 & 0.850 & 0.924 & 0.948 & 0.967 & \textbf{1.000} & \textbf{1.000} & 0.938 & 0.968 & \textbf{1.000} \\
11 & CES0600000008 & \textbf{1.000} & \textbf{1.000} & \textbf{1.000} & \textbf{1.000} & \textbf{1.000} & \textbf{1.000} & \textbf{1.000} & \textbf{1.000} & 0.472 \\
12 & WPSFD49207 & \textbf{1.000} & \textbf{1.000} & \textbf{1.000} & \textbf{1.000} & \textbf{1.000} & \textbf{1.000} & \textbf{1.000} & \textbf{1.000} & 0.334 \\
13 & PPIACO & \textbf{1.000} & \textbf{1.000} & \textbf{1.000} & \textbf{1.000} & \textbf{1.000} & \textbf{1.000} & \textbf{1.000} & \textbf{1.000} & 0.262 \\
14 & AMDMNOx & \textbf{1.000} & \textbf{1.000} & \textbf{1.000} & \textbf{1.000} & \textbf{1.000} & \textbf{1.000} & \textbf{1.000} & \textbf{1.000} & 0.743 \\
15 & HOUST & \textbf{1.000} & \textbf{1.000} & \textbf{1.000} & \textbf{1.000} & \textbf{1.000} & \textbf{1.000} & 0.953 & \textbf{1.000} & 0.171 \\
16 & S\&P 500 & \textbf{1.000} & 0.920 & 0.876 & 0.980 & \textbf{1.000} & \textbf{1.000} & 0.936 & \textbf{1.000} & $-$ \\
17 & EXUSUKx & \textbf{1.000} & \textbf{1.000} & \textbf{1.000} & \textbf{1.000} & \textbf{1.000} & \textbf{1.000} & \textbf{1.000} & \textbf{1.000} & 0.403 \\
18 & TB3SMFFM & 0.539 & \textbf{1.000} & \textbf{1.000} & \textbf{1.000} & 0.652 & \textbf{1.000} & 0.765 & 0.932 & $-$ \\
19 & T5YFFM & \textbf{1.000} & \textbf{1.000} & \textbf{1.000} & \textbf{1.000} & \textbf{1.000} & \textbf{1.000} & 0.874 & \textbf{1.000} & 0.413 \\
20 & AAAFFM & \textbf{1.000} & \textbf{1.000} & \textbf{1.000} & \textbf{1.000} & \textbf{1.000} & \textbf{1.000} & 0.959 & \textbf{1.000} & 0.682 \\
\bottomrule
\end{tabular}
\caption{MCS $p$-values for squared error (SE) based on $T_{max}$ (4-step ahead). Entries with `$-$' indicate exclusion from $\widehat{\mathcal{M}}^*_{90\%}$.}
\label{tab:mcs_se_4}
\end{table}

\begin{table}[H]
\small
\centering
\setlength{\tabcolsep}{3pt}
\begin{tabular}{rlccccccccc}
\toprule
\# & Variable & DFSV & DFSVL & DFSVM & DFSVML & FSV & FSVL & FSVM & FSVML & LSVVAR \\
\midrule
1 & GDPC1 & \textbf{1.000} & \textbf{1.000} & 0.745 & \textbf{1.000} & \textbf{1.000} & \textbf{1.000} & \textbf{1.000} & \textbf{1.000} & 0.829 \\
2 & PCECTPI & 0.137 & $-$ & 0.154 & $-$ & \textbf{1.000} & \textbf{1.000} & \textbf{1.000} & \textbf{1.000} & 0.929 \\
3 & FEDFUNDS & \textbf{1.000} & \textbf{1.000} & 0.978 & 0.986 & 0.996 & \textbf{1.000} & 0.404 & \textbf{1.000} & \textbf{1.000} \\
4 & PCECC96 & \textbf{1.000} & \textbf{1.000} & \textbf{1.000} & \textbf{1.000} & \textbf{1.000} & 0.978 & 0.906 & \textbf{1.000} & 0.645 \\
5 & CMRMTSPLx & 0.996 & \textbf{1.000} & 0.730 & \textbf{1.000} & \textbf{1.000} & \textbf{1.000} & \textbf{1.000} & \textbf{1.000} & 0.718 \\
6 & INDPRO & 0.357 & \textbf{1.000} & 0.955 & \textbf{1.000} & 0.985 & \textbf{1.000} & \textbf{1.000} & \textbf{1.000} & \textbf{1.000} \\
7 & CUMFNS & 0.463 & \textbf{1.000} & 0.409 & \textbf{1.000} & 0.607 & 0.998 & 0.744 & \textbf{1.000} & \textbf{1.000} \\
8 & UNRATE & \textbf{1.000} & \textbf{1.000} & \textbf{1.000} & \textbf{1.000} & \textbf{1.000} & \textbf{1.000} & \textbf{1.000} & \textbf{1.000} & 0.440 \\
9 & PAYEMS & \textbf{1.000} & \textbf{1.000} & \textbf{1.000} & \textbf{1.000} & \textbf{1.000} & \textbf{1.000} & \textbf{1.000} & \textbf{1.000} & 0.394 \\
10 & CES0600000007 & 0.574 & \textbf{1.000} & 0.977 & \textbf{1.000} & \textbf{1.000} & \textbf{1.000} & \textbf{1.000} & \textbf{1.000} & 0.872 \\
11 & CES0600000008 & $-$ & $-$ & $-$ & $-$ & \textbf{1.000} & 0.187 & 0.996 & 0.112 & \textbf{1.000} \\
12 & WPSFD49207 & \textbf{1.000} & 0.891 & \textbf{1.000} & 0.996 & \textbf{1.000} & \textbf{1.000} & \textbf{1.000} & \textbf{1.000} & 0.601 \\
13 & PPIACO & \textbf{1.000} & \textbf{1.000} & \textbf{1.000} & \textbf{1.000} & \textbf{1.000} & \textbf{1.000} & \textbf{1.000} & \textbf{1.000} & 0.615 \\
14 & AMDMNOx & 0.659 & \textbf{1.000} & 0.641 & 0.888 & \textbf{1.000} & \textbf{1.000} & \textbf{1.000} & \textbf{1.000} & 0.940 \\
15 & HOUST & 0.917 & \textbf{1.000} & 0.558 & \textbf{1.000} & \textbf{1.000} & \textbf{1.000} & 0.944 & \textbf{1.000} & $-$ \\
16 & S\&P 500 & \textbf{1.000} & \textbf{1.000} & 0.952 & \textbf{1.000} & \textbf{1.000} & \textbf{1.000} & \textbf{1.000} & \textbf{1.000} & 0.258 \\
17 & EXUSUKx & \textbf{1.000} & 0.937 & \textbf{1.000} & \textbf{1.000} & \textbf{1.000} & \textbf{1.000} & 0.996 & \textbf{1.000} & \textbf{1.000} \\
18 & TB3SMFFM & $-$ & $-$ & $-$ & $-$ & \textbf{1.000} & \textbf{1.000} & 0.294 & \textbf{1.000} & $-$ \\
19 & T5YFFM & \textbf{1.000} & \textbf{1.000} & 0.881 & \textbf{1.000} & \textbf{1.000} & \textbf{1.000} & 0.511 & \textbf{1.000} & 0.971 \\
20 & AAAFFM & \textbf{1.000} & \textbf{1.000} & 0.984 & \textbf{1.000} & 0.949 & \textbf{1.000} & 0.175 & \textbf{1.000} & \textbf{1.000} \\
\bottomrule
\end{tabular}
\caption{MCS $p$-values for the negative log predictive likelihood (LPL) based on $T_{max}$ (4-step ahead). Entries with `$-$' indicate exclusion from $\widehat{\mathcal{M}}^*_{90\%}$.}

\label{tab:mcs_lpl_4}
\end{table}

\end{document}